\pdfoutput=1
\documentclass[preprint,11pt,nopreprintline]{elsarticle}
\usepackage[a4paper,margin=2.5cm]{geometry}
\usepackage[T1]{fontenc}
\usepackage{textcomp}
\usepackage{amsmath,amssymb}
\usepackage{bm}
\usepackage{upgreek}
\usepackage{booktabs}
\usepackage{longtable}
\usepackage{array}
\usepackage{graphicx}
\usepackage{url}
\usepackage{xurl}
\usepackage{listings}
\usepackage{tikz}
\usetikzlibrary{arrows.meta,positioning,fit,backgrounds,calc}
\usepackage{xcolor}
\usepackage[colorlinks=true,allcolors=blue]{hyperref}
\newcommand{\costPerPointFile}{99}

\newcommand{\costRatio}{53}
\newcommand{\costShareLargest}{0.028}
\newcommand{\costShareFile}{0.042}
\newcommand{\costShareClass}{0.007}
\newcommand{\costShareFileGPU}{0.061}
\newcommand{\costShareLargestGPU}{0.124}
\newcommand{\costEngineSpeedup}{2.1}
\newcommand{\costShareSilane}{0.083}
\newcommand{\costShareWorstHost}{0.09}
\newcommand{\costShareWorst}{0.13}

\lstdefinestyle{code}{
  basicstyle=\ttfamily\scriptsize, backgroundcolor=\color{black!3},
  frame=single, columns=fullflexible, keepspaces=true,
  showstringspaces=false, breaklines=false}
\lstdefinestyle{yaml}{
  basicstyle=\ttfamily\small, backgroundcolor=\color{black!3},
  frame=single, columns=fullflexible, keepspaces=true,
  showstringspaces=false, breaklines=true}

\newenvironment{unbroken}%
  {\par\smallskip\noindent\begin{minipage}{\linewidth}}%
  {\end{minipage}\par}

\begin{document}

\begin{frontmatter}

\title{Surface chemistry as data for level-set simulation of etching and
deposition}

\author[iue]{Lado Filipovic\corref{cor1}}
\ead{filipovic@iue.tuwien.ac.at}
\author[iue]{Tobias Reiter}
\author[iue]{Philipp Haslhofer}
\cortext[cor1]{Corresponding author.}
\affiliation[iue]{organization={CDL for Multi-Scale Process Modeling of
  Semiconductor Devices and Sensors, Institute for Microelectronics, TU Wien},
  addressline={Gu\ss{}hausstra\ss{}e 27-29},
  postcode={1040},
  city={Vienna},
  country={Austria}}

\begin{abstract}
Level-set simulators of etching and deposition move the surface with a velocity
that depends on adsorbate coverages. These coverages are usually computed from balances
solved and coded by hand for each chemistry, whereas voxel-based Monte Carlo
profile simulators read the reaction mechanism as data. For the open-source level-set simulator ViennaPS, this work presents a surface model whose rate laws, Jacobian and velocity are derived from a reaction file and evaluated with ray-traced fluxes. Steady-state coverages are obtained with Newton's method and transient
coverages with a conservative integrator. One compiled model is applied to eight chemistries for chemical vapour deposition, selective epitaxy, plasma etching,
sputtering and atomic layer deposition. It reproduces a hand-written silicon etch
model to seven significant digits at a single point and within the Monte Carlo
noise in a masked trench. In a high-aspect-ratio cavity, alumina conformality is determined by a trimethylaluminium reaction probability within the measured range.
\end{abstract}


\end{frontmatter}

\section*{Program summary}
\begin{description}
  \setlength{\itemsep}{1pt}\setlength{\parskip}{0pt}
  \item[Program title:] ViennaChem, with the \texttt{SurfaceChemistry} model
    of ViennaPS
  \item[Developer's repository link:] \sloppy
    \url{https://github.com/ViennaTools/ViennaChem} and
    \url{https://github.com/ViennaTools/ViennaPS/tree/cpc-deposit}
  \item[Licensing provisions:] GPLv3
  \item[Programming language:] Python 3 for the front end, C++20 for the surface model, CUDA for the optional GPU flux engine
  \item[Supplementary material:] None
  \item[Nature of problem:] Feature-scale process simulation on a level-set
    surface needs a surface model that converts the particle fluxes arriving at
    each point of the moving surface into a local interface velocity. The model
    has to track the coverages of the adsorbed species that mediate the
    reactions. Each coverage is determined by a balance between the steps that
    create and consume that species. In level-set simulators, this balance is usually solved analytically for each chemistry and the closed form is
    implemented in code. Closed forms exist only for a few coupled species, the
    chemistry is fixed when the code is compiled, and processes without a steady
    state are excluded.
  \item[Solution method:] The chemistry is written as a reaction file that
    lists the species with their phases and the surface reactions with their
    rate constants. The free-site exponents, mass-action rate laws,
    stoichiometry, analytic Jacobian and surface velocity are derived from the
    reactions and evaluated at each surface point by a generic surface model.
    Where the surface equilibrates quickly, the coverage balance is solved by a
    damped Newton iteration that keeps the coverages between zero and one.
    Where it does not, the same balance is integrated in time by a conservative
    modified Patankar-Euler scheme with adaptive sub-steps. The velocity is computed from the rates of the reactions that create or consume a solid and is used to advect the
    level set.
  \item[Additional comments including restrictions and unusual features:]
    ViennaPS depends on ViennaLS, ViennaRay, ViennaCS, ViennaCore, VTK and Embree, and optionally on OptiX for the GPU flux engine. The front end uses PyYAML and
    SymPy, and the Python evaluator uses only the standard library. In the ray
    tracer, a traced species is absorbed at a surface hit with the probability
    that it reacts there according to its rate laws. Its source distribution is
    a cosine. Rate constants have the modified Arrhenius form and depend on temperature and material only, so barriers that vary with coverage are not supported in this version. The
    rate laws are mean field in the coverages. The rate of a step that consumes
    two adsorbates of one species is therefore proportional to the square of that coverage, unless an \texttt{order} entry sets another exponent. Correlations between neighbouring sites, such as isolated
    unpaired sites at low coverage, are not represented. No species flows during
    a purge in a cyclic process, and only thermal steps proceed.
\end{description}

\section{Introduction}
\label{sec:intro}

Feature-scale process simulation predicts how the topography of a single
device feature evolves during etching or deposition. The surface is commonly
represented implicitly as the zero level set of a signed distance function that
evolves under a Hamilton-Jacobi equation~\cite{Osher1988,Osher2003}. The particle fluxes that reach the surface can be computed by Monte Carlo ray tracing, because at typical process pressures the mean free path is long compared with the feature size and transport inside the feature is therefore nearly ballistic~\cite{Cale1991}. The level-set method and Monte Carlo ray tracing are both available in open-source codes, among them ViennaPS~\cite{Reiter2025}, on which this work builds.

\begin{figure}[!htbp]
\centering
\begin{minipage}{\linewidth}
\noindent{\small (a)}\par\smallskip
\centering
\begin{tikzpicture}[font=\small, >=Stealth,
    box/.style={rounded corners=3pt, align=center, inner sep=4pt,
                minimum height=10mm, text width=27mm, draw, thick},
    geo/.style={box, draw=geoC, fill=geoF},
    tra/.style={box, draw=traC, fill=traF},
    che/.style={box, draw=cheC, fill=cheF},
    hot/.style={box, draw=cheC, fill=cheF, very thick},
    ar/.style={->, thick, black!65},
    lb/.style={font=\scriptsize, black!65, inner sep=3pt}]

  \definecolor{geoC}{RGB}{40,105,175}   \definecolor{geoF}{RGB}{228,239,250}
  \definecolor{traC}{RGB}{198,105,35}   \definecolor{traF}{RGB}{253,238,226}
  \definecolor{cheC}{RGB}{20,120,75}    \definecolor{cheF}{RGB}{225,243,232}

  \node[geo] (g1) at (0,0)      {level-set\\surface};
  \node[geo] (g2) at (4.2,0)    {surface points\\{\scriptsize normal, material}};
  \node[tra] (t1) at (8.4,0)    {ray tracing};
  \node[tra] (t2) at (8.4,-2.2) {fluxes $\Gamma_g$\\at each point};
  \node[hot] (c1) at (4.2,-2.2) {\textbf{surface model}\\$\theta_i,\ v_n$};

  \node[che] (f1) at (0,-4.7)   {reaction file\\{\scriptsize YAML}};
  \node[che] (m1) at (4.2,-4.7) {mechanism data\\{\scriptsize arrays}};
  \node[che] (e1) at (8.4,-4.7) {Python evaluator\\{\scriptsize reference}};

  \draw[ar] (g1) -- (g2) node[lb,midway,above] {sample};
  \draw[ar] (g2) -- (t1);
  \draw[ar] (t1) -- (t2);
  \draw[ar] (t2) -- (c1);
  \draw[ar] (c1.west) -- (0,-2.2) -- (g1.south);
  \node[lb,anchor=north] at (1.5,-2.32) {advect by $v_n$};
  \draw[ar] (f1) -- (m1) node[lb,midway,above] {compile};
  \draw[ar] (m1) -- (c1) node[lb,pos=0.22,right] {read};
  \draw[ar] (m1) -- (e1) node[lb,midway,above] {read};

  \draw[black!40,dashed] (-1.6,-3.45) -- (10.0,-3.45);
  \node[font=\footnotesize\bfseries,geoC,anchor=south east] at (10.0,-3.42)
       {ViennaPS};
  \node[font=\footnotesize\bfseries,cheC,anchor=north east] at (10.0,-3.48)
       {ViennaChem};

  \node[font=\scriptsize,anchor=west] at (-1.5,-5.75)
       {\textcolor{geoC}{\rule{6pt}{6pt}}~geometry};
  \node[font=\scriptsize,anchor=west] at (2.4,-5.75)
       {\textcolor{cheC}{\rule{6pt}{6pt}}~surface chemistry};
  \node[font=\scriptsize,anchor=west] at (6.6,-5.75)
       {\textcolor{traC}{\rule{6pt}{6pt}}~particle transport};
\end{tikzpicture}
\end{minipage}

\bigskip
\begin{minipage}{\linewidth}
\noindent{\small (b)}\par\smallskip
\centering
\begin{tikzpicture}[
    font=\small, >=Stealth,
    b/.style={draw=black!55, rounded corners=2pt, align=center,
              inner sep=4pt, minimum height=8mm, text width=20mm},
    hb/.style={b, draw=hum, very thick, fill=white},
    tb/.style={b, draw=tool, very thick, fill=white},
    lbl/.style={font=\scriptsize, midway, below, inner sep=3.5pt},
    hlbl/.style={lbl, hum}, tlbl/.style={lbl, tool},
    ha/.style={->, hum, thick}, ta/.style={->, tool, thick}]
  \definecolor{hum}{RGB}{176,42,42}
  \definecolor{tool}{RGB}{20,110,70}

  \node[b, fill=white] (ch) at (0.7,-0.35) {chemistry};

  \node[font=\bfseries\footnotesize,anchor=west] at (3.1,1.95) {manual derivation};
  \node[hb] (c2) at (4.6,1.05)  {closed form $\theta(\Gamma)$};
  \node[hb] (c3) at (8.2,1.05)  {model source};
  \node[tb] (c4) at (11.8,1.05) {simulator};
  \draw[ha] (ch) -- (c2);
  \node[hum,font=\scriptsize,anchor=south east] at (2.75,0.28) {solve};
  \draw[ha] (c2) -- (c3) node[hlbl] {code};
  \draw[ta] (c3) -- (c4) node[tlbl] {build};
  \node[font=\footnotesize,anchor=west] at (13.05,1.05) {per chemistry};

  \node[font=\bfseries\footnotesize,anchor=west] at (3.1,-0.85) {derivation from chemistry};
  \node[hb] (r1) at (4.6,-1.75)  {reaction file};
  \node[tb] (r2) at (8.2,-1.75)  {compiled data};
  \node[tb] (r3) at (11.8,-1.75) {one\\simulator};
  \draw[ha] (ch) -- (r1);
  \node[hum,font=\scriptsize,anchor=north east] at (2.75,-1.00) {write};
  \draw[ta] (r1) -- (r2) node[tlbl] {compile};
  \draw[ta] (r2) -- (r3) node[tlbl] {read};
  \node[font=\footnotesize,anchor=west] at (13.05,-1.75) {any chemistry};

  \begin{scope}[on background layer]
    \foreach \n in {c2,c3,c4,r1,r2}{
      \draw[black!25,rounded corners=2pt,fill=white]
        ([shift={(0.10,0.22)}]\n.south west) rectangle
        ([shift={(0.10,0.22)}]\n.north east);
      \draw[black!40,rounded corners=2pt,fill=white]
        ([shift={(0.05,0.11)}]\n.south west) rectangle
        ([shift={(0.05,0.11)}]\n.north east);}
  \end{scope}

  \draw[black!30] (3.0,-0.30) -- (14.9,-0.30);

  \node[hum,font=\scriptsize,anchor=west] at (2.60,-3.15)
    {\textcolor{hum}{\rule{7pt}{7pt}}~written by a person, per chemistry};
  \node[tool,font=\scriptsize,anchor=west] at (9.00,-3.15)
    {\textcolor{tool}{\rule{7pt}{7pt}}~produced by the tools};
\end{tikzpicture}
\end{minipage}
\caption{Surface chemistry supplied as data to a level-set simulation.
(a) One time step, coloured by geometry, transport and surface chemistry. The level set is sampled onto surface points, and the incident fluxes at each point are computed by ray tracing. From these fluxes, the surface model computes the normal velocity $v_n$ that advects the level set. The surface model evaluates the data that ViennaChem compiles once from a reaction file. The geometry and the ray tracer are those of ViennaPS, and the particle types and their absorption probabilities are derived from these data. As a reference for the C++ surface model, a Python evaluator solves the coverage equations from these data with independent code. (b) Two routes from a chemistry to a level-set
simulation, by manual derivation (top) and from a reaction file (bottom).
Stacked outlines mark what has to be produced again for each new chemistry.
Solving the coverage balance analytically and implementing the result~\cite{BelenSF6O2,Bobinac2023} requires a new derivation and new source code, and the simulator has to be recompiled. With the mechanism supplied as data, only the reaction file and the arrays compiled from it are new, and the compiled simulator is the same for all chemistries.}
\label{fig:concept}
\end{figure}
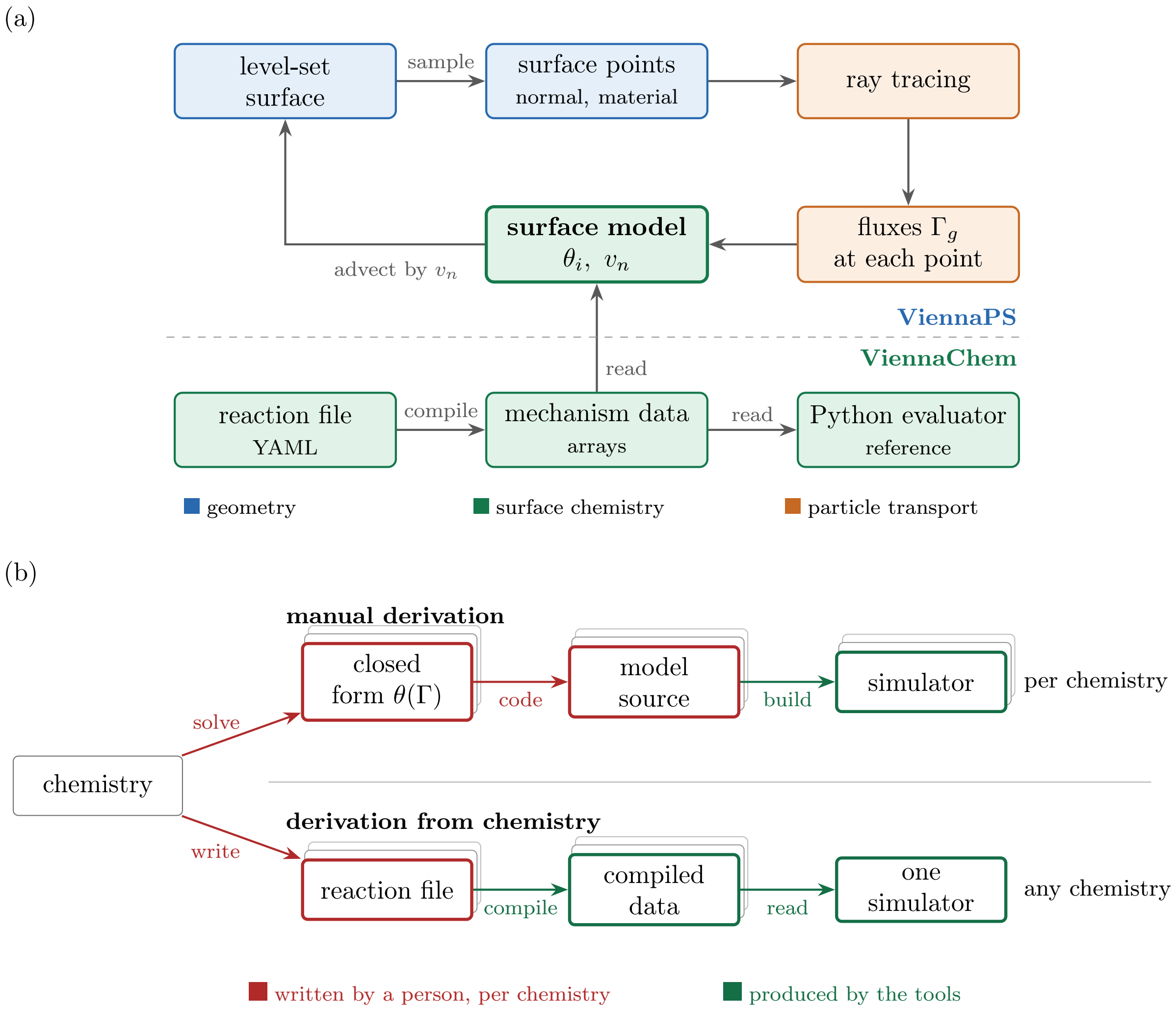

In a level-set simulator, the fluxes and the moving surface are connected by a
surface model, which converts the incident fluxes at a surface point into the
normal velocity $v_n$ of the interface (Fig.~\ref{fig:concept}a). In most chemistries, the velocity depends on the state of the surface, described by the
fractional coverages $\theta_i$ of the adsorbed species $i$ that take part in
the reactions. Each coverage is determined by a balance between the reactions that create and consume the corresponding adsorbate. The velocity is then computed
from the reactions that add or remove solid material.

Feature-scale simulators differ in how they treat the surface chemistry.
In process emulation, the surface is moved by geometric operations whose parameters are calibrated against measured cross-sections, without a model of the chemistry. Voxel-based virtual
fabrication emulates a complete process flow in this way at a small fraction of
the cost of a physical simulation~\cite{Holzer2009}, and it is used in logic and
memory development to predict structural variation and its effect on
metrology~\cite{Hargrove2023,Sobieski2019}. Geometric advection applies the same idea to level-set surfaces~\cite{Klemenschits2021}. Because the parameters of
an emulation describe shapes, a change in the chemistry requires a new
calibration. Monte Carlo profile models start from the chemistry, given as a list of surface reactions with probabilities, and each simulated particle that strikes the surface undergoes one of these reactions.
The Kushner group has developed this approach since the 1990s. Coupled to the reactor-scale fluxes of a hybrid plasma equipment model~\cite{Kushner2009}, its Monte Carlo feature-profile model predicted polysilicon etch profiles in an inductively coupled reactor~\cite{Hoekstra1997}. Fluorocarbon mechanisms of the same form described silicon and SiO$_2$ etching with polymer passivation~\cite{Zhang2000,Zhang2001}, and the method has since been applied to
high-aspect-ratio SiO$_2$ etching~\cite{Huang2019} and to the transients of
quasi-atomic-layer etching~\cite{Huard2018}.

In a Monte Carlo profile model, the list of reactions with their probabilities
defines the update of the surface directly. When a particle strikes a cell, one of the reactions is selected and the cell is changed accordingly. A level-set surface moves with a velocity computed from a rate per unit area and time.
This rate depends on the coverages, which have to be known at each surface
point before the velocity can be evaluated. In level-set topography simulators, the coverage balance is usually solved analytically, and the closed form is
implemented in code once for each chemistry. For the two coupled coverages of a silicon etch in fluorine and oxygen, the closed form is already a long expression~\cite{Belen2005,BelenSF6O2,Bobinac2023}, and its implementation in ViennaPS has separate cases for the limits in which one of the fluxes vanishes. Other hand-written models of such simulators are likewise derived for particular chemistries~\cite{Klemenschits2018,Aguinsky2022}. Atomic layer
deposition (ALD) is usually described by purpose-written models as well. Models
that combine a single Langmuir coverage and a prescribed growth per cycle with
Knudsen diffusion reproduce the measured conformality in high-aspect-ratio
structures with acceptable accuracy~\cite{Aguinsky2023,Reiter2024}.

Hand-written models work well for the chemistries they were written for, but
they are difficult to extend to larger or different chemistries. Closed forms
are available for two coupled species and rarely beyond, and chemistries with six or seven coverages are impractical to treat this way. The chemistry is also
fixed when the simulator is compiled. Adding an intermediate, making a step reversible or changing the order of a rate law requires a developer of the simulator to derive and implement the balance again and to recompile the code (Fig.~\ref{fig:concept}b).

In reactor-scale simulation, the corresponding problem was solved decades ago by treating the mechanism as data. Surface Chemkin
introduced a formalism in which the heterogeneous reactions, site types and
bulk phases are listed in an input file and the rate expressions are
constructed from that file~\cite{Coltrin1991}. Detailed mechanisms for specific systems, such as the metal-organic chemical vapour deposition (MOCVD) of GaAs~\cite{Mountziaris1991}, fit this formalism, because they consist of gas-phase and surface reactions with rate parameters on distinct site types. The open-source code Cantera reads such mechanisms from input files in the same way~\cite{Cantera2023}. The front-end process simulator FLOOPS separates the physical models from the grid and the numerics, and their parameters and model selections are set in scripts that users edit without recompiling the code~\cite{Law1998}. Mechanisms in Chemkin form have also been
coupled to feature-scale transport. Labun et al. transferred a
tetraethoxysilane mechanism validated at reactor scale to a feature-scale
profile simulation and compared the result with measured step
coverage~\cite{Labun2000}. The autonomous optimization of a reduced SiO$_2$
etching mechanism by Kr\"uger et al. shows how strongly profile predictions
depend on the reaction probabilities, yields, rate coefficients and threshold energies~\cite{Krueger2024}.

To the best of the authors' knowledge, mechanisms have not yet been supplied as data to
a level-set topography simulator. Level-set simulators are used even so, because at the same grid spacing they
describe the surface more precisely than a voxel model, which represents it by its filled cells. A level set
represents the surface as a smooth interface whose position is resolved below the grid spacing and whose normal is the gradient of the signed distance function~\cite{Osher1988,Osher2003}. This surface has no roughness from discrete
cells, and its normal gives the angle of incidence on which ion yields and ion reflection depend. Its sub-grid
position also allows a film to grow by a small fraction of the grid spacing in each step, meaning that even a coarse grid gives a reasonable representation of the advected surface. In a technology computer-aided design (TCAD)
flow and in design-technology co-optimization (DTCO), the geometry produced by process simulation is passed on to
device simulation. Techniques such as marching cubes~\cite{Lorensen1987} convert the continuum level set straightforwardly into the
meshed surface that a device simulator requires, whereas a set of filled cells is difficult to smooth into such a
geometry.

Without mechanisms as data, however, the surface chemistry of a level-set simulation is restricted to what can be
solved and coded by hand, although MOCVD, ALD and fluorocarbon etching involve several adsorbates, often on more
than one site type. A mechanism for such a process, whether validated at reactor scale or built from
first-principles barriers, then has to be reduced before a level-set simulator can use it, and the reduction can
lose a reaction channel, as the GaAs case below shows. Supplying mechanisms as data to a level-set simulator
combines the chemical detail of Monte Carlo profile models with these properties of the level set.

This combination requires that the coverage balance and its Jacobian be derived from these data and that the resulting rates, evaluated with ray-traced fluxes, determine a continuum surface velocity. The formulation must also represent the reaction types of plasma processing, including adsorption, desorption, reversible surface steps and ion-driven removal, and extend to processes that never reach a steady state. Closed-form surface models assume stationary coverages. This approximation is appropriate when the surface equilibrates quickly compared with the etch or deposition rate. Within an ALD half-cycle, the approximation fails, because whether a dose saturates depends on $\mathrm{d}\theta/\mathrm{d}t$ over a finite dose time. The coverages then have to be integrated through the cycle.

The surface model described here meets these requirements. It combines
the mechanism input of reactor-scale codes with the ray-traced fluxes and
continuum surface velocity of a level-set topography simulator. The chemistry
is written as a reaction file that lists the species with their phases and the
surface reactions with their rate constants. A Python package, ViennaChem, derives the free sites, mass-action rate laws, stoichiometry and surface velocity from this file, and a surface model in ViennaPS evaluates the rate laws, their analytic Jacobian and the velocity at each surface point. A damped Newton iteration solves for the coverages where the surface equilibrates quickly, and a conservative modified Patankar-Euler scheme integrates the same balance in time through the stages of a cyclic process. Because the surface model, the surface representation, the ray tracing and the time stepping of ViennaPS contain no chemistry-specific terms, a reaction file can be used in two and three dimensions and from the C++ or the Python interface, with CPU flux transport and, for reaction files within the size limits of the GPU code, with GPU flux transport. The implementation is tested with eight chemistries for chemical vapour deposition, selective epitaxy, diamond
growth, MOCVD, physical sputtering, ion-enhanced etching, simultaneous etching
and polymer deposition, and ALD. With the corresponding reaction file, the surface model reproduces the hand-written SF$_6$/O$_2$ silicon etch model of ViennaPS to seven significant digits at a single surface point, and in a masked trench the etched depths of the two models agree within the statistical noise of the ray tracing. In a lateral high-aspect-ratio cavity, alumina ALD is simulated with a trimethylaluminium (TMA) reaction probability from the range measured in such structures. The film saturates near the entrance and thins along the cavity, as in the thickness profiles by which conformality is measured.

\section{Surface chemistry written as a reaction file}
\label{sec:format}

The chemistry enters a simulation as a reaction file in YAML format (Fig.~\ref{fig:concept}a). ViennaChem derives the mass-action rate laws and the stoichiometry from this file and writes them as compiled data, in which each rate law is stored as indices and integer exponents. The surface model of ViennaPS reads these data and evaluates the rate laws at every surface point. The derivation is carried out once, when a simulation is set up, and the cost per surface point is that of solving for the coverages. ViennaChem also contains a Python evaluator that reads the same data and solves the coverage equations. The evaluator shares no code with the C++ model and is used as its reference.

\subsection{Polysilicon from silane as an example}
\label{sec:example}

The derivation from a reaction file is shown first for polysilicon deposition from silane, a chemistry with three reactions and two coverages. Silane adsorbs dissociatively on two free
sites and leaves an adsorbed silyl, $\mathrm{SiH_3^\ast}$, on one and an
adsorbed hydrogen, $\mathrm{H^\ast}$, on the other. The silyl decomposes into
one silicon atom of the solid, a second adsorbed hydrogen and a hydrogen
molecule, and adsorbed hydrogen desorbs recombinatively. At $900$\,K and an incident silane flux of $10^{18}$\,cm$^{-2}$s$^{-1}$, the reaction file reads

\begin{unbroken}
\begin{lstlisting}[style=yaml]
name: silane_polySi_LPCVD
temperature: 900.0        # K

species:
  SiH4:  {phase: gas,     flux: 1.0e3}          # Gamma in 1e15 /cm^2 /s
  H2:    {phase: gas,     reemit_only: true}
  SiH3*: {phase: surface}
  H*:    {phase: surface}
  Si:    {phase: solid,   rho: 5.02}            # 1e22 atoms/cm^3

# Only the reactions and their rate constants. Free sites, rate laws, coverage
# balance, steady-state relations and the growth rate are all derived. Free
# sites (*) are inferred from site conservation, so they are not written here.
surface:
  - {eq: "SiH4 -> SiH3* + H*",    type: adsorption, s0: 1.0,    Ea: 0.6}
  - {eq: "SiH3* -> Si + H* + H2", type: surface,    k0: 6.8e12, Ea: 1.5}
  - {eq: "2H* -> H2",             type: desorption, k0: 6.8e12, Ea: 2.0}
\end{lstlisting}
\end{unbroken}

\noindent
The free sites are not written in the file. From the adsorbates on the two sides of each equation, ViennaChem counts the sites that the reaction consumes or releases and stores the adsorption as $\mathrm{SiH_4} + 2\ast \rightarrow \mathrm{SiH_3^\ast} + \mathrm{H^\ast}$ with a free-site exponent of two. The
rate laws become
\begin{equation}
  r_1 = s_0(T)\, \Gamma_{\mathrm{SiH_4}}\, \theta_\ast^2 ,
  \qquad
  r_2 = k_2(T)\, \theta_{\mathrm{SiH_3^\ast}} ,
  \qquad
  r_3 = k_3(T)\, \theta_{\mathrm{H^\ast}}^2 ,
  \label{eq:silanerates}
\end{equation}
where $s_0(T)$ is the sticking coefficient, the probability that a silane
molecule arriving at a bare surface adsorbs, $\Gamma_{\mathrm{SiH_4}}$ the
incident silane flux, $\theta_\ast = 1 - \theta_{\mathrm{SiH_3^\ast}} -
\theta_{\mathrm{H^\ast}}$ the free fraction, and $k_2(T)$ and $k_3(T)$ the rate
constants of the two thermal steps. The rates, like the fluxes, are numbers of events per unit area and time, given in units of
$10^{15}$\,cm$^{-2}$s$^{-1}$. The rate constants $k_2$ and $k_3$ have the same unit, because they are the rates of
the two steps on a surface fully covered by their reactant. Divided by the density $S_0$ of surface sites, they give
the number of events per site per second. With the default $S_0 = 10^{15}$\,cm$^{-2}$, $k_2/S_0$ in s$^{-1}$ has the
same numerical value as $k_2$ in $10^{15}$\,cm$^{-2}$s$^{-1}$, and the prefactor $6.8 \times 10^{12}$ of the file
corresponds to an attempt frequency of $6.8 \times 10^{12}$\,s$^{-1}$ per site (Section~\ref{sec:rates}). The free fraction is squared in $r_1$ and $\theta_{\mathrm{H^\ast}}$ in $r_3$, because the adsorption occupies two sites and the desorption consumes two adsorbed hydrogen atoms. The
stoichiometry gives the balances
\begin{equation}
  S_0\, \frac{\mathrm{d}\theta_{\mathrm{SiH_3^\ast}}}{\mathrm{d}t} = r_1 - r_2 ,
  \qquad
  S_0\, \frac{\mathrm{d}\theta_{\mathrm{H^\ast}}}{\mathrm{d}t} = r_1 + r_2 - 2 r_3 ,
  \label{eq:silanebalance}
\end{equation}
and since only the second reaction creates solid, the velocity is $v_n = r_2 /
\rho_{\mathrm{Si}}$ with the number density $\rho_{\mathrm{Si}}$ of silicon. With $r_2$ in
$10^{15}$\,cm$^{-2}$s$^{-1}$ and $\rho_{\mathrm{Si}}$ in $10^{22}$\,cm$^{-3}$, the units of the file, this quotient is
the velocity in nm/s.
All four entries of the Jacobian are nonzero, because both coverages enter $r_1$
through the free fraction even though neither is a reactant of the adsorption. Setting Eq.~(\ref{eq:silanebalance}) to zero at the flux and temperature of the file gives $\theta_{\mathrm{SiH_3^\ast}} = 1.3299
\times 10^{-5}$, $\theta_{\mathrm{H^\ast}} = 9.1590 \times 10^{-2}$ and $v_n =
7.1779 \times 10^{-2}$\,nm/s. On a blanket surface, with fluxes from the ray
tracer and the surface advanced by the level set, four runs at $5000$ rays per
surface point give growth rates between $7.1732$ and $7.1793 \times 10^{-2}$\,nm/s, from $0.06\,\%$ below to $0.02\,\%$ above the single-point value.

Larger chemistries are written with further entries of the same format (Table~\ref{tab:entries}). With constants given per material, a step can be confined to selected materials, and an \texttt{order} entry sets rate-law exponents that differ from the stoichiometry. A reversible step becomes two one-way reactions. Each site type has its own free fraction, and the threshold yield of an ion-driven step is evaluated during the ray tracing (Sections~\ref{sec:rates} to \ref{sec:materials}). In a hand-written model, the developer has to include each of these items in the derivation. For the GaAs chemistry discussed below, with seven adsorbates
and thirty reactions, the Jacobian is a $7 \times 7$ matrix whose entries are
sums over the thirty rate laws. A new species can change every row of this matrix, but the exponents from which it is assembled are derived anew whenever the file is compiled.

\subsection{Reaction files}
\label{sec:files}

A reaction file is a YAML document in three parts (Table~\ref{tab:entries}). The top level contains a name, the surface temperature, the units of pressure and activation energy, the site types with their densities. In the species block, each
species is given one of four phases, which determines how it enters a
rate law (Table~\ref{tab:phases}). A gas species that reacts at the surface has an incident flux, and its transport to the surface is computed by ray tracing from a cosine source. An ion also has a flux, together with the parameters of its energy and angular distributions and of its reflection at the surface. Its directions are sampled from a
power-cosine distribution and its energies from a Gaussian. A surface species is an adsorbate that occupies one site, and its name usually ends with an asterisk. A solid has a number density, and its creation or removal moves the
surface. The reactions are listed under the key \texttt{surface}, each with its
equation, its type and its constants and, where needed, with per-material
values, an \texttt{order} entry or an ion yield.

A gas species is traced when
it appears on the left-hand side of a reaction. A species that appears only as
a product leaves the surface and is given neither an incident flux nor a particle in the ray tracer. In the files, such species are marked with an optional \texttt{reemit\_only} flag for readability. A traced species is given either an incident flux or a
partial pressure, which ViennaChem converts into the wall-collision flux.
Incident fluxes are given in units of $10^{15}$\,cm$^{-2}$s$^{-1}$, solid
densities in $10^{22}$ atoms or formula units per cm$^3$, site densities in
cm$^{-2}$, temperatures in kelvin, times in seconds and ion energies in
electronvolts. Activation energies may be given in electronvolts or in
kilocalories per mole, where $1$\,kcal/mol $= 4.184$\,kJ/mol.

When a file is read, ViennaChem counts the
atoms of each element on the two sides of each equation, weighted by the
stoichiometric coefficients, and rejects the file if the counts differ. Species names are read as chemical formulas, so that \texttt{SiH4} is counted as one silicon and four hydrogen atoms. For a
species whose name is not a formula, the composition is given in an
\texttt{atoms} entry. Free sites contain no atoms and are excluded from the atom balance. Ions are excluded as well, because they are treated as projectiles that deliver energy and leave neutralised.

ViennaChem converts each reaction into a column of the stoichiometric matrix
and derives the exponents of its rate law, the temperature dependence of its
rate constant and the number of atoms or formula units it adds to or removes from each solid
(Table~\ref{tab:derived}). In the compiled data, the rate law of each reaction is stored as integer exponents. From the stored arrays alone, a program without computer algebra can reconstruct the coverage balance and its Jacobian. The compiled data contain the version number of their format, and ViennaPS rejects data with an unknown version.

\subsection{Surface sites and rate laws}
\label{sec:rates}

Every adsorbate occupies one site of one type. Each site type has a density of $S_0$ sites per unit area. For a single site type, the site-type entry can be omitted, since all adsorbates then share a default
type with $S_0 = 10^{15}$\,cm$^{-2}$. The fraction $\theta_{\ast \alpha}$ of
the sites of type $\alpha$ that are free is given by site conservation,
\begin{equation}
  \theta_{\ast \alpha} \;=\; 1 - \sum_{i \,:\, \alpha(i) = \alpha} \theta_i ,
  \label{eq:closure}
\end{equation}
where the sum runs over the adsorbates $i$ whose site type $\alpha(i)$ is
$\alpha$. Coverages on one type therefore do not affect the free fraction of
any other type. Free sites are not written in a reaction file. For each site type, ViennaChem counts the adsorbates on the two
sides of an equation and adds the difference as free sites to the side with
fewer adsorbates. The number of free sites on the left-hand side becomes the free-site exponent of the rate law.

Each reaction has a mass-action rate law built from the incident fluxes, the free fractions of the site types and the coverages,
\begin{equation}
  r_j = k_j \prod_g \Gamma_g^{\,n_{gj}}
        \prod_\alpha \theta_{\ast \alpha}^{\,n_{\ast \alpha j}}
        \prod_i \theta_i^{\,n_{ij}},
  \label{eq:massaction}
\end{equation}
where $r_j$ is the rate of reaction $j$ per unit area, $k_j$ its rate constant,
$\Gamma_g$ the flux of gas species $g$ arriving at the surface point and
$\theta_i$ the coverage of adsorbate $i$. The exponents $n_{gj}$, $n_{\ast
\alpha j}$ and $n_{ij}$ are the stoichiometric coefficients on the left-hand
side of the reaction. Where a published rate law is not of mass-action form,
the exponents are set to other integer values with an \texttt{order} entry, and the stoichiometry is kept as written. Rate constants have the modified Arrhenius form
\begin{equation}
  k_j = k_{0j}\, T^{\beta_j} \exp\!\left(-\frac{E_{a,j}}{k_B T}\right),
  \label{eq:arrhenius}
\end{equation}
with prefactor $k_{0j}$, temperature exponent $\beta_j$, activation energy $E_{a,j}$, surface temperature $T$ and Boltzmann constant $k_B$. A reversible step has a second set of constants and is expanded into a forward and a backward reaction when the file is compiled, and the compiled data contain only one-way reactions.

The rate $r_j$ is the number of reaction events per unit area and time and is
expressed in the flux unit, $10^{15}$\,cm$^{-2}$s$^{-1}$. Coverages and free
fractions are dimensionless, so the unit of $k_j$ depends on the number of
incident fluxes in its rate law,
\begin{equation}
  [k_j] = \left( 10^{15}\,\mathrm{cm^{-2}\,s^{-1}} \right)^{1 - \sum_g n_{gj}} .
  \label{eq:kunits}
\end{equation}
In a step with one gas reactant, such as an adsorption or the reaction of an
incident particle with an adsorbate, $k_j$ is dimensionless. It is the
probability that an incident particle reacts when the coverage and free-site
factors of the rate law are one, and for an adsorption it is the sticking
coefficient $s_0$ on a bare surface. In a step among adsorbates, such as a surface
reaction, a desorption or a step in which adsorbates form or remove a solid,
$k_j$ is the rate per unit area that the step would have if these factors were one. A step that consumes one adsorbate with a first-order rate
constant $k^{(1)}$ per occupied site has $k_j = k^{(1)} S_0$, where $S_0$ is
the density of the site type of that adsorbate. For the default site density of
$10^{15}$\,cm$^{-2}$, $k_j$ in the flux unit is numerically equal to $k^{(1)}$
in s$^{-1}$. A step between two adsorbates
with a second-order rate constant $k^{(2)}$ has $k_j$ equal to $k^{(2)}$ times
the densities of both site types involved. An ion-driven step has $k_j = 1$, because its rate is given by the
yield-weighted ion flux described below. The prefactor $k_{0j}$ has the unit of
$k_j$ divided by $\mathrm{K}^{\beta_j}$. The meaning and unit of every rate constant of the eight cases are given with their reaction files in
Appendices~\ref{note:sf6o2} to \ref{note:al2o3}.

Some sources report an adsorption rate constant $k_{\mathrm{ads}}$ in cm/s in place of a sticking coefficient. It is converted by equating the
two expressions for the adsorption rate, $k_{\mathrm{ads}} n = s_0 \Gamma$,
with the wall-collision flux $\Gamma = n \bar{v} / 4$ and the mean thermal speed
$\bar{v} = \sqrt{8RT/\pi M}$~\cite{Kolasinski2012}, which gives
\begin{equation}
  s_0 = \frac{4 k_{\mathrm{ads}}}{\bar{v}} \propto T^{-1/2} .
  \label{eq:sticking}
\end{equation}
Here, $n$ is the number density of the gas above the surface, $R$ the gas
constant and $M$ the molar mass of the species. The conversion fixes $\beta =
-\tfrac{1}{2}$ in Eq.~(\ref{eq:arrhenius}), so that the sticking coefficient
includes the temperature dependence of the mean speed.

\subsection{Ion-driven steps}
\label{sec:ions}

An ion-driven step has a yield in place of a rate constant,
\begin{equation}
  Y = A \max\!\left(\sqrt{E} - \sqrt{E_{\mathrm{th}}},\, 0\right) f(\vartheta) ,
  \label{eq:yield}
\end{equation}
where $Y$ is the number of reaction events per incident ion, $A$ the yield
prefactor in eV$^{-1/2}$, $E$ the ion energy, $E_{\mathrm{th}}$ the threshold
energy and $\vartheta$ the angle between the ion trajectory and the surface
normal. The square-root threshold form is the low-energy dependence that
Steinbr\"uchel established for both physical and ion-enhanced
yields~\cite{Steinbruchel1989}. An ion-enhanced reaction is insensitive to the
angle until the incidence becomes too oblique for the ion to deliver its energy
efficiently, whereas physical sputtering peaks away from normal incidence,
because an obliquely incident ion deposits more of its energy close to the
surface. The angular factor therefore takes one of two forms, the first for
ion-enhanced reactions and the second for sputtering,
\begin{equation}
  f(\vartheta) =
  \begin{cases}
    1, & \vartheta \le 60^\circ \\[2pt]
    \max\!\left( 3 - \dfrac{6\vartheta}{\pi},\; 0 \right), & \vartheta > 60^\circ
  \end{cases}
  \qquad\text{and}\qquad
  f(\vartheta) = \max\!\left( \left( 1 + B \sin^2\vartheta \right) \cos\vartheta,\; 0 \right),
  \label{eq:angular}
\end{equation}
where $B$ determines both the angle at which the sputtering yield peaks and how
steeply it rises towards that angle (Fig.~\ref{fig:yield}). The angular factor is chosen according to the reaction type given in the file, and both of its forms are those of the plasma-etching models of ViennaPS. The ion-enhanced form is constant up to $60^\circ$ and falls to zero towards grazing incidence, as in the SF$_6$/O$_2$ model of Belen et al.~\cite{BelenSF6O2}, and this dependence was later measured in situ for silicon etching~\cite{Belen2006}. The other form has the off-normal maximum of physical sputtering yields~\cite{Belen2005}. Similar yield
functions are used for oblique ion beam etching~\cite{Reiter2024IBE}, and the
angular dependence determines feature-scale effects such as the erosion of mask
corners and the resulting taper~\cite{Bobinac2023}.

The yield cannot be written in mass-action
form, because the ion energy is drawn from a distribution and decreases with
every reflection, and $\vartheta$ depends on the point of impact. The yield is therefore evaluated during the ray tracing. Each ion contributes to the flux at
its point of impact in proportion to its yield, and this yield-weighted flux
enters Eq.~(\ref{eq:massaction}) in the same way as an incident flux.

\subsection{Per-material constants}
\label{sec:materials}

The constants of any step can be given per material. Either a single number multiplies the prefactor of the step on that material, with the barrier unchanged, or one or more of the prefactor, the barrier and the temperature exponent are given for that material and replace the corresponding values of the step. A \texttt{default} entry applies to all other materials, and a prefactor of zero switches the step off, so a reaction can
be confined to selected materials. Materials are identified by the names stored with the topography. Because the material at a surface point is updated at every advection step, the constants of a growing film apply as soon as the film covers the substrate. If the density of a solid is also given per material, a mask and a substrate removed at the same reaction rate recede at different speeds. Per-material values of the parameters $A$ and $E_{\mathrm{th}}$ of an ion yield make a sputtering step selective. A sticking coefficient given per material changes the transport as well as the kinetics, since the same value enters the absorption probability of the ray tracer.

\subsection{Surface velocity}
\label{sec:velocity}

The surface velocity is computed from the rates of the reactions that create or consume a solid,
\begin{equation}
  v_n = \sum_j \frac{n_{S,j}\, r_j}{\rho_{S(j)}} ,
  \label{eq:velocity}
\end{equation}
where $n_{S,j}$ is the number of atoms or formula units of solid $S(j)$ created by reaction $j$
and $\rho_{S(j)}$ the density of that solid, which may depend on the material.
The count $n_{S,j}$ is positive when the solid is a product and negative when it
is a reactant, so that $v_n > 0$ corresponds to deposition and $v_n < 0$ to
etching. Because each contribution is divided by the density of its own solid, the sum gives the net motion of a surface on which a polymer is deposited while the substrate is etched.

\section{Coverage solvers}
\label{sec:solvers}

\subsection{Steady-state coverages}
\label{sec:steady}

At steady state, the balance of each adsorbate gives one equation,
\begin{equation}
  F_i(\bm{\theta}) \;=\; S_0\, \frac{\mathrm{d}\theta_i}{\mathrm{d}t}
    \;=\; \sum_j \nu_{ij}\, r_j(\bm{\theta}) \;=\; 0 ,
  \qquad i = 1,\dots,N_\theta ,
  \label{eq:steady}
\end{equation}
where $N_\theta$ is the number of adsorbates, $\bm{\theta}$ the vector of their
coverages, $\nu_{ij}$ the net stoichiometric coefficient of adsorbate $i$ in
reaction $j$, and $S_0$ the density of the site type occupied by adsorbate $i$.
At given rate constants $k_j$, the site density scales a whole row of the system and leaves
the steady state unchanged, but it sets the time scale of the transient integration. With the
free fractions given by Eq.~(\ref{eq:closure}), the $N_\theta$ coverages are the
only unknowns and the system is square. A reaction file may also contain no adsorbates, $N_\theta = 0$, and the velocity is then computed from the fluxes alone.

Each rate law is split into a factor $A_j$ that does not depend on the
coverages, a free-site factor $\Phi_j$ and a coverage factor $\Theta_j$,
\begin{equation}
  r_j \;=\; A_j\, \Phi_j\, \Theta_j ,
  \qquad
  A_j = k_j \prod_g \Gamma_g^{\,n_{gj}} ,
  \quad
  \Phi_j = \prod_\alpha \theta_{\ast \alpha}^{\,n_{\ast \alpha j}} ,
  \quad
  \Theta_j = \prod_i \theta_i^{\,n_{ij}} .
  \label{eq:ratesplit}
\end{equation}
The factor $A_j$ contains the rate constant and the incident fluxes of the
reaction, including the yield-weighted ion flux of an ion-driven step. A
coverage $\theta_k$ enters a rate law in two ways, as an explicit factor in
$\Theta_j$ when the reaction involves adsorbate $k$, and through
Eq.~(\ref{eq:closure}) in the free fraction of its site type $\tau = \alpha(k)$,
which affects every reaction whose rate law contains that free fraction. The Jacobian
of Eq.~(\ref{eq:steady}),
\begin{equation}
  J_{ik} \;=\; \frac{\partial F_i}{\partial \theta_k}
         \;=\; \sum_j \nu_{ij}\, \frac{\partial r_j}{\partial \theta_k} ,
  \label{eq:jacobian}
\end{equation}
therefore contains both contributions,
\begin{equation}
  \frac{\partial r_j}{\partial \theta_k} = A_j \left(
    n_{kj}\, \theta_k^{\,n_{kj}-1} \prod_{i \neq k} \theta_i^{\,n_{ij}}
    \prod_\alpha \theta_{\ast \alpha}^{\,n_{\ast \alpha j}}
    \;-\; n_{\ast \tau j}\, \theta_{\ast \tau}^{\,n_{\ast \tau j}-1}
    \prod_{\alpha \neq \tau} \theta_{\ast \alpha}^{\,n_{\ast \alpha j}}
    \prod_i \theta_i^{\,n_{ij}} \right) .
  \label{eq:drate}
\end{equation}
The first term is obtained with the product rule over the coverage factors and the second with the chain rule through Eq.~(\ref{eq:closure}). Because the reaction
type enters the rate law only through $A_j$, Eq.~(\ref{eq:drate}) holds for
every reaction type. The exponents are stored as integers, so a term with a zero
exponent vanishes identically, and only the reactions that involve $\theta_k$
contribute to column $k$. Forming the products over the remaining factors keeps the derivative finite at $\theta_k = 0$, where the iteration starts on a bare surface.

\begin{figure}[!htbp]
\centering
\begin{tikzpicture}[scale=3.9,>=Stealth,font=\small]
  \definecolor{fea}{RGB}{226,236,247}
  \definecolor{stp}{RGB}{176,42,42}
  \definecolor{dmp}{RGB}{20,110,70}

  \fill[fea] (0,0) -- (1,0) -- (0,1) -- cycle;
  \draw[black!55] (0,0) -- (1,0) -- (0,1) -- cycle;
  \draw[->,black!70] (-0.04,0) -- (1.20,0) node[right,font=\footnotesize] {$\theta_1$};
  \draw[->,black!70] (0,-0.04) -- (0,1.16) node[above,font=\footnotesize] {$\theta_2$};
  \node[font=\footnotesize,black!60,anchor=west] at (0.09,0.11)
       {$\theta_1+\theta_2\le 1$};
  \node[font=\scriptsize,anchor=north] at (1.0,-0.02) {$1$};
  \node[font=\scriptsize,anchor=east] at (-0.02,1.0) {$1$};

  \coordinate (m) at (0.22,0.30);
  \coordinate (d) at (0.34,0.42);
  \coordinate (x) at (0.46,0.54);
  \coordinate (u) at (0.62,0.70);

  \draw[stp,thick,dashed,->] (m) -- (u);
  \draw[dmp,very thick,->] (m) -- (d);
  \draw[black!45,thin] (x) -- ++(0.14,-0.06)
       node[anchor=west,font=\scriptsize,black!55,inner sep=1pt]
       {bound reached here};

  \fill[stp] (u) circle (0.014);
  \node[stp,font=\footnotesize,anchor=south west,inner sep=2.5pt] at (u)
       {$+\,\bm{\delta}$};
  \draw[black!55,fill=white,thin] (x) circle (0.013);
  \fill[dmp] (d) circle (0.014);
  \node[dmp,font=\footnotesize,anchor=east,inner sep=3pt] at (0.295,0.455)
       {$+\,\lambda\bm{\delta}$};
  \fill[black] (m) circle (0.014);
  \node[font=\footnotesize,anchor=north,inner sep=3pt] at (0.235,0.275)
       {$\bm{\theta}^{(m)}$};
\end{tikzpicture}
\caption{The damped Newton step of Eqs.~(\ref{eq:newton}) to
(\ref{eq:dampsite}) for two coverages on one site type. The shaded region is
the physical simplex. Its edges on the axes are the bounds where a coverage
becomes negative, Eq.~(\ref{eq:damping}), and its hypotenuse is the bound where
the site type is full, Eq.~(\ref{eq:dampsite}). The full step
$\bm{\delta}$ crosses the hypotenuse, and $\lambda$ shortens it to half the
distance to the crossing. With several site types, the feasible region is the product of one such simplex per type.}
\label{fig:simplex}
\end{figure}
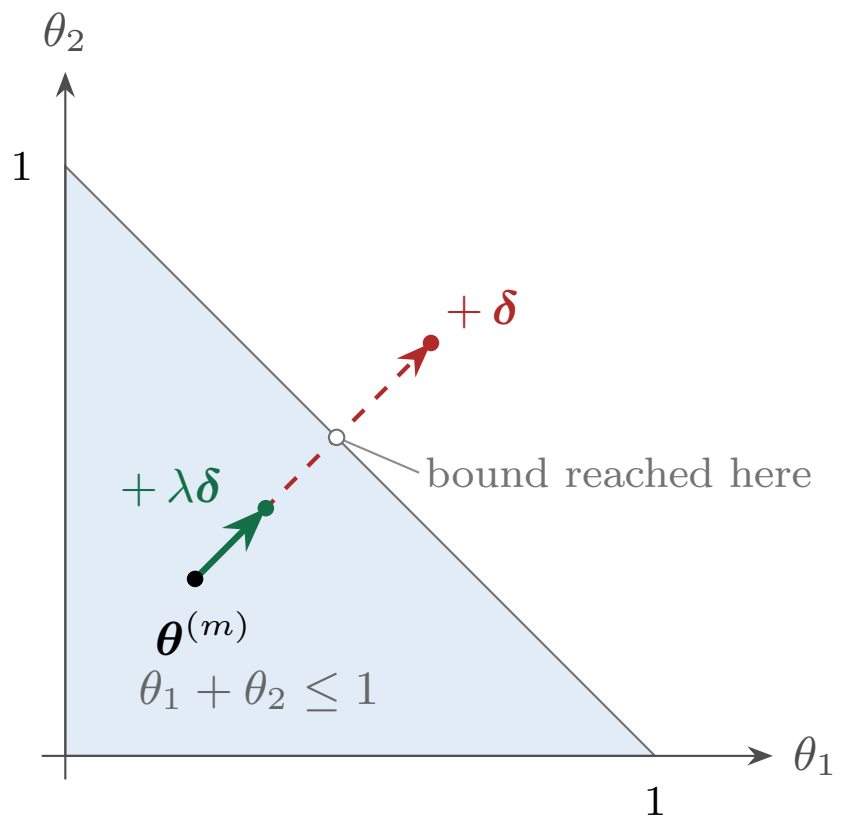

The coverages are found with a damped Newton iteration~\cite{Deuflhard2011}. From the iterate $\bm{\theta}^{(m)}$, the Newton step and the next iterate are computed as
\begin{equation}
  \mathbf{J}\, \bm{\delta} \;=\; -\mathbf{F} ,
  \qquad
  \bm{\theta}^{(m+1)}
    \;=\; \Pi\!\left( \bm{\theta}^{(m)} + \lambda\, \bm{\delta} \right) ,
  \label{eq:newton}
\end{equation}
where $\mathbf{F}$ is the vector of residuals, $\mathbf{J}$ its Jacobian,
$\bm{\delta}$ the Newton step, $m$ the iteration index, $\lambda \le 1$ a
damping factor and $\Pi$ a projection that clamps each coverage to $[0,1]$.
Since $N_\theta$ is small, the linear system is solved directly by Gaussian
elimination with partial pivoting. The damping factor keeps every coverage non-negative and the occupancy of each site type below one, except when a coverage that is already zero has a negative step component, which $\Pi$ sets back to zero (Fig.~\ref{fig:simplex}). It starts from
\begin{equation}
  \lambda \;=\; \min\left\{ 1,\;
    \min_{i \in \mathcal{A}} \frac{\theta_i}{2\,|\delta_i|} \right\} ,
  \label{eq:damping}
\end{equation}
where $\delta_i$ is the component of $\bm{\delta}$ for adsorbate $i$ and the
set $\mathcal{A} = \{\, i : \theta_i > 0,\; \theta_i + \delta_i < 0 \,\}$
contains the coverages that a full step would make negative. Coverages that are already zero are excluded from $\mathcal{A}$, since they
would otherwise set $\lambda = 0$ and stop the iteration, and the projection
$\Pi$ keeps them at zero. The site types are
then taken in turn, and for each type $\alpha$
\begin{equation}
  \lambda \;\leftarrow\; \min\left\{ \lambda,\;
    \frac{1 - \Sigma_\alpha}{2\,\sigma_\alpha} \right\}
  \quad \text{if} \quad \sigma_\alpha > 0 \ \text{and}\
  \Sigma_\alpha + \lambda\,\sigma_\alpha > 1 ,
  \label{eq:dampsite}
\end{equation}
where $\Sigma_\alpha = \sum_{i : \alpha(i)=\alpha} \theta_i$ is the occupied
fraction of the type and $\sigma_\alpha = \sum_{i : \alpha(i)=\alpha} \delta_i$
its change over a full step. The bound of a site type is applied only if the step, as already shortened, would still overfill that type. With the factor of two, the step ends halfway to a violated bound,
and the next Jacobian is evaluated away from the boundary of the simplex. The iteration stops when $\max_i |\theta_i^{(m+1)} -
\theta_i^{(m)}|$ falls below $10^{-13}$, with a limit of $500$ iterations.

The coverage balance is solved independently at every surface point, and only the factors $A_j$ differ from point to point. With per-material constants, every reaction that involves a species can be
switched off on a given material. The residual of that species is then zero,
its row of $\mathbf{J}$ is empty and the matrix is singular. The species is
neither produced nor consumed there, and its row is replaced by
\begin{equation}
  J_{ii} = 1 , \qquad F_i = \theta_i ,
  \label{eq:pin}
\end{equation}
so that $\theta_i$ is set to zero in a single iteration when the Newton step is not damped, while the other rows are unchanged.

\subsection{Transient coverages}
\label{sec:transient}

Where the coverages do not reach a steady state, they are integrated in time,
\begin{equation}
  \frac{\mathrm{d}\theta_i}{\mathrm{d}t}
    = \frac{1}{S_0}\sum_j \nu_{ij}\, r_j , \label{eq:ode}
\end{equation}
which is Eq.~(\ref{eq:steady}) divided by the site density, without setting the
right-hand side to zero. The rate laws and the stoichiometry are those of the
steady-state solver, and a reaction file can be used with either solver without
changes. The modified Patankar-Euler scheme of Burchard et al.~\cite{Burchard2003} is used, and each reaction is written as a transfer of occupancy
between the states of a site type. Within one site type, reaction $j$ removes
occupancy from the states with $\nu_{ij} < 0$ and adds it to the states with
$\nu_{ij} > 0$ in the proportions given by the stoichiometry. The free state
$\ast$ of the type is included as a further state, whose stoichiometric
coefficient balances the occupancy of the two sides. With the gain and loss
weights $g_{ij} = \max(\nu_{ij}, 0)$ and $l_{kj} = \max(-\nu_{kj}, 0)$, extended
to the free state, the transfer from state $k$ to state $i$ has the rate
coefficient
\begin{equation}
  q_{ik} = \frac{1}{S_{0}}\sum_j
      \frac{g_{ij}\, l_{kj}}{\sum_m g_{mj}}\;\frac{r_j}{\theta_k} ,
  \label{eq:transfer}
\end{equation}
the occupancy transferred from state $k$ to state $i$ per unit time and per unit
occupancy of $k$. The indices $i$, $k$ and $m$ run over the states of one site
type, including its free state. The normalisation by $\sum_m g_{mj}$ makes the occupancy removed from the consumed states equal to the occupancy added to the produced ones. In all cases considered here, the rate of every reaction that consumes state $k$ contains $\theta_k$ as a factor, so the quotient $r_j/\theta_k$ remains finite. A step of length $h$ is then obtained from the linear system
\begin{equation}
  \Big( \mathbf{I} + h\,\mathrm{diag}(\mathbf{L}) - h\,\mathbf{q} \Big)\,
    \boldsymbol{\theta}(t + h) = \boldsymbol{\theta}(t) ,
  \qquad
  L_k = \sum_i q_{ik} ,
  \label{eq:patankar}
\end{equation}
where $\boldsymbol{\theta}(t)$ and $\boldsymbol{\theta}(t + h)$ contain the
coverages and the free fraction of the site type at the start and the end of
the step, $\mathbf{I}$ is the identity, $\mathbf{q}$ the matrix of the
coefficients of Eq.~(\ref{eq:transfer}), and $\mathrm{diag}(\mathbf{L})$ the
diagonal matrix of the total loss coefficients $L_k$. Every column of the system matrix sums to one, and the matrix is an M-matrix that is strictly diagonally dominant by columns. For any $h$, the occupancy of the site type is therefore conserved and the solution stays non-negative. The scheme is first order in $h$ and linearly implicit. Its linear system has one row for each coverage and for the free fraction of each site type, at most twelve rows in the cases considered here, and like the Newton step it is solved by Gaussian elimination with partial pivoting.

The interval over which the coverages are advanced is divided into steps of this scheme, and the length of each of these sub-steps is set by the net rate of change of the coverages,
\begin{equation}
  h = \min\!\left( \Delta t - t,\;
      \frac{\epsilon}{\max_i \left| \sum_k q_{ik} \theta_k
                      - L_i \theta_i \right|} \right) ,
  \label{eq:substep}
\end{equation}
where $\Delta t$ is the length of this interval, $t$ the time already elapsed within it and $\epsilon$ the largest change of any coverage allowed in one sub-step. The last sub-step is shortened to end exactly at $\Delta t$. The maximum is taken over the coverages, excluding the free states. A fast intermediate at its quasi-steady value has a large loss coefficient but a small net rate, and Eq.~(\ref{eq:patankar}) is accurate in this regime. The sub-step length is therefore not limited by the short lifetime of such an intermediate. The number of sub-steps scales as $1/\epsilon$, and the
transient calculations use $\epsilon = 2 \times 10^{-4}$ unless stated
otherwise. To guard against round-off errors, the coverages are clamped to $[0,1]$ after each sub-step, and a site type whose occupancy exceeds one is rescaled.

\section{Coupling to the flux transport}
\label{sec:transport}

The fluxes at the surface points are computed with the Monte Carlo ray tracer
ViennaRay (\url{https://github.com/ViennaTools/ViennaRay}), version 4.5.1, on the CPU or on the GPU.
The coverage balance is solved on the CPU with either flux engine. At the start of a process, ViennaPS builds the ray-tracing engine from the geometry and from the list of particle types to be traced. The particle types, the interaction of each with the surface and the flux to which it contributes are derived from the reaction file. Each traced gas species is represented by one particle type. The engine is kept for the whole process, and in a cyclic process the list of particle types is therefore the union over all reaction files added to the process. A species that does not flow in the current stage is switched off by setting its source flux to zero, so a change of stage on the GPU engine requires no data transfer and no recompilation of the ray-tracing programs.

A traced neutral of gas species $g$ is absorbed at a surface hit with the probability that it reacts there. This probability is the sum of the rate laws of Eq.~(\ref{eq:massaction}) that consume the species, divided by its incident flux,
\begin{equation}
  s_{\mathrm{eff},g} = \sum_{j \in \mathcal{R}_g} k_j(\text{material})
    \prod_\alpha \theta_{\ast \alpha}^{\,n_{\ast \alpha j}}
    \prod_i \theta_i^{\,n_{ij}} ,
  \label{eq:reemission}
\end{equation}
where $\mathcal{R}_g$ is the set of reactions in which $g$ is the only gas reactant and enters the rate law with exponent one, $k_j$ is evaluated for the material at the hit, and the free
fractions and coverages are those of the surface point. The sum is limited to
one. For an adsorption onto free sites, it reduces to the sticking coefficient
times the free-site factor of that step. For a step in which the gas consumes
an adsorbate, such as the ligand exchange of an ALD dose, it is proportional to
the coverage of that adsorbate, so a surface on which the adsorbate is used up
reflects the gas. Particles that are not absorbed are re-emitted diffusely. Through the absorption probability, the coverages affect the transport, and the ray tracer contains no chemistry-specific code.

With ion-driven steps, the ray tracer has one additional particle type for the ions. An ion that strikes the surface at an angle of incidence below
\texttt{thetaRMin} is absorbed, and above this angle the probability of
reflection rises linearly to one at \texttt{thetaRMax}. A reflected ion loses
energy according to its angle of incidence, with the loss set by
\texttt{inflectAngle} and \texttt{n\_l} as in the ion models of ViennaPS~\cite{Reiter2024IBE}. The reflected ion leaves in a cone around the specular direction, with a half-angle of $90^\circ$ minus the angle of incidence but no smaller than $90^\circ$ minus \texttt{minAngle}. An ion whose energy falls below the lowest threshold of its yields is stopped. At each ion hit,
Eq.~(\ref{eq:yield}) is evaluated for every ion-driven reaction, and the
yield-weighted contribution of the ion is added to the flux of that reaction.

\section{Cyclic processes}
\label{sec:cyclic}

Each half-cycle of an ALD process is written as a reaction file of its own and
added to the process under a name. Because the coverages at the end of one
stage are the initial condition of the next, all reaction files of a process must describe the same surface with the same surface species. Before the process starts, ViennaPS compares the surface species of each added file with those of the first and rejects a file whose surface species differ. Rate constants, gas species and reactions may differ between the files, and a surface species need not be consumed by any reaction. Such a coverage accumulates until its site type is full, and the occupied sites are then no longer available to any step that consumes a free site of that type, as for the hydroxylated aluminium of the alumina model (Appendix~\ref{note:al2o3}).

A cycle is defined as a list of stages, each with a name, a duration, a time step, the species that flow during it and the name of its reaction file (Appendix~\ref{note:solvers}). The stages are run in the order of the list, and the list is repeated once per cycle. The coverages are advanced by integrating Eq.~(\ref{eq:ode}) with the time step of the current stage, which the integrator subdivides according to Eq.~(\ref{eq:substep}). The fluxes are recomputed by ray tracing at every time step. During each stage, the source fluxes of the listed species are switched on and all others are set to zero, and a reaction whose gas reactant is absent then has zero rate. A purge is a stage with an empty species list. Adsorption then stops while the thermal steps continue, and material that has left the surface does not return.

In a cyclic process, the surface is kept fixed during a cycle and advected once
at its end, by the thickness deposited at each point during the cycle,
\begin{equation}
  \Delta n_{\mathrm{cycle}} = \int_{\mathrm{cycle}} v_n\, \mathrm{d}t ,
  \label{eq:growthintegral}
\end{equation}
where $\Delta n_{\mathrm{cycle}}$ is the distance by which a point advances
along its normal and the integral runs over all stages of the cycle, including
the purges. The integral is accumulated on the sub-steps of Eq.~(\ref{eq:substep}) inside the integrator, because the velocity depends on coverages that can fall by orders of magnitude within a small fraction of a pulse. On each sub-step, the rate of every solid-forming step is evaluated with the end-of-sub-step values of the coverages it consumes, as in Eq.~(\ref{eq:patankar}), so that the film grown in a pulse equals the occupancy converted during the pulse. The accumulated thicknesses are applied as the velocity field of the advection for one unit of time, and each point moves by its own $\Delta n_{\mathrm{cycle}}$. A point inside a feature that receives too little flux to saturate advances less than a point on the open field.

\begin{figure}[!htbp]
\centering
\begin{tikzpicture}[scale=1.62,>=Stealth]
  \definecolor{oldc}{RGB}{120,120,120}
  \definecolor{newc}{RGB}{42,120,214}
  \definecolor{acc}{RGB}{190,30,45}

  \draw[oldc,dashed,thick]
    plot[smooth,tension=0.7] coordinates
    {(0,1.30) (1.1,1.44) (2.2,1.32) (3.3,0.92) (4.4,0.58) (5.5,0.52)
     (6.6,0.68) (7.6,0.88)};
  \draw[newc,thick]
    plot[smooth,tension=0.7] coordinates
    {(0,1.56) (1.1,1.70) (2.2,1.58) (3.3,1.18) (4.4,0.84) (5.5,0.78)
     (6.6,0.94) (7.6,1.14)};

  \foreach \x/\y in {0.35/1.35, 1.45/1.42, 2.55/1.22, 3.65/0.80,
                     4.75/0.55, 5.85/0.55, 6.95/0.75}{
    \fill[white] (\x,\y) circle (2.9pt);
    \draw[oldc,thick] (\x,\y) circle (2.9pt);}

  \foreach \x/\y in {0.2/1.59, 1.0/1.69, 1.8/1.66, 2.6/1.48,
                     4.2/0.90, 5.0/0.79, 5.8/0.80, 6.6/0.94, 7.4/1.11}
    \fill[newc] (\x,\y) circle (2.7pt);

  \foreach \x/\y in {2.55/1.22, 3.65/0.80, 4.75/0.55}
    \draw[acc,->,semithick] (\x,\y) -- (3.42,1.15);
  \fill[newc] (3.42,1.15) circle (2.7pt);
  \draw[acc,very thick] (3.42,1.15) circle (5.6pt);

  \node[newc,anchor=west,font=\scriptsize] at (7.7,1.14) {new points};
  \node[oldc,anchor=west,font=\scriptsize] at (7.7,0.86) {previous points};
  \node[anchor=west,font=\small] at (4.9,1.74)
    {$\theta_i = \dfrac{\sum_k w_k\, \theta_i^{\,\mathrm{old}}}{\sum_k w_k},
      \ \ w_k = \dfrac{1}{d_k}$};
  \node[anchor=north west,font=\scriptsize,acc] at (2.3,0.42)
    {the $k=3$ nearest previous points};
\end{tikzpicture}
\caption{Transfer of a coverage field between the surface points of
successive cycles. One cycle advances the surface by a small fraction of the
grid spacing, so the two point clouds sample nearly the same surface at
different positions and with different numbers of points. The coverage at each
new point is the inverse-distance average over the three nearest previous
points, where $d_k$ is the distance to previous point $k$.}
\label{fig:transfer}
\end{figure}
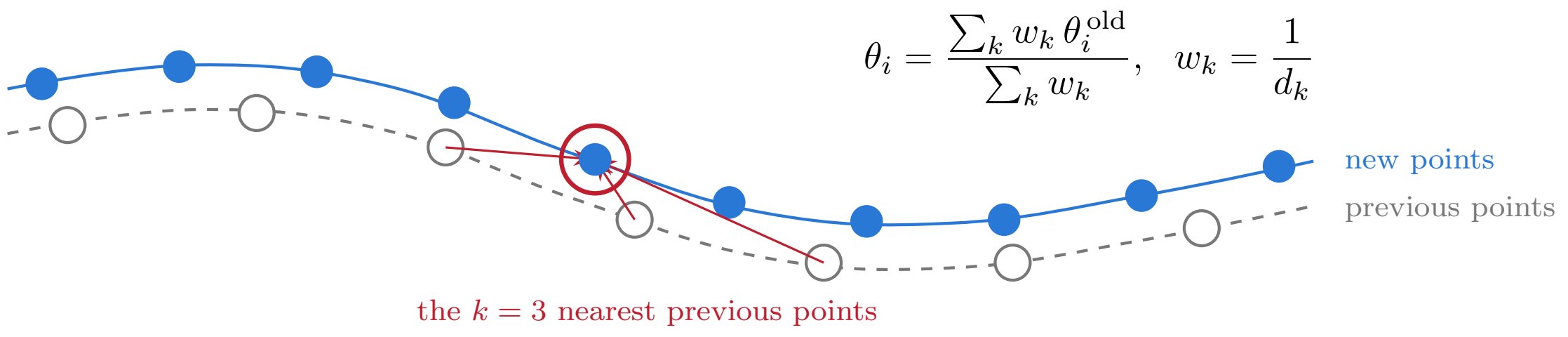

At the end of each cycle, the level set is sampled again, and the new surface points differ from the previous ones in
number and position. Each coverage at a new point is interpolated as the
inverse-distance average over the three nearest points of the previous cloud,
which are found with a k-d tree (Fig.~\ref{fig:transfer}). In this way, the surface termination left by the previous cycle, on which the self-limiting behaviour of the next dose depends, is carried over to the new points. Since one cycle moves the surface by a small fraction of the grid spacing, an average over three neighbours smooths the sampling noise and the interpolation remains local.

\section{Verification and test cases}
\label{sec:results}

\subsection{Verification and sources of the constants}
\label{sec:provenance}

The implementation is verified at three levels. First, the Jacobian assembled from the stored exponents by Eq.~(\ref{eq:drate}) is compared at random admissible coverages with the derivative of the same rate laws computed symbolically with SymPy~\cite{Meurer2017} and with central differences of the residual. No hand-derived reference is required for this check of the Jacobian or for the atom balance of every equation, which is checked when a file is read. Second, the C++ surface model is compared with the Python evaluator, which solves the equations above from the same compiled data with separate solver code that uses only the Python standard library. Agreement between the two shows that the compiled data contain all the information needed to evaluate the chemistry. Third, the results are compared with external sources. Such a comparison tests the reactions as written in the file, the derived rate laws and the coverage balances together, and it also tests the compilation of the free sites, \texttt{order} entries and reversible steps into the stored exponents. The constants of five reaction files are transcribed from published mechanisms or from hand-written ViennaPS models, namely the standard diamond growth model of May and Mankelevich~\cite{May2008}, the GaAs MOCVD mechanism of Mountziaris and Jensen~\cite{Mountziaris1991}, the SF$_6$/O$_2$ etch model based on Belen et al.~\cite{BelenSF6O2}, the fluorocarbon etch model based on La Magna and Garozzo~\cite{LaMagna2003}, and an Al$_2$O$_3$ ALD mechanism assembled from Puurunen's review~\cite{Puurunen2005} with barriers from Travis and Adomaitis~\cite{Travis2014} and Sperling et al.~\cite{Sperling2020}. In the other three reaction files, standard reaction forms are combined with illustrative constants chosen to test particular entries of the format.

\subsection{Simulation set-up}
\label{sec:setup}

The eight cases are run with two example programs of ViennaPS, one that solves the coverage balance at steady state and advances the surface for a given process time, and one that runs a stage list. Table~\ref{tab:cases} gives the temperature, the geometry and the variations of each run and the reference it is compared with. The other conditions, including the incident fluxes or pressures and the rate constants, are those of the reaction files (Appendices~\ref{note:sf6o2} to \ref{note:al2o3}). The commands, grids and ray counts of the runs are given in the README of the example directory and in the reproduction scripts archived with the data (Data availability).

The ray tracer draws new random seeds in every run. Each number quoted for the profiles of the two-dimensional cases of Table~\ref{tab:cases} is the mean of three runs, which differ by at most $0.18$\,nm in every quoted thickness and depth. The trench depths of the ray-count sweep are single runs. For the cylindrical holes, the quoted numbers are also means of three runs, with ranges of up to $0.4\,\%$ in depth and $4.2$\,nm in the radius at half depth. The moving-surface values of Table~\ref{tab:mechanisms} are taken on a blanket surface on a $2$\,nm grid at $5000$ rays per surface point, and the alumina value on the open field of the $0.2$\,s cavity run.

In the ALD case, twenty cycles are simulated, each consisting of a TMA dose, a $0.5$\,s purge, a water exposure of $10$\,s and a second $0.5$\,s purge. The TMA dose lasts $0.02$, $0.05$, $0.1$, $0.14$, $0.2$, $0.28$, $0.4$ or $0.8$\,s, and the run with the $0.2$\,s dose is repeated with a water exposure of $30$\,s. All cavity runs begin on a dehydroxylated surface, with bare oxygen bridges on the anion sites and bare aluminium on the cation sites. The coverages are integrated over twelve time steps per pulse and the surface is advected once per cycle. The cavity runs use the GPU flux engine with line elements. In a cavity whose walls absorb nothing, this engine gives the source flux to within $1\,\%$ in every $10\,\upmu$m of the cavity. The film thickness is read along a cut at half the gap height and measured from the cavity entrance up to $0.5\,\upmu$m before its closed end. The half-thickness depth is taken relative to the saturated thickness of $22.6$\,\AA\ ($2.26$\,nm), and the open-field growth is the mean film thickness on the top surface of the structure.

\begin{table}[tbp]
\centering
\footnotesize
\begin{tabular}{@{}p{22mm}c p{48mm} p{44mm}@{}}
\toprule reaction file & $T$ (K) & geometry and variations & reference \\
\midrule
\texttt{silane\_inputs} & $900$, $1100$ & trench $40 \times 180$\,nm, $22$\,nm film,
  at $k_\mathrm{ads}$ and $100$ and $1000$ times it, and at $1100$\,K
  & \texttt{silane}, given as flux and $s_0$ \\
\texttt{sige\_stack}    & $900$  & Si/SiGe superlattice pillar, $8$\,nm film
  & the same file on each material \\
\texttt{diamond}        & $900$ to $1400$ & fin $60 \times 90$\,nm, and the
  reverse constant off and $\times 10$
  & Eq.~(\ref{eq:diamondR}) \\
\texttt{gaas\_cvd}      & $800$ to $1300$ & trench $80 \times 120$\,nm,
  $20$\,nm film, and a methyl sweep
  & \texttt{gaas\_reversible} \\
\texttt{ar\_sputter}    & $300$  & masked substrate at $100$ and $200$\,eV
  mean ion energy
  & $N_\theta = 0$, velocity direct \\
\texttt{sf6o2}          & $300$  & $0.4\,\upmu$m hole, $160$\,s, four feed
  gases, and a masked trench $60$\,nm wide, $0.5$\,s, at $125$ to $8000$ rays
  & \texttt{SF6O2Etching} in the same hole and trench \\
\texttt{polymer\_etch}  & $300$  & masked SiO$_2$, $10$\,s, at two
  fluorocarbon fluxes
  & \texttt{FluorocarbonEtching}~\cite{LaMagna2003} \\
\texttt{al2o3}          & $450$  & $20$ cycles in a lateral cavity at eight dose times
  & $1.1$\,\AA/cycle~\cite{Travis2014} \\
\bottomrule
\end{tabular}
\caption{The eight cases with their temperatures, geometries and references.
Each case is run by one of the two example programs of ViennaPS, which take the compiled reaction file and the geometry as command-line arguments. \texttt{SF6O2Etching} and \texttt{FluorocarbonEtching} are the
hand-written models of ViennaPS.}
\label{tab:cases}
\end{table}

\subsection{Verification of the derived equations}
\label{sec:verify}

The verification has three levels (Section~\ref{sec:provenance}), and the third, the comparison with external sources, is the subject of Sections~\ref{sec:sf6o2} to \ref{sec:ald}. At the first level, the Jacobian assembled by Eq.~(\ref{eq:drate}) is checked for the eight reaction files of the seven chemistries in Table~\ref{tab:mechanisms} that have coverages. At ten random admissible coverage vectors per file, its largest relative difference from the symbolic derivative of the rate laws is $1.3 \times 10^{-16}$ over $2660$ entries. Central differences of the residual, taken at twenty such vectors, agree with the assembled Jacobian to a relative difference of at most $1.3 \times 10^{-10}$ over $5320$ entries, the round-off limit of the finite differences. At the second level, the single-point values of the C++ model and the Python evaluator for the steady-state cases agree in all six significant digits shown in Table~\ref{tab:mechanisms}. Moving-surface runs of ViennaPS on a blanket surface with ray-traced fluxes give rates within one per cent of these values for all cases. In the cases without ions, this difference is due only to the error of the moving-surface calculation. For the largest case, \texttt{gaas\_cvd}, the run in the table and five repetitions of it at $5000$ rays per surface point give rates between $0.53\,\%$ below and $0.10\,\%$ above the single-point value, and the other cases without ions differ by at most $0.13\,\%$. In a case with ions, the moving-surface calculation also samples the energy spread and the angular distribution of the ion source, whereas the single-point value is evaluated at the mean energy and normal incidence. For argon sputtering, whose ion source has an energy spread of $10$\,eV and a cosine exponent of $200$, the tabulated run and five further runs give rates between $0.62$ and $0.75\,\%$ below the single-point value. Averaging the yield over the energy and angular distributions of the source lowers it by $0.72\,\%$, and with a monoenergetic source at normal incidence, five runs differ from the single-point value by at most $0.12\,\%$.

\begin{table}[htbp]
\centering
\footnotesize
\setlength{\tabcolsep}{4pt}
\begin{tabular}{@{}clp{34mm}rrl@{}}
\toprule
& reaction file & capability & single point & moving surface & agreement with the source \\
\midrule
1 & \texttt{silane\_inputs} & three reactions, both input forms
  & $7.17808 \times 10^{-2}$ & $7.17856 \times 10^{-2}$ & $0.003\,\%$ vs \texttt{silane} \\
2 & \texttt{sige\_stack} & per-material rates
  & $7.17786 \times 10^{-2}$ & $7.17481 \times 10^{-2}$ & form only~\cite{Coltrin1984} \\
3 & \texttt{diamond} & a reversible step
  & $7.27846 \times 10^{-1}$ & $7.26912 \times 10^{-1}$ & $0.26\,\%$ on $R$~\cite{May2008} \\
4 & \texttt{gaas\_cvd} & two site types, 30 reactions
  & $1.88735 \times 10^{1}$ & $1.88333 \times 10^{1}$ & $2 \times 10^{-4}\,\%$ vs reduced~\cite{Mountziaris1991} \\
5 & \texttt{ar\_sputter} & no adsorbate, $N_\theta = 0$
  & $-1.10117$ & $-1.09375$ & form only~\cite{Steinbruchel1989} \\
6 & \texttt{sf6o2} & threshold ion yields, hand-written reference
  & $-5.80683 \times 10^{1}$ & $-5.80138 \times 10^{1}$ & exact~\cite{BelenSF6O2} \\
7 & \texttt{polymer\_etch} & two solids at once
  & $-6.24144$ & $-6.29661$ & exact without polymer~\cite{LaMagna2003} \\
8 & \texttt{al2o3} & transient, cyclic
  & $1.1299$\,\AA/cyc & $1.1297$\,\AA/cyc & $3\,\%$ vs $1.1$\,\AA/cyc~\cite{Travis2014} \\
\bottomrule
\end{tabular}
\caption{The eight cases, each solved at a single point and on a moving surface. Rates are in nm/s on a blanket surface with unobstructed flux, where the two solutions of a case without ions differ only by the error of the moving-surface calculation. The moving-surface values were computed with $5000$ rays per surface point. The alumina row gives the growth per cycle, from the limit cycle without transport and from the open field of the $0.2$\,s cavity run of Fig.~\ref{fig:pillarhall}. Cases 3, 4 and 6 use published constants, and cases 7 and 8 published constants with one illustrative value each. For these cases, the last column gives the agreement with the source (Section~\ref{sec:provenance}). The other cases combine reaction forms from the literature with illustrative constants, and their citations refer to the reaction forms.}
\label{tab:mechanisms}
\end{table}

The transient solver is needed where the coverages do not reach a steady state,
such as during the pulses of an ALD cycle. The time-dependent equations are stiff, because a model built from individual surface steps contains intermediates whose lifetimes are orders of magnitude shorter than a pulse. Errors in site conservation also accumulate from cycle to cycle. The
modified Patankar-Euler scheme used here conserves the occupancy of each site
type and keeps the coverages non-negative for any step length (Section~\ref{sec:transient}). For a minimal model with an intermediate of lifetime $1.2$\,ms, explicit Euler is
stable only for steps shorter than $2.4$\,ms. At longer steps, the intermediate coverage turns negative and is clamped to
zero at every second step, and it is zero at the end of the pulse. With the conservative scheme, the intermediate coverage stays positive at all six fixed step lengths tested, and both schemes converge to the same value (Table~\ref{tab:euler}). Because the coverages on which the velocity depends can fall by orders of magnitude within a small fraction of a pulse, the growth of a cycle is accumulated on the sub-steps of the integrator. Evaluating the velocity once per time step and
multiplying by the step length overestimates the thickness deposited in the co-reactant pulse of a minimal cycle of two reaction files by a factor of up to $40$, whereas the accumulated integral is the same for all five step lengths tested, from one to $150$ time steps per pulse (Fig.~\ref{fig:cycle}
and Table~\ref{tab:growthint}).

\subsection{A hand-written etch model reproduced from a reaction file}
\label{sec:sf6o2}

ViennaPS contains a hand-written model of silicon etching in SF$_6$/O$_2$ whose closed-form coverages are taken from Belen et al.~\cite{Belen2005,BelenSF6O2}. This model consists of a C++ class and the plasma-etching surface model and particle classes it uses. Without comments and blank lines, these classes have about $690$ lines of code and the reaction file for this chemistry has $40$ lines. The file contains all constants of the class and no additional terms (Appendix~\ref{note:sf6o2}). It has two coverages, fluorine and oxygen, and nine reactions, of which seven act on silicon and seven on the mask. The four steps with threshold ion yields are the ion-enhanced etching of silicon, the ion-enhanced release of fluorine on the mask, the sputtering and the removal of oxygen. At a single point with the fluxes given in the file, the published balance and the equations derived from the reaction file give the same result, \begin{equation*} \theta_F = 4.905587 \times 10^{-1} , \qquad \theta_O = 1.471532 \times 10^{-1} , \qquad v_n = -5.806831 \times 10^{1}\ \mathrm{nm/s} , \end{equation*} to seven significant digits. This agreement relies on the \texttt{order} entries of the two steps that consume fluorine on silicon.
The chemical etch consumes four fluorine atoms per silicon atom and the
ion-enhanced step two, but both rates are first order in $\theta_F$. Exponents
taken from the stoichiometry would make them fourth and second order.

To include the transport in the comparison, both implementations were run with
the same example program, geometry and process time. On a blanket surface, the moving-surface etch rate obtained with the file differs by $0.09\,\%$ from the single-point value that both implementations give (Table~\ref{tab:mechanisms}).  In a masked trench at the highest ray count, $8000$ rays per surface point, the etched depths from the file and from the class are $29.212$ and $29.217$\,nm, respectively. Over six doublings of the ray count, the difference between the two depths has no systematic sign. It scatters about a trend that falls with the inverse square root of the ray count, as Monte Carlo noise does. This trend reaches $0.03\,\%$ at the highest ray count (Fig.~\ref{fig:conv}). The two implementations therefore agree within the noise of the transport.

\begin{figure}[!htbp]
\centering
\includegraphics[width=0.98\textwidth]{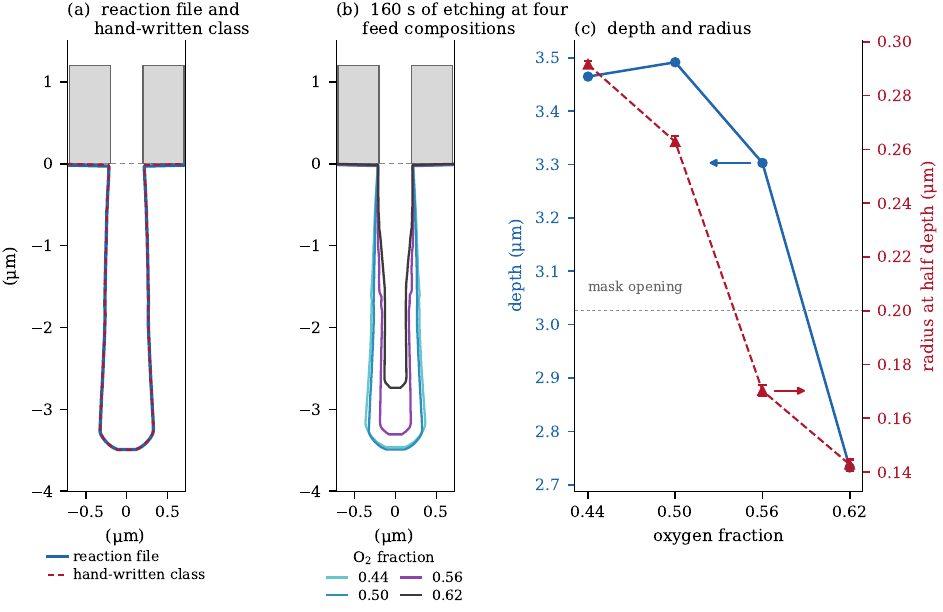}
\caption{Silicon etched in SF$_6$/O$_2$ in a cylindrical hole $0.4\,\upmu$m in
diameter through a $1.2\,\upmu$m mask, close to the structure of Bobinac et
al.~\cite{Bobinac2023}.
(a) Profiles after $160$\,s at an oxygen fraction of $0.50$ from the reaction file and from the hand-written class of ViennaPS, with the same geometry and ray count. (b) The
profile at each of the four feed-gas compositions of Bobinac et al., labelled by
the oxygen fraction $y_{\mathrm{O_2}}$. (c) Depth and radius at half depth against the oxygen fraction, with the radius of the mask opening marked. The profile bows when the radius at half depth lies above this mark and tapers inward when it lies below. Panels (a) and (b) show the first of three runs. The points are means
of three runs and the bars give their range.}
\label{fig:sf6o2}
\end{figure}

A cylindrical hole $0.4\,\upmu$m in diameter through a $1.2\,\upmu$m mask, close to the structure of Bobinac et al.~\cite{Bobinac2023}, is etched for $160$\,s with the file and with the class, and the resulting profiles overlap (Fig.~\ref{fig:sf6o2}a). Over three runs each, the mean depths from the file and from the class are $3.491$ and $3.492\,\upmu$m and the mean radii at half depth $263$ and $262$\,nm, respectively, and the ranges of the repeated runs overlap. Bobinac et al. also give four feed-gas compositions for a chamber at $25$\,mTorr ($3.33$\,Pa)
with an inductive coil power of $800$\,W and a bias of
$-120$\,V~\cite{Bobinac2023}. Over these compositions, the oxygen fraction rises
from $0.44$ to $0.62$ as the fluorine flux falls and the oxygen flux rises,
with the ion flux held at $10^{16}$\,cm$^{-2}$s$^{-1}$ (Fig.~\ref{fig:sf6o2}b).
The depth is $3.46$ and $3.49\,\upmu$m at the two lowest oxygen fractions and falls to $3.30$ and $2.73\,\upmu$m at the two highest, respectively. The radius at half depth falls from $292$ through $263$ and $170$ to $143$\,nm
(Fig.~\ref{fig:sf6o2}c). Measured against the $200$\,nm radius of the mask
opening, the profile changes from bowed at the two lower oxygen fractions to inward tapered at the two higher ones as the sidewall passivation increases. Belen et al. measured this progression for silicon etching in SF$_6$/O$_2$~\cite{BelenSF6O2}, and the simulations with the reaction file reproduce it.

The ion-neutral synergy observed by Coburn and Winters~\cite{Coburn1979} is also reproduced. Setting either the neutral or the ion flux to zero, with the reaction file otherwise unchanged, separates the two channels. Neutrals alone etch at $3.58 \times 10^{-2}$\,nm/s and ions alone at $0.445$\,nm/s, while together they etch at $58.07$\,nm/s, $121$ times the sum of the separate rates.

\subsection{Published mechanisms beyond their nominal conditions}
\label{sec:published}

\begin{figure}[!htbp]
\centering
\includegraphics[width=0.78\linewidth]{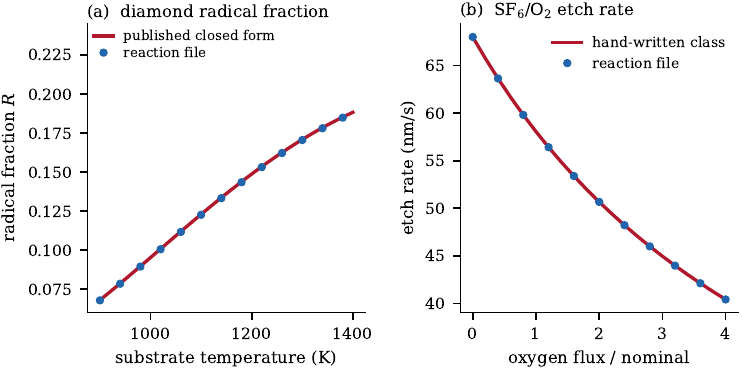}
\caption{Reaction files compared with published expressions.
(a) Radical fraction of a diamond surface and Eq.~(\ref{eq:diamondR}) from $900$ to $1400$\,K~\cite{May2008}. (b) SF$_6$/O$_2$ etch rate for oxygen fluxes from zero to four times the nominal value, from the reaction file and from the hand-written class, which implements the balance of Belen et al.~\cite{BelenSF6O2}. The reaction file was written for
the nominal point only.}
\label{fig:verify}
\end{figure}

The constants of five of the eight reaction files are taken from the literature or from hand-written ViennaPS models, and the last column of
Table~\ref{tab:mechanisms} gives the agreement with each source. Two of the sources give results for more than one operating point, and with these the corresponding reaction files can be tested beyond the conditions for which they were written. One of these sources is the standard growth model of diamond chemical vapour deposition in the form given by May and Mankelevich~\cite{May2008}, which contains a reversible hydrogen abstraction. The free fraction computed from the reaction file is the fraction of surface radical sites, and
over $900$ to $1400$\,K it agrees with the closed form
\begin{equation}
  R = \left[ 1 + 0.3 \exp(3430/T)
      + 0.1 \exp(-4420/T) \frac{[\mathrm{H_2}]}{[\mathrm{H}]} \right]^{-1} ,
  \label{eq:diamondR}
\end{equation}
in which $[\mathrm{H_2}]/[\mathrm{H}]$ is the ratio of the gas concentrations above the surface and the last term is due to the reverse of the abstraction (Fig.~\ref{fig:verify}a). The second source is the hand-written SF$_6$/O$_2$ class, and the etch rate from the reaction file agrees with that of the class as the oxygen flux is varied from zero to four times its nominal value, although the file was written for the nominal point only (Fig.~\ref{fig:verify}b).

\begin{figure}[!htbp]
\centering
\includegraphics[width=0.96\textwidth]{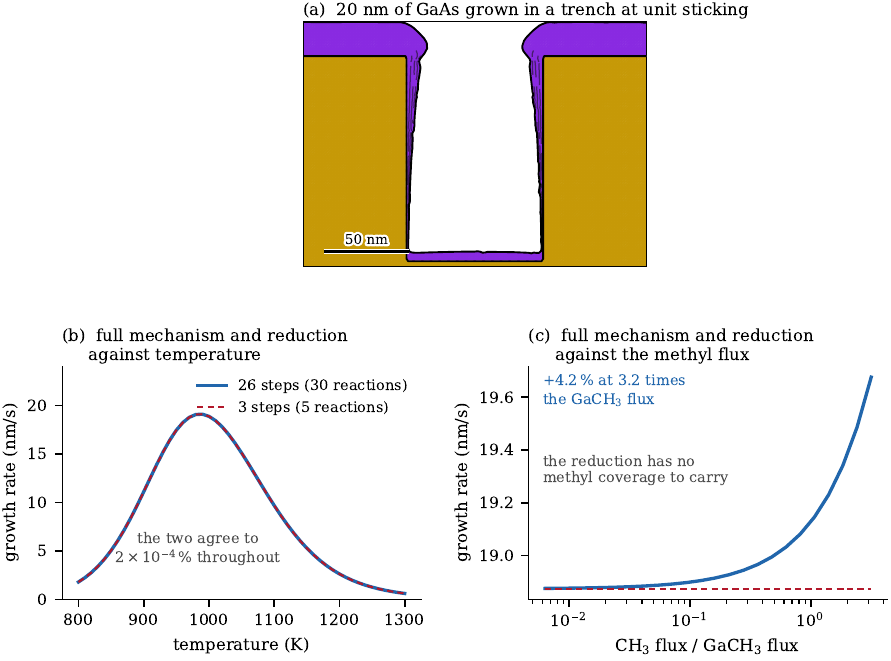}
\caption{GaAs from TMG and arsine with the mechanism of Mountziaris and
Jensen~\cite{Mountziaris1991}, thirty reactions with seven coverages on two
site types. (a) With unit sticking, the flux is absorbed near the opening and the deposit overhangs at the mouth. The film on the floor reaches $28\,\%$ of the thickness on the field. GaCH$_3$ that desorbs again is not traced further. (b) Growth rate of the full model and of its reduction to the three steps that the same authors identified as dominant, as a function of temperature. (c) Both with methyl added to the gas phase. Methyl opens a growth channel that the reduction cannot represent.}
\label{fig:gaas}
\end{figure}

The GaAs mechanism of Mountziaris and Jensen~\cite{Mountziaris1991} is used to test the format on a larger system. Transcribed from their Table~II, it compiles to thirty reactions with seven coverages and two solids. The coverages lie on two site types, the cation and anion positions of the surface, and the solids are GaAs and carbon incorporated on an arsenic site. By sensitivity analysis, the same authors identified three dominant surface steps, the reversible adsorption of GaCH$_3$ and of AsH and their reaction to GaAs. A reaction file reduced to these steps compiles to five one-way reactions, because both adsorption steps are reversible. Without methyl in the gas phase, the full and the reduced file agree on the growth rate to $2 \times 10^{-4}\,\%$ between $800$ and $1300$\,K. In both, the rate has its maximum near $990$\,K (Fig.~\ref{fig:gaas}b). The coverages of adsorbed GaCH$_3$ and AsH stay close to the equilibrium of adsorption and desorption, and above the maximum their product falls faster than the rate constant of their reaction to GaAs rises. In the full model, methyl released by the pyrolysis of trimethylgallium (TMG) removes the hydrogen from an adsorbed AsH and leaves a bare arsenic atom. The bare arsenic reacts with adsorbed GaCH$_3$ to form GaAs, with a barrier of $20$\,kcal/mol, compared with $29.3$\,kcal/mol for the main growth step. Because the reduced file contains no methyl, in the gas phase or as a coverage, and no bare arsenic, its rate does not respond to methyl. With a methyl flux of $3.16$ times the GaCH$_3$ flux, the growth rate of the full model rises by $4.25\,\%$ (Fig.~\ref{fig:gaas}c).

In a trench $80$\,nm wide and $120$\,nm deep, the GaAs film is $5.6$\,nm thick on the floor and $3.1$\,nm on the sidewall at half depth when $20$\,nm have grown on the field (Fig.~\ref{fig:gaas}a). This profile is caused by the unit sticking of the adsorption steps, although only $2.5 \times 10^{-3}$ of the incident GaCH$_3$ is incorporated into the film. With unit sticking, a GaCH$_3$ molecule is absorbed at a wall with a probability equal to the free cation fraction, and GaCH$_3$ that desorbs again leaves the surface as a product and is not traced further (Appendix~\ref{note:gaas}).

\subsection{Deposition, etching and sputtering with one surface model}
\label{sec:chem}

\begin{figure}[!htbp]
\centering
\includegraphics[width=0.98\textwidth]{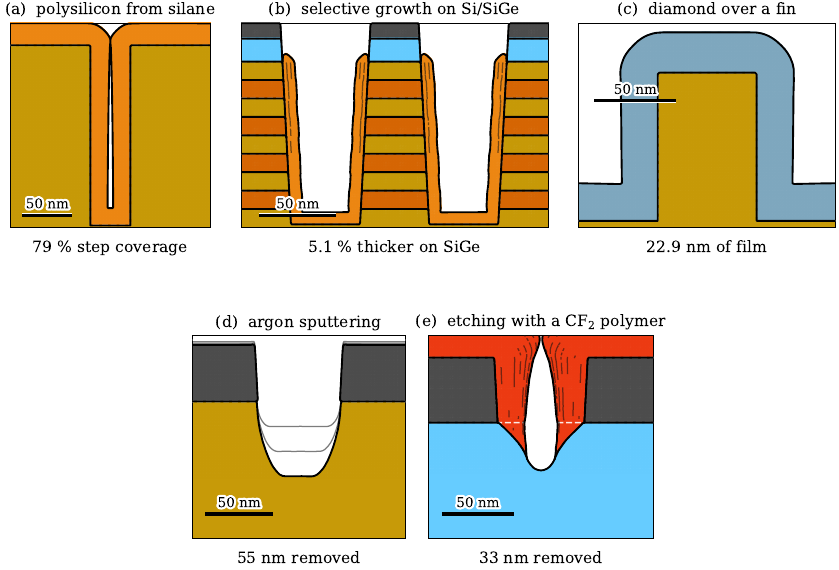}
\caption{Five further chemistries, each given by one reaction file. (a) Polysilicon from silane in a trench of aspect ratio $4.5$ at the adsorption rate constant given in the file. The closing film leaves a void along the axis. (b) Selective epitaxy on a Si/SiGe superlattice pillar under an oxide cap, with rate constants given per material. (c) Diamond over a silicon fin, with a reversible abstraction step. (d) Argon sputtering of silicon through a mask at a mean ion energy of $100$\,eV, a model without adsorbates. (e) SiO$_2$ etched while a fluorocarbon polymer is deposited as a second solid, at a CF$_2$ flux of $100 \times 10^{15}$\,cm$^{-2}$s$^{-1}$. Each panel shows the first of three runs, and the number under it is the mean of the three. Figs.~\ref{fig:case1} to \ref{fig:case7} show the variations of each case.}
\label{fig:chem}
\end{figure}

The five reaction files of Fig.~\ref{fig:chem} contain further reaction types. The variations below were made by changing values in the reaction files or options of the example programs, with the compiled surface model left unchanged (Appendices~\ref{note:silane} to \ref{note:polymer}).

In polysilicon deposition from silane, the adsorption can be given as a rate
constant $k_{\mathrm{ads}}$ in cm/s, which ViennaChem converts into a sticking
coefficient (Section~\ref{sec:rates}). A value of $1.9256 \times 10^{4}$\,cm/s with a barrier of $0.6$\,eV corresponds to a sticking coefficient of $4.37 \times 10^{-4}$ at $900$\,K, and the film in a
trench of aspect ratio $4.5$ reaches a step coverage of $79\,\%$ (Fig.~\ref{fig:chem}a). Because the sticking coefficient enters the absorption probability of the ray tracer as well as the rate law, raising $k_{\mathrm{ads}}$ by two and by three orders of magnitude lowers the step coverage to $66$ and $55\,\%$, respectively, and widens the void that the closing film leaves along the trench axis (Fig.~\ref{fig:case1}). Per-material constants make the same
chemistry selective. On a superlattice of silicon and silicon germanium under
an oxide, the growth rate on bare SiGe is $6.47$ times that on Si and zero on the oxide. At the end of the run, the film on the SiGe bands is $5.1\,\%$ thicker than on the Si bands, because a SiGe band grows at the SiGe rate only until the deposit covers it (Fig.~\ref{fig:chem}b and Fig.~\ref{fig:case2}).

In the diamond model, removing the reverse abstraction or multiplying its
constant by ten gives radical fractions of $0.1601$ and $0.0895$, respectively, at $1200$\,K, against $0.1605$ and $0.0896$ from Eq.~(\ref{eq:diamondR}). Through the radical fraction, the film grown on the field in $33.6$\,s changes from $22.9$\,nm to $24.8$ and $13.7$\,nm, respectively (Fig.~\ref{fig:chem}c and Fig.~\ref{fig:case3}).

Physical sputtering of silicon by argon ions through a mask involves no adsorbates and has no coverages to solve for. Doubling the mean ion energy from $100$ to $200$\,eV deepens the cavity by a factor of $1.75$, in agreement with the threshold form of the yield at the substrate threshold of $20$\,eV. The selectivity to the mask falls from $15$ to $12$, because the mask has the
higher threshold (Fig.~\ref{fig:chem}d and Fig.~\ref{fig:case5}).

In fluorocarbon etching of SiO$_2$, a polymer is deposited while the oxide is
etched. The sticking coefficients, the chemical etch and the ion yields are those of the fluorocarbon model of ViennaPS, which is based on the work of La Magna and Garozzo~\cite{LaMagna2003}, and the rate of polymer formation is illustrative. On an oxide surface free of polymer, the reaction file and this model give the same etch rate. The
hand-written model treats the polymer as a coverage that may exceed unity and
clamps the resulting velocity. In the reaction file, the polymer is a solid of its own, and the velocity is the sum of the contributions of the two solids, with signs given by the side of each equation on which they appear (Section~\ref{sec:velocity}). At the fluxes given in the file, a $60$\,nm opening in a mask is etched to a depth of $63$\,nm in $10$\,s. With a twentyfold fluorocarbon flux, polymer covers the mask and narrows the opening to a slit, and the trench reaches a depth of $33$\,nm (Fig.~\ref{fig:chem}e and Fig.~\ref{fig:case7}). As the fluorocarbon flux is raised further with the rest of the file unchanged, the open-field velocity changes sign near $170 \times 10^{15}$\,cm$^{-2}$s$^{-1}$, and at $200 \times 10^{15}$\,cm$^{-2}$s$^{-1}$ the surface grows at $0.98$\,nm/s.

\subsection{Atomic layer deposition in a high-aspect-ratio cavity}
\label{sec:ald}

ALD of Al$_2$O$_3$ from TMA and water tests the parts of
the formulation that the steady-state cases do not use, namely the transient
coverages, the stage list of a cycle, the growth integrated over a cycle and
the transfer of coverages between cycles. The two half-cycles are written as separate reaction files. Both files describe the same surface with two site types, the
anion and cation positions of the lattice (Appendix~\ref{note:al2o3}). The reactions are taken from Puurunen's review of this process~\cite{Puurunen2005}. The TMA half-cycle contains ligand exchange with a
surface hydroxyl and a dissociative channel on a bare oxygen bridge, with
barriers from Travis and Adomaitis~\cite{Travis2014}. In the water half-cycle, water removes the surface methyl groups with a barrier taken from Sperling et al.~\cite{Sperling2020}, and two hydroxylated aluminium species condense into one formula unit of Al$_2$O$_3$.
The TMA reaction probability of $10^{-3}$ per incident molecule is the logarithmic midpoint of the range that Arts et al. measured in lateral high-aspect-ratio structures~\cite{Arts2019}. The kinetics of Sperling et al. give a reaction probability of $2.7 \times 10^{-5}$ for water on a methylated surface at $450$\,K. A cycle consists of four stages, a TMA dose, a purge, a water dose and a second purge, each with its own duration and reaction file. The state at the end of one stage is the initial condition of the next (Section~\ref{sec:cyclic}).

\begin{figure}[!htbp]
\centering
\includegraphics[width=0.74\linewidth]{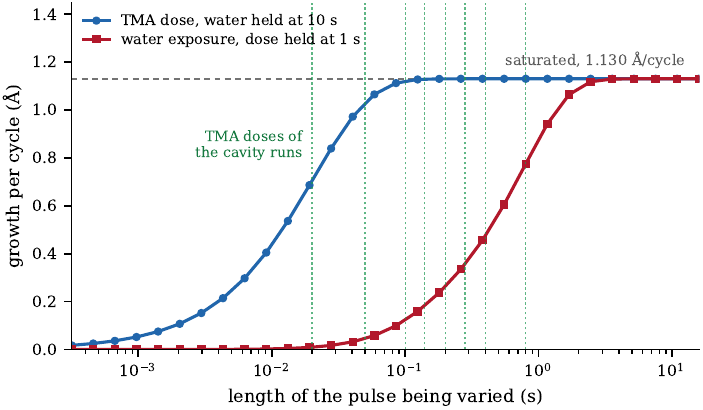}
\caption{Growth per cycle of the Al$_2$O$_3$ model as a function of the length of each half-cycle pulse. The TMA dose is varied at a water exposure of $10$\,s, and the water exposure at a TMA dose of $1$\,s. At $100$\,mTorr of each reactant, both curves saturate at $1.130$\,\AA\ per cycle and reach $99\,\%$ of this value after $0.095$\,s for the TMA dose and $2.5$\,s for the water exposure. The saturated value is determined by the site density, the stoichiometry and the film density. Green lines mark the eight TMA doses of the cavity runs.}
\label{fig:al2o3sat}
\end{figure}

For a fully hydroxylated starting surface at $450$\,K, the growth per cycle is $1.130$\,\AA. The surface returns to full hydroxyl termination at the end of every cycle, and the process reaches a limit cycle. Travis and Adomaitis obtained $1.1$\,\AA\ for the same surface and temperature after a saturating TMA exposure of $2$\,s at $2$\,Pa~\cite{Travis2014}. At a partial pressure of $100$\,mTorr of each reactant, the saturation curves
of the two half-cycles reach $99\,\%$ of the plateau after a TMA dose of
$0.095$\,s and a water exposure of $2.5$\,s (Fig.~\ref{fig:al2o3sat}).

\begin{figure}[!htbp]
\centering
\includegraphics[width=0.92\textwidth]{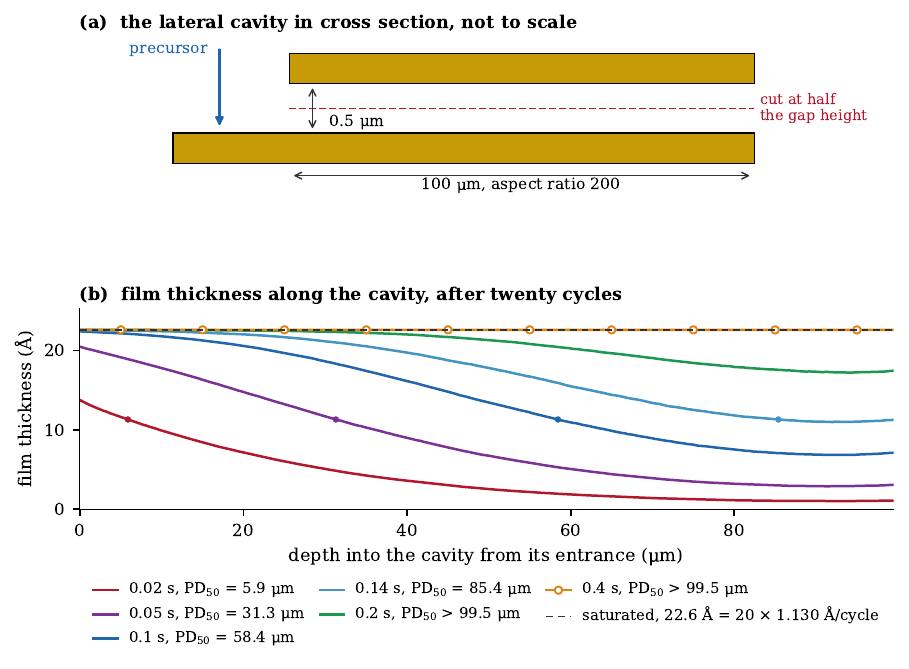}
\caption{Al$_2$O$_3$ from TMA and water in a lateral cavity of aspect ratio
$200$. (a) The structure, with the cut along which the film is measured. (b)
The film thickness along the cavity after twenty cycles at six of the eight TMA doses,
each followed by a water exposure of $10$\,s, against the depth from the
cavity entrance, with the half-thickness penetration depth PD$_{50}$ marked
where it lies inside the cavity. The saturated thickness is the growth per cycle of Table~\ref{tab:mechanisms} multiplied by the number of cycles. The profile of the $0.4$\,s dose coincides with it along the whole cavity and is marked by symbols.}
\label{fig:pillarhall}
\end{figure}

With the same two files, twenty cycles were simulated in a lateral cavity $0.5\,\upmu$m high and $100\,\upmu$m long (Fig.~\ref{fig:pillarhall}a), within the range of the lateral high-aspect-ratio test structures of Gao et al.~\cite{Gao2015}. Such structures are used to measure
conformality and to extract sticking coefficients from the thickness profile
after a dose~\cite{Arts2019,Cremers2019}, with diffusion-reaction models such as that of Ylilammi et al.~\cite{Ylilammi2018}. The TMA dose is varied from $0.02$ to $0.8$\,s. Each dose is followed by a water exposure of $10$\,s, four times the saturation time of the water half-cycle. The runs start from a dehydroxylated surface of bare oxygen bridges and bare
aluminium. On the open top surface of the structure, the film of the $0.2$\,s
dose grows by $1.12975$\,\AA\ per cycle over the twenty cycles, against $1.12982$\,\AA\ from the Python evaluator for the same cycles without
transport. Over all eight doses, including the $0.02$ and $0.05$\,s doses that do not saturate the open field, the two differ by at most $0.03\,\%$. With the
largest coverage change per sub-step raised from $\epsilon = 2 \times 10^{-4}$
to $10^{-3}$ in Eq.~(\ref{eq:substep}), the $0.2$\,s run and the evaluator give $1.12969$ and $1.12973$\,\AA, respectively. Compared with the finer tolerance, both values change by less than $0.01\,\%$, and the moving-surface value stays within $0.01\,\%$ of the evaluator.

For doses of $0.14$\,s and longer, the film saturates at the cavity entrance at $22.6$\,\AA, twenty times the open-field growth per cycle, and thins along the cavity (Fig.~\ref{fig:pillarhall}b). At $0.1$, $0.05$ and $0.02$\,s, it reaches $99$, $89$ and $59\,\%$ of this thickness at the entrance, respectively. The depth from the entrance at which the film has half its saturated thickness is $5.9$, $31.3$, $58.4$ and $85.4\,\upmu$m at $0.02$, $0.05$, $0.1$ and $0.14$\,s, respectively. From $0.2$\,s on, the film exceeds half its saturated thickness up to the closed end, and from $0.4$\,s on it lies within $0.1\,\%$ of saturation along the whole cavity. The profiles are set by the TMA dose, since tripling the water exposure to $30$\,s changes the profile of the $0.2$\,s dose by at most $0.8\,\%$ of the saturated thickness.

A saturated plateau followed by a decaying front is the profile measured in lateral high-aspect-ratio structures and described by the model of Ylilammi et al.~\cite{Ylilammi2018}. In the simulation, this shape is a consequence of the coupling of the rate laws to the transport. A TMA molecule that hits a wall reacts and is absorbed with a probability proportional to the summed coverage of hydroxyls and bare oxygen bridges. Where these are used up, the molecule is re-emitted and travels further into the cavity by diffuse reflection (Section~\ref{sec:transport}). The front is broad because a TMA molecule reacts with a probability of $10^{-3}$ per collision with a hydroxylated wall and collides about a thousand times before it reacts. At $0.1$\,s, the film at the closed end still has about a third of its saturated thickness. The position of the front depends on the TMA reaction probability given in the reaction file.

\section{Computational cost}
\label{sec:cost}

Solving the coverage balance of a reaction file with Newton's method costs more
per point than evaluating a closed form derived by hand. Measured in isolation,
the coverage calculation per point of the SF$_6$/O$_2$ chemistry takes
$\costPerPointFile$\,ns with the reaction file, about $\costRatio$ times as long as
with the closed form. The run time of a
feature-scale simulation grows by much less, because each step is dominated by
ray tracing and the coverage balance is solved in parallel over the surface points.
With the CPU flux engine, solving for the coverages takes at most
$\costShareWorstHost\,\%$ of the run time for each of the seven steady-state cases, and the share does not grow with the number of reactions. For \texttt{gaas\_cvd}, with thirty reactions and seven coverages, the share is $\costShareLargest\,\%$, less than the $\costShareSilane\,\%$ for the three reactions of \texttt{silane\_inputs}, because a larger reaction file also has more traced species and its transport time grows at least as fast. The GPU engine is the least favourable case, because the
transport runs on the GPU while the coverage balance is still solved on the CPU. The \texttt{sf6o2} run is $\costEngineSpeedup$ times faster there, and for each case, solving for the coverages takes at most $\costShareWorst\,\%$ of the shorter GPU run.

Table~\ref{tab:cost} gives the share of the run time spent solving for the coverages of the seven steady-state cases, with the CPU and with the GPU flux engine. The \texttt{ar\_sputter} case has no coverages and therefore no balance to solve. The SF$_6$/O$_2$ chemistry exists as a reaction file and as a hand-written closed form. On the CPU engine, solving for its coverages takes $\costShareFile\,\%$ of the run with the reaction file, compared with $\costShareClass\,\%$ with the closed form. On the GPU engine, the share is $\costShareFileGPU\,\%$ for
\texttt{sf6o2} and $\costShareLargestGPU\,\%$ for \texttt{gaas\_cvd}. Solving
for the coverages on the GPU would shorten the GPU runs by at most these
shares.

All timings were measured on an Intel Core Ultra 7 165H CPU with $16$
physical and $22$ logical cores and an NVIDIA RTX A500 Laptop GPU, with the
operating system Linux 6.17.0-40-generic (x86\_64, glibc~2.39), Python~3.12.3,
\texttt{OMP\_NUM\_THREADS=16} and \texttt{OMP\_WAIT\_POLICY=active}, which keeps
idle threads waiting actively between parallel loops. Each timing is the median of seven repetitions in
one geometry, a trench $80$\,nm wide and $120$\,nm deep on a $1.5$\,nm grid at
$2000$ rays per surface point, advanced by $5$\,nm. The share of the coverage calculation is the time spent solving for the coverages during a run divided by the wall time of the same run. The cost per point was measured by solving for the coverages at $200\,000$ points with the transport left out, for the reaction file and for the closed form.

\begin{table}[htbp]
\centering
\footnotesize
\begin{tabular}{@{}lccrr@{}}
\toprule
reaction file & reactions & coverages & CPU (\%) & GPU (\%) \\
\midrule
\texttt{silane\_inputs} & 3 & 2 & $0.083$ & $0.102$ \\
\texttt{sige\_stack} & 3 & 2 & $0.077$ & $0.101$ \\
\texttt{diamond} & 4 & 1 & $0.017$ & $0.040$ \\
\texttt{ar\_sputter} & 1 & 0 & $0.000$ & $0.000$ \\
\texttt{sf6o2} & 9 & 2 & $0.042$ & $0.061$ \\
\texttt{sf6o2}, hand-written & 9 & 2 & $0.007$ & $0.021$ \\
\texttt{polymer\_etch} & 7 & 2 & $0.046$ & $0.119$ \\
\texttt{gaas\_cvd} & 30 & 7 & $0.028$ & $0.124$ \\
\bottomrule
\end{tabular}
\caption{Share of the run time of a feature-scale simulation spent solving the coverage balance, with the CPU and with the GPU flux engine. Each row is the median of 7 runs in one geometry, a trench $80$\,nm wide and $120$\,nm deep on a $1.5$\,nm grid at $2000$ rays per surface point, advanced by $5$\,nm. The \texttt{sf6o2} chemistry appears twice, solved from the reaction file and from the hand-written closed form of ViennaPS. The \texttt{ar\_sputter} case has no adsorbates and therefore no balance to solve. With the GPU engine, only the flux calculation runs on the GPU. Solving for the coverages is still done on the CPU and takes a larger share of a shorter run. The timings were measured on an Intel Core Ultra 7 165H with 16 OpenMP threads, one per physical core, and on an NVIDIA RTX A500 Laptop GPU. The OpenMP threads wait actively between parallel loops.}
\label{tab:cost}
\end{table}

\section{Discussion}
\label{sec:discussion}

A level-set simulator that takes its surface chemistry as data can match the accuracy of a hand-written model. The SF$_6$/O$_2$ model of ViennaPS, derived by hand and implemented for this one chemistry, is reproduced from a $40$-line reaction file, both at a single point and in the etched depths of a masked trench. Ray tracing dominates the run time, and solving the coverage balance of the reaction file takes at most $\costShareWorst\,\%$ of it. Without recompilation, the surface model computes the velocity for reaction files of one to thirty reactions with up to seven coverages on two site types, including ion-driven steps and the simultaneous deposition and etching of two solids. It also integrates ten coverages through ALD cycles defined by two files. The constants of three of the eight cases are illustrative and were chosen to test individual entries of the format.

Supplying the mechanism as data separates the development of a chemistry from
that of the simulator. An intermediate, a reversible step or a second site type
is added in the reaction file, and the coverage balance, its Jacobian and the coupling to the transport are derived from the file. A mechanism can therefore be used in full. In the GaAs case, all thirty reactions are required, since the three-step reduction agrees with them to $2 \times 10^{-4}\,\%$ without methyl in the gas phase but misses a growth channel once methyl is present. Because profile predictions depend strongly on the reaction probabilities, yields and rate coefficients~\cite{Krueger2024}, the entries of a reaction file are also the parameters to vary in calibration, sensitivity analysis and uncertainty quantification. Barriers computed from first principles or yields
from molecular dynamics enter a feature-scale calculation through the same
entries.

The coupling between the chemistry and the transport is derived from the same rate laws as the coverage balance. A traced species is absorbed at a wall with its local reaction probability and re-emitted where the termination it reacts with is used up. In the alumina cavity, the thickness profiles computed with this coupling have the shape measured in lateral high-aspect-ratio structures~\cite{Ylilammi2018}. The TMA reaction probability of the file lies in
the range that Arts et al. extracted from such thickness profiles~\cite{Arts2019}.
A value obtained in this way can be written into the reaction file and used in other geometries with the rest of the file.

The formulation has the limits of mean-field kinetics and of the transport it
is coupled to. The rate of a step that consumes two adsorbates of one species depends on the square of the coverage of that species, the probability that two randomly chosen sites are both occupied. Correlations between neighbouring sites are not represented. Rate constants depend on temperature and material, whereas barriers on real surfaces can also vary with coverage. For the water half-cycle of
alumina, Sperling et al. report an effective activation energy that rises from $18$ to $35$\,kJ/mol as the methyl coverage decreases~\cite{Sperling2020}. A coverage-dependent barrier would add to the Jacobian a term with the same
structure as the existing terms, assembled from the stored exponents and the
coverage coefficients of the barrier. The coverage balance is local to each surface point, so lateral surface diffusion has to be added at the process level. Steric blocking of neighbouring sites by bulky adsorbates, as by the small-molecule inhibitors of area-selective ALD, is an extension of the rate laws left to follow-up work.

Products that leave the surface are not traced and do not redeposit, and during a purge only thermal steps proceed. A reactant that adsorbs and desorbs again, such as GaCH$_3$ in the GaAs model, is likewise not traced after it desorbs. Its transport is then set by the sticking coefficient, which is much larger than the net incorporation probability. Particles are traced along straight trajectories. Neutrals are launched from a cosine source and re-emitted diffusely, and ions are launched from a power-cosine source and reflected only at glancing incidence. Photon-driven steps, feature charging, gas-phase collisions and
continuum diffusion inside the feature would change the flux calculation, while
the surface model uses only the flux arriving at each point.

Two further extensions would keep the reaction file as the source of the chemistry and change only the data used by the solver or the transport. The incident fluxes
and the ion energy and angular distributions are currently given as numbers and
distribution parameters in the file. Equipment-scale simulation, for example
the hybrid plasma equipment model of the Kushner group~\cite{Kushner2009},
computes these quantities. Tabulated distributions from such a model would bring reactor-scale results into the feature-scale calculation without changes
to the solver, because the ion yield is evaluated after the energy and the
direction of each ion have been sampled. Since species and materials are
identified by name until the file is compiled, sticking coefficients, solid
densities and ion-yield parameters could also be read from databases indexed by
material and species, together with their sources and uncertainties. An entry in the reaction file would override the database value.

\appendix
\makeatletter
\gdef\thesection{\@Alph\c@section}
\def\@seccntformat#1{\appendixname~\csname the#1\endcsname\@seccntDot\hskip 0.5em}
\makeatother
\counterwithin*{figure}{section}
\counterwithin*{table}{section}

\section{The reaction-file format}
\label{note:format}

Table~\ref{tab:entries} lists the entries that a reaction file may contain, and Table~\ref{tab:phases} gives the four phases of a species with their role in the rate law of Eq.~(\ref{eq:massaction}). The quantities derived from a file, by ViennaChem when the file is compiled or by the surface model from the compiled data, are listed in Table~\ref{tab:derived}. A reaction file contains only the entries needed for its chemistry.
The silane file of Section~\ref{sec:example} uses nine of the nineteen rows of Table~\ref{tab:entries}, and the SF$_6$/O$_2$ file of Appendix~\ref{note:sf6o2} uses fourteen of them.

\begin{table}[htbp]
\centering
\footnotesize
\begin{tabular}{@{}llp{99mm}@{}}
\toprule
block & entry & meaning \\
\midrule
top level & \texttt{name}, \texttt{temperature} & the name and the surface temperature \\
        & \texttt{energyUnit}                   & eV or kcal/mol for every barrier \\
        & \texttt{pressureUnit}                 & the unit of any partial pressure \\
        & \texttt{siteTypes}                    & the site types and their densities \\
\midrule
species & \texttt{phase}                        & \texttt{gas}, \texttt{ion}, \texttt{surface} or \texttt{solid} \\
        & \texttt{flux} or \texttt{pressure}    & the supply of a traced gas, and the flux of an ion \\
        & \texttt{atoms} & the composition of a name that is not a formula \\ & \texttt{reemit\_only} & an optional marker of a product that is not traced, ignored when the file is compiled \\
        & \texttt{site}                         & the site type an adsorbate occupies \\
        & \texttt{rho}                          & the density of a solid, one value or per material \\
        & \texttt{meanEnergy} \dots             & the ion energy distribution and its reflection parameters \\
\midrule
reaction & \texttt{eq}                          & the equation, from which the stoichiometry and the free sites are derived \\
        & \texttt{type}                         & \texttt{adsorption}, \texttt{desorption},
                                                  \texttt{surface}, \texttt{sputtering}
                                                  or \texttt{ion\_enhanced} \\
        & \texttt{s0}, \texttt{k\_ads} or \texttt{k0} & the prefactor, as sticking,
                                                  adsorption rate constant or rate constant \\
        & \texttt{Ea}, \texttt{beta}            & the barrier and the temperature exponent
                                                  of Eq.~(\ref{eq:arrhenius}) \\
        & \texttt{k0\_rev}, \texttt{Ea\_rev}, \texttt{beta\_rev}
                                                & the reverse constants of a reversible step \\
        & \texttt{order}                        & exponents that differ from the stoichiometry \\
        & \texttt{yield}                        & $A$, $E_{\mathrm{th}}$ and $B$ of
                                                  Eqs.~(\ref{eq:yield}) and (\ref{eq:angular}),
                                                  and $A$ and $E_{\mathrm{th}}$ per material \\
        & \texttt{materials} & the prefactor, barrier and temperature exponent per material \\
\bottomrule
\end{tabular}
\caption{Entries of a reaction file. The first column gives the block in which
an entry is written, at the top level of the file or within a species or
reaction entry.}
\label{tab:entries}
\end{table}

\begin{table}[htbp]
\centering
\begin{tabular}{@{}lp{80mm}l@{}}
\toprule
phase & entries & role in the rate law \\
\midrule
\texttt{gas}     & \texttt{flux} or \texttt{pressure}, and the sticking of
                   the step that consumes it
                 & factor $\Gamma_g^{\,n_{gj}}$ \\
\texttt{ion}     & \texttt{flux}, \texttt{meanEnergy},
                   \texttt{sigmaEnergy}, \texttt{exponent} and the reflection parameters \texttt{inflectAngle}, \texttt{n\_l},
                   \texttt{minAngle}, \texttt{thetaRMin}, \texttt{thetaRMax}
                 & an ion yield, Eq.~(\ref{eq:yield}) \\
\texttt{surface} & \texttt{site}, and the density of that site type
                 & factor $\theta_i^{\,n_{ij}}$ \\
\texttt{solid}   & \texttt{rho}, one value or one per material
                 & the surface velocity, Eq.~(\ref{eq:velocity}) \\
\bottomrule
\end{tabular}
\caption{The four phases of a species with their entries and their role in the
mass-action rate law, Eq.~(\ref{eq:massaction}).}
\label{tab:phases}
\end{table}

\begin{table}[htbp]
\centering
\begin{tabular}{@{}ll@{}}
\toprule
given in the file & derived from it \\
\midrule
species and their phases       & the coverage vector and its site assignment \\
site types and densities       & the free fraction of each type \\
each reaction equation         & the stoichiometry $\nu_{ij}$ and the exponents \\
                               & the free-site exponents $n_{\ast \alpha j}$ \\
                               & the rate law, Eq.~(\ref{eq:massaction}) \\
rate constants                 & $k_j(T)$ per material, Eq.~(\ref{eq:arrhenius}) \\
steps with a gas reactant      & the traced particles and their absorption
                                 probability, Eq.~(\ref{eq:reemission}) \\
solid densities                & the surface velocity, Eq.~(\ref{eq:velocity}) \\
                               & the sign of the motion, from the side on which the solid appears \\
all of the above               & the balance $F_i$ and its Jacobian $J_{ik}$,
                                 Eqs.~(\ref{eq:steady}) and (\ref{eq:jacobian}) \\
\bottomrule
\end{tabular}
\caption{Quantities given in a reaction file and quantities derived from it,
by ViennaChem when the file is compiled or by the surface model from the compiled data.}
\label{tab:derived}
\end{table}

The constants of a step can be given per material in two forms, a single number
that multiplies the prefactor of the step on that material, and a mapping that gives one or more of the prefactor, the barrier and the temperature exponent and replaces the corresponding values of the step. The selective epitaxy case of Appendix~\ref{note:sige} uses both,

\begin{unbroken}
\begin{lstlisting}[style=yaml]
  - {eq: "SiH4 -> SiH3* + H*", type: adsorption, s0: 1.0, Ea: 0.60,
     materials: {default: 0.0,
                 SiGe: {s0: 0.80, Ea: 0.45},
                 Si:   {s0: 1.00, Ea: 0.60},
                 PolySi: 1.0}}
\end{lstlisting}
\end{unbroken}

\noindent
which describes adsorption on silicon and, with a lower barrier, on silicon germanium, and adsorption on the deposited polysilicon at the silicon rate. The \texttt{default} prefactor of zero applies to all other materials, such as an oxide, and no film grows on them.

\begin{figure}[!htbp]
\centering
\includegraphics[width=0.92\linewidth]{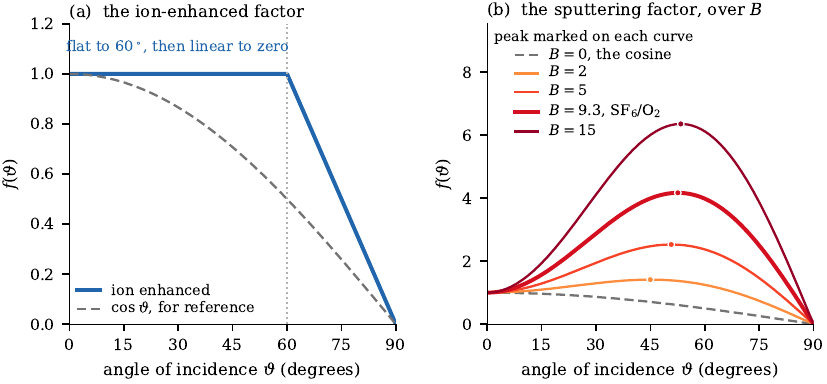}
\caption{The angular factors $f(\vartheta)$ of Eq.~(\ref{eq:angular}).
(a) The ion-enhanced form, constant up to $60^\circ$ and falling linearly to
zero at grazing incidence, with the cosine for reference. (b) The sputtering form for several values of $B$. It reduces to $\cos \vartheta$ for $B = 0$, and for $B > 1/2$ its maximum moves away from normal incidence, to $45^\circ$ for $B = 2$ and to $53^\circ$ for $B = 9.3$, the value in the SF$_6$/O$_2$ reaction file of Appendix~\ref{note:sf6o2}. At $B = 9.3$, the maximum is $4.2$ times the value at normal incidence.}
\label{fig:yield}
\end{figure}

The reaction type of an ion-driven step determines which of the two angular factors of Eq.~(\ref{eq:angular}) is used, and Fig.~\ref{fig:yield} shows both.

\section{Numerical behaviour of the transient solver}
\label{note:solvers}

Two minimal cases show the behaviour of the transient solver, a model with a short-lived intermediate and a cycle of two reaction files. In the first, the intermediate is an adsorbate that a fast surface reaction consumes and converts into a second species. Table~\ref{tab:euler} compares the modified Patankar-Euler scheme of Eqs.~(\ref{eq:transfer}) and (\ref{eq:patankar}) with explicit Euler for this model over a single pulse. Explicit Euler is stable only for steps shorter than twice the lifetime of the intermediate, $h < 2 / L_k$, here $2.4$\,ms. At larger steps, the intermediate coverage becomes negative and is clamped to zero, although the exact coverage is positive. The computed coverage then alternates between zero and $1.79 \times 10^{-6}$ from one step to the next and is zero at the end of the pulse. With the modified Patankar-Euler scheme, the intermediate coverage is positive at all six step sizes, and both schemes converge to the same value as the step size decreases.

\begin{table}[htbp]
\centering
\begin{tabular}{@{}rrr@{}}
\toprule
$h$ (s) & $\theta_{A^\ast}$, explicit Euler & $\theta_{A^\ast}$, modified Patankar-Euler \\
\midrule
$1 \times 10^{-2}$ & $0$                     & $1.7773 \times 10^{-6}$ \\
$5 \times 10^{-3}$ & $0$                     & $1.5983 \times 10^{-6}$ \\
$2 \times 10^{-3}$ & $1.3941 \times 10^{-6}$ & $1.5045 \times 10^{-6}$ \\
$1 \times 10^{-3}$ & $1.4202 \times 10^{-6}$ & $1.4754 \times 10^{-6}$ \\
$5 \times 10^{-4}$ & $1.4336 \times 10^{-6}$ & $1.4612 \times 10^{-6}$ \\
$1 \times 10^{-4}$ & $1.4445 \times 10^{-6}$ & $1.4501 \times 10^{-6}$ \\
\bottomrule
\end{tabular}
\caption{The intermediate coverage $\theta_{A^\ast}$ after a $1$\,s pulse,
integrated at six step sizes. The model consists of $\mathrm{G} \rightarrow A^\ast$ with $s_0 = 0.1$ at an incident flux $\Gamma_{\mathrm{G}} = 100 \times 10^{15}$\,cm$^{-2}$s$^{-1}$, followed by $A^\ast \rightarrow B^\ast$ and by $B^\ast \rightarrow S$, which deposits the solid $S$ and releases the site, with $k_0 = 10^{13}$\,s$^{-1}$ and $E_a = 1.0$ and $1.6$\,eV, respectively, at $500$\,K. The lifetime of $A^\ast$ is then $1.2$\,ms, compared with a pulse length of $1$\,s. The converged value is $1.4473 \times 10^{-6}$,
which explicit Euler approaches from below and modified Patankar-Euler from
above.}
\label{tab:euler}
\end{table}

\begin{figure}[!htbp]
\centering
\includegraphics[width=0.94\linewidth]{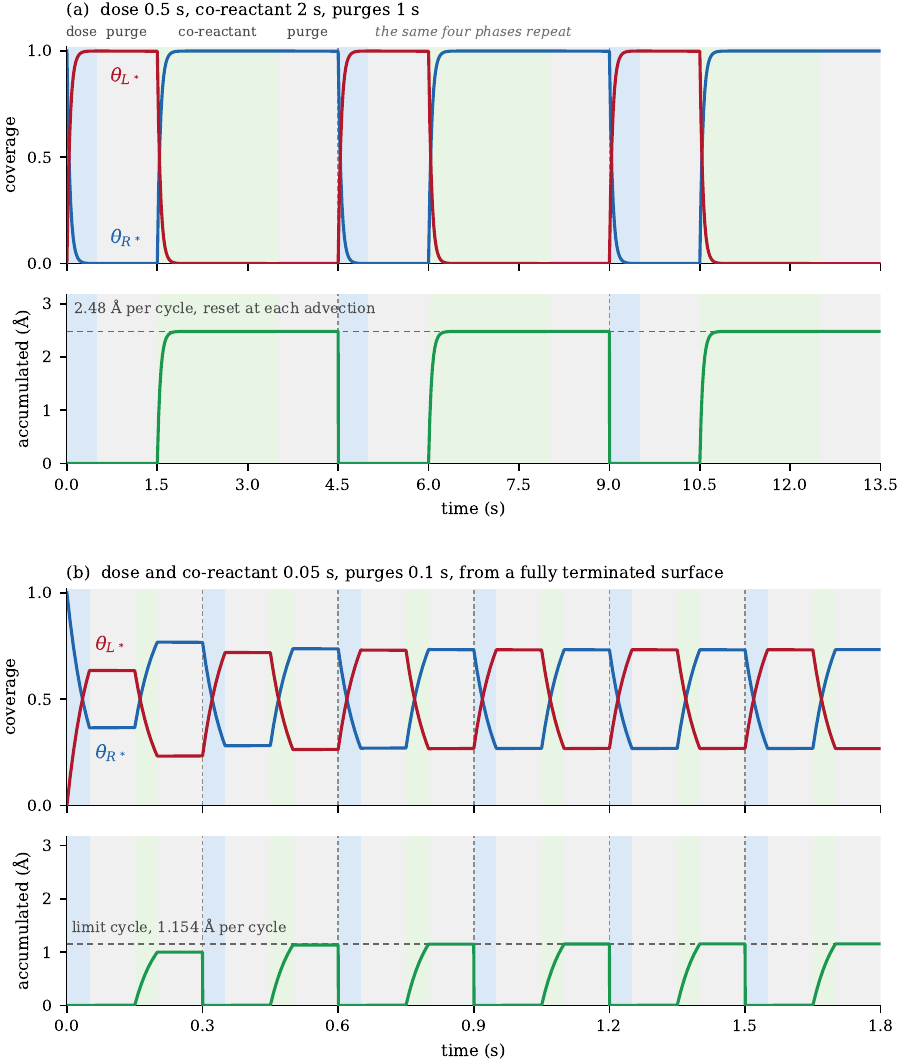}
\caption{A process with two reaction files. In the dose, a precursor consumes the reactive termination and leaves a fragment, $\mathrm{P} + R^\ast \rightarrow L^\ast$, and in the co-reactant stage the fragment is converted back while the solid is deposited, $\mathrm{C} + L^\ast \rightarrow S + R^\ast$. The stages are shaded, the dose in blue, the co-reactant stage in green and the purges in grey. (a) Three cycles with pulses of many time constants, with the stages labelled over the first cycle. The upper panel shows the coverages, which are continuous across the cycle boundaries and do not change during the purges. In the lower panel, the thickness accumulated according to Eq.~(\ref{eq:growthintegral}) reaches $2.48$\,\AA\ per cycle and is reset by each advection. (b) Six cycles with pulses of about one time constant, from a fully terminated surface. The coverages stay away from their limits, and the growth per cycle settles onto the limit cycle of $1.154$\,\AA.}
\label{fig:cycle}
\end{figure}

For the second case, Fig.~\ref{fig:cycle}(a) shows three cycles of a minimal pair of reaction files. During the dose, a precursor consumes the reactive surface
termination and leaves a fragment in its place. During the co-reactant
exposure, the fragment is converted back into the termination and the solid is
deposited. Each half-cycle is self-limiting, because it consumes a surface species that only the other supplies. Both reactions proceed with the time constant $S_0/(s_0 \Gamma) = 0.050$\,s, so the dose of $0.5$\,s and the co-reactant pulse of $2$\,s both saturate. The coverages do not change during the purges and are continuous across the cycle boundaries. Growth occurs
only during the co-reactant stage, and the accumulated thickness is reset by
each advection. The four-stage cycle is set up in the C++ interface of
ViennaPS as

\begin{unbroken}
\begin{lstlisting}[style=code]
model->addMechanism("dose",       doseMechanism);
model->addMechanism("coreactant", coreactantMechanism);
model->setAtomicLayerProcess();

AtomicLayerProcessParameters cycle;
//          name          duration  time step   flowing species  mechanism
//                           [s]        [s]
cycle.addStage("dose",        0.5,      0.05,   {"P_flux"},      "dose");
cycle.addStage("purge",       1.0,      0.25,   {},              "dose");
cycle.addStage("coreactant",  2.0,      0.10,   {"C_flux"},      "coreactant");
cycle.addStage("purge",       1.0,      0.25,   {},              "coreactant");
process.setParameters(cycle);
\end{lstlisting}
\end{unbroken}

\noindent
where the time step is the interval at which the fluxes are recomputed by ray
tracing and the coverages advanced, and the integrator subdivides it
adaptively according to Eq.~(\ref{eq:substep}). The rate constants remain in
the reaction files, and the stage list contains only the process recipe.

Figure~\ref{fig:cycle}(b) shows the same two files with both pulses shortened to $0.05$\,s, about one
time constant, starting from a fully terminated surface. Neither half-cycle saturates, and the fragment left at the end
of each co-reactant pulse is carried into the next dose. Over the first four cycles, the termination at the start of
the dose falls from $1$ to $0.768$, $0.737$ and $0.733$, and the growth per cycle rises from $1.000$ to $1.133$,
$1.151$ and $1.153$\,\AA. For this model, the limit cycle has a closed form. The dose converts the fraction $1 - \exp(-a)$ of the
termination and the co-reactant pulse the fraction $1 - \exp(-b)$ of the fragment, where $a$ and $b$ are the lengths of
the two pulses divided by the time constant $S_0/(s_0 \Gamma)$. The termination at the start of each dose is then
$x^\ast = [1 - \exp(-b)]/[1 - \exp(-a-b)]$, and the growth per cycle is $[1 - x^\ast \exp(-a)][1 - \exp(-b)]$ times one site
layer, $S_0/\rho = 2.48$\,\AA. The closed form
gives $x^\ast = 0.7326$ and $1.1539$\,\AA\ per cycle, and the integrator $0.7326$ and $1.1537$\,\AA, a relative
difference of $2 \times 10^{-4}$ in the growth.

Within a pulse, the velocity depends on coverages that can fall by orders of
magnitude, since a step that consumes an adsorbate may remove it within a small
fraction of the pulse. In the co-reactant pulse of this cycle, evaluating $v_n$ once per time step and multiplying by the step length applies the high rate early in the pulse to the whole step and overestimates the deposited thickness by a factor that depends on the step length (Table~\ref{tab:growthint}). Accumulated on the sub-steps of the integrator (Section~\ref{sec:cyclic}), the thickness is the same for each of these step lengths, from one to $150$ time steps per pulse.

\begin{table}[htbp]
\centering
\begin{tabular}{@{}rrr@{}}
\toprule
time steps per pulse & $\Delta n$ (nm) & ratio to Eq.~\eqref{eq:growthintegral} \\
\midrule
$1$ & $9.9992 \times 10^{0}$ & $40.32$ \\
$3$ & $3.3331 \times 10^{0}$ & $13.44$ \\
$15$ & $7.1547 \times 10^{-1}$ & $2.89$ \\
$30$ & $4.5110 \times 10^{-1}$ & $1.82$ \\
$150$ & $2.8299 \times 10^{-1}$ & $1.14$ \\
\bottomrule
\end{tabular}
\caption{Thickness deposited in one co-reactant pulse of the minimal cycle of Fig.~\ref{fig:cycle} when $v_n$ is evaluated once per time step and multiplied by the step length. The ratio compares this thickness with the integral accumulated on the sub-steps of the integrator, which is $2.4798 \times 10^{-1}$\,nm at each of these step lengths.}
\label{tab:growthint}
\end{table}

\section{Silicon etching in SF$_6$/O$_2$, \texttt{sf6o2}}
\label{note:sf6o2}

The reaction file below takes all its constants from the hand-written ViennaPS class, which implements the balance of Belen et al.~\cite{BelenSF6O2}. The sticking coefficients, $0.7$ for fluorine and $1$
for oxygen, are dimensionless probabilities per incident atom on a bare
surface. The prefactors of the chemical etch, $k_{0j} = 75$, and of the thermal desorption of oxygen, $k_{0j} = 0.04$, are rates per unit area at unit coverage in $10^{15}$\,cm$^{-2}$s$^{-1}$, which for the default site density equal rates per site in s$^{-1}$. The four ion-driven steps have $k_j = 1$ on the materials they act on and the yield prefactors $7.0$, $0.0337$ and $3.0$\,eV$^{-1/2}$ with thresholds of $15$, $20$ and $10$\,eV, for the two ion-enhanced steps, the sputtering of silicon and the removal of oxygen, respectively. The ion entry specifies the energy and angular distribution over which the yields are evaluated. The solid has two densities, so that silicon and mask removed at the same rate recede at velocities that differ by a factor of one hundred.

\begin{unbroken}
\begin{lstlisting}[style=code]
name: sf6o2_silicon_etch
temperature: 300.0        # K, the model has no thermal barrier
species:
  F:      {phase: gas, flux: 2.5714286e3}   # etchantFlux / s0, 1e15 /cm2 /s
  O:      {phase: gas, flux: 1.0e2}         # passivationFlux / s0, s0 = 1
  Ar_ion: {phase: ion, flux: 1.2e1,      # ionFlux
           meanEnergy: 100.0, sigmaEnergy: 10.0, exponent: 500.0,
           inflectAngle: 89.0, n_l: 10.0, minAngle: 80.0,
           thetaRMin: 70.0, thetaRMax: 90.0}
  SiF4:   {phase: gas, reemit_only: true}     # the chemical etch product
  SiF2:   {phase: gas, reemit_only: true}     # the ion-enhanced etch product
  Si_s:   {phase: gas, reemit_only: true, atoms: {Si: 1}}   # a sputtered atom
  F2:     {phase: gas, reemit_only: true}     # fluorine lost on the mask
  F*:     {phase: surface}               # eCoverage
  O*:     {phase: surface}               # pCoverage
  Si:     {phase: solid, rho: {default: 5.02, Mask: 500.0}}
surface:
  - {eq: "F -> F*", type: adsorption, s0: 0.7, Ea: 0.0}
  - {eq: "O -> O*", type: adsorption, s0: 1.0, Ea: 0.0}
  - {eq: "4F* + Si -> SiF4", type: surface, k0: 7.5e1, Ea: 0.0,
     order: {F*: 1},
     materials: {default: 0.0, Si: 1.0}}          # substrate only
  - {eq: "4F* -> 2F2", type: surface, k0: 7.5e1, Ea: 0.0,
     order: {F*: 1},
     materials: {default: 0.0, Mask: 1.0}}        # mask only
  - {eq: "Ar_ion + 2F* -> F2", type: ion_enhanced,
     order: {F*: 1},
     yield: {A: 7.0, Eth: 15.0},
     materials: {default: 0.0, Mask: 1.0}}        # mask only
  - {eq: "Ar_ion + 2F* + Si -> SiF2", type: ion_enhanced,
     order: {F*: 1},
     yield: {A: 7.0, Eth: 15.0},
     materials: {default: 0.0, Si: 1.0}}          # substrate only
  - {eq: "Ar_ion + Si -> Si_s", type: sputtering,
     yield: {A: 0.0337, Eth: 20.0, B: 9.3,
             materials: {Mask: {A: 0.0139, Eth: 20.0}}}}
  - {eq: "Ar_ion + O* -> O", type: ion_enhanced,
     order: {O*: 1},
     yield: {A: 3.0, Eth: 10.0}}
  - {eq: "O* -> O", type: surface, k0: 4.0e-2, Ea: 0.0}
\end{lstlisting}
\end{unbroken}

The chemical and the ion-enhanced etch are confined to the substrate by their material blocks. In each block, the \texttt{default} value of zero switches the step off everywhere, and the entry for \texttt{Si} switches it on again for silicon. The hand-written class, however, solves the same coverage balance on the mask as on silicon. On the mask, the chemical and the ion-enhanced term consume fluorine, but the mask is removed only by sputtering. The file therefore contains two further steps that release this fluorine as F$_2$. They have the rate laws of the chemical and the ion-enhanced etch and are confined to the mask by the same kind of block. As in the class, the ion-enhanced yield is evaluated on every material, and the material blocks restrict the etch to silicon and the release of fluorine to the mask. Physical sputtering removes both materials. Its yield prefactors differ between them and set the selectivity together with the two densities.

In the four steps that consume fluorine, \texttt{order} entries make the rates first order in the fluorine coverage, as in the rate law of Belen et al. Because a surface silicon atom binds four fluorine atoms before it leaves as $\mathrm{SiF_4}$, each event of the chemical etch still consumes four fluorine atoms in the coverage balance. With the yields converted into fluxes,
the balances are
\begin{equation}
  \frac{\mathrm{d}\theta_F}{\mathrm{d}t} = r_1 - 4r_3 - 2r_4 ,
  \qquad
  \frac{\mathrm{d}\theta_O}{\mathrm{d}t} = r_2 - r_6 - r_7 ,
  \label{eq:sf6o2balance}
\end{equation}
with the two adsorptions $r_1 = s_{0,F}\,\Gamma_F\,\theta_\ast$ and $r_2 =
s_{0,O}\,\Gamma_O\,\theta_\ast$, the chemical etch $r_3 = k_3\,\theta_F$, the
ion-enhanced steps $r_4 = \Gamma_{\mathrm{ie}}\,\theta_F$ and $r_6 =
\Gamma_{\mathrm{p}}\,\theta_O$, the sputtering $r_5 = \Gamma_{\mathrm{sp}}$ and
the thermal desorption of oxygen $r_7 = k_7\,\theta_O$. Here, $\Gamma_{\mathrm{ie}}$, $\Gamma_{\mathrm{sp}}$ and $\Gamma_{\mathrm{p}}$ are
the yield-weighted ion fluxes of the three ion-driven steps on silicon. Since silicon is a reactant in three steps, the velocity $v_n = -(r_3 + r_4 + r_5)/\rho_{\mathrm{Si}}$ is negative. On the mask, the two steps that release F$_2$ take the place of $r_3$ and $r_4$ in Eq.~(\ref{eq:sf6o2balance}), and the mask recedes only by sputtering.

The simulation is run with quarter symmetry on a $20$\,nm grid in a cylindrical hole $0.4\,\upmu$m in diameter through a $1.2\,\upmu$m mask (Fig.~\ref{fig:sf6o2}). Bobinac et al.~\cite{Bobinac2023} used a similar structure for this chemistry, with a mask opening that narrows from $0.4$ to $0.35\,\upmu$m.
The fluxes of the four feed-gas compositions in their Table~1 are used as printed, because they are incident fluxes and the sticking coefficient of $0.7$ enters separately through their Eq.~(4), as in the file.

\begin{figure}[!htbp]
\centering
\includegraphics[width=0.80\textwidth]{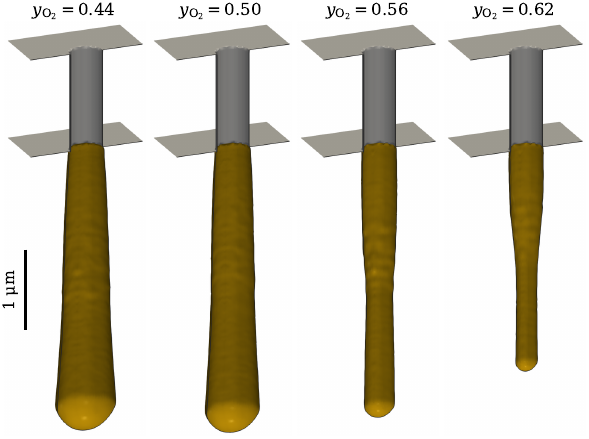}
\caption{Silicon surfaces etched in SF$_6$/O$_2$ for $160$\,s at the four oxygen fractions $y_{\mathrm{O_2}}$ of Fig.~\ref{fig:sf6o2}b, from the first of the three runs at each composition. Each run computes one quarter of the hole, which is mirrored into the full hole. The half of the hole behind a cut through its axis is shown, with the mask in grey, silicon in gold and the section in the cut plane in dark grey. All four holes are drawn to one scale in parallel projection, and the bar gives $1\,\upmu$m of depth.}
\label{fig:sf6o23d}
\end{figure}

Fig.~\ref{fig:sf6o23d} shows the etched surfaces at the four feed-gas compositions in three dimensions, with the computed quarter mirrored into the full hole and the hole cut open through its axis. Each profile in Fig.~\ref{fig:sf6o2} gives the largest radius of such a surface at each depth.

\begin{figure}[!htbp]
\centering
\includegraphics[width=0.86\textwidth]{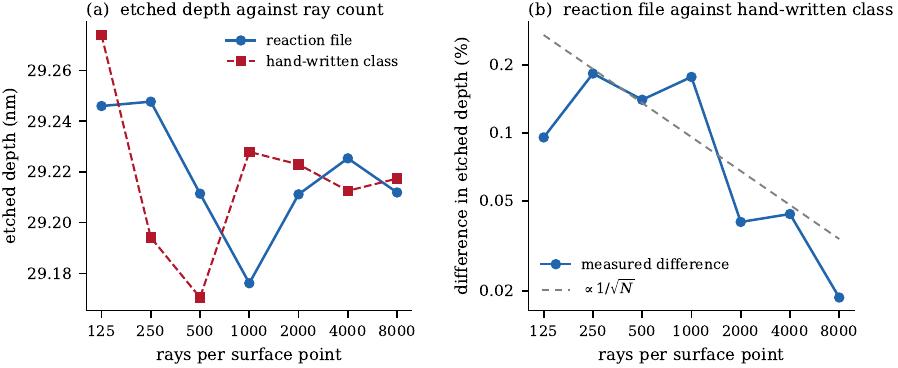}
\caption{Reaction file and hand-written class in a masked trench at
increasing ray count. (a) Etched depth from both. (b) Difference between the two depths, which scatters about an $N^{-1/2}$ trend fitted to all points. Monte Carlo noise scales as $N^{-1/2}$ with the number $N$ of rays per surface point.}
\label{fig:conv}
\end{figure}

\section{Polysilicon from silane, \texttt{silane\_inputs}}
\label{note:silane}

Case~1 of Table~\ref{tab:mechanisms} is polysilicon deposition from silane with the three reactions of Section~\ref{sec:example}, and it tests the conversions of a partial pressure into a flux and of an adsorption rate constant into a sticking coefficient (Sections~\ref{sec:files} and \ref{sec:rates}). The silane supply is given as a partial pressure of $4.84$\,mTorr and the adsorption as a rate constant $k_{\mathrm{ads}} = 1.9256 \times 10^{4}$\,cm/s with a barrier of $0.6$\,eV. ViennaChem converts the pressure into the incident flux, and with Eq.~(\ref{eq:sticking}) it converts $k_{\mathrm{ads}}$ into the sticking coefficient, the dimensionless probability that a silane molecule adsorbs on a bare surface. The
two thermal steps have $k_{0j} = 6.8 \times 10^{12}$ with barriers of $1.5$ and $2.0$\,eV, respectively, and their $k_j$ are rates per unit area at unit coverage in
$10^{15}$\,cm$^{-2}$s$^{-1}$, Eq.~(\ref{eq:kunits}). With the default site density of $10^{15}$\,cm$^{-2}$, the first is numerically equal to the decomposition rate of one silyl in s$^{-1}$ and the second to the number of hydrogen molecules that desorb per site per second from a fully hydrogenated surface.

\begin{unbroken}
\begin{lstlisting}[style=code]
name: silane_polySi_LPCVD_inputs
temperature: 900.0        # K
species:
  SiH4:  {phase: gas,     pressure: 4.8403}     # mTorr, instead of flux: 1.0e3
  H2:    {phase: gas,     reemit_only: true}
  SiH3*: {phase: surface}
  H*:    {phase: surface}
  Si:    {phase: solid,   rho: 5.02}            # 1e22 atoms/cm^3
surface:
  - {eq: "SiH4 -> SiH3* + H*",    type: adsorption, k_ads: 1.9256e4, Ea: 0.6}
  - {eq: "SiH3* -> Si + H* + H2", type: surface,    k0: 6.8e12,      Ea: 1.5}
  - {eq: "2H* -> H2",             type: desorption, k0: 6.8e12,      Ea: 2.0}
\end{lstlisting}
\end{unbroken}

The growth rate agrees to $0.003\,\%$ with that of the same chemistry written with flux and sticking coefficient. The difference is due to the rounding of the constants in the file.

\begin{figure}[!htbp]
\centering
\includegraphics[width=0.78\textwidth]{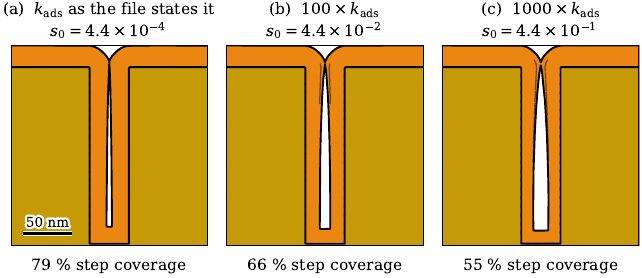}
\caption{Polysilicon from silane in a trench of aspect ratio $4.5$, at the adsorption rate constant of the reaction file (a) and at $100$ (b) and $1000$ (c) times this value. The substrate is shown in gold and the film in orange.  The step coverage is measured $140$\,nm below the opening and averaged over three runs with different random seeds. The
void left along the axis in (a) widens as the sticking coefficient rises.}
\label{fig:case1}
\end{figure}

The simulation is run in a trench $40$\,nm wide and $180$\,nm deep on a $1$\,nm grid until $22$\,nm of film has grown on
the open field, and the surface is saved after each quarter of the process
time. Figure~\ref{fig:case1}(a) shows the result for the file above. The film reaches $17.4$\,nm on the sidewall $140$\,nm below the opening, a step coverage of $79\,\%$. It closes over the mouth and leaves a narrow void along the axis. The conformality is determined by the sticking coefficient and has no parameter of its own. Through Eq.~(\ref{eq:sticking}), the listed $k_\mathrm{ads}$ corresponds to a sticking coefficient of $4.37 \times 10^{-4}$ at $900$\,K. At this value, a silane molecule is re-emitted many times inside the trench before it reacts.

In Figs.~\ref{fig:case1}(b) and (c), $k_\mathrm{ads}$ is raised by two and by three orders of magnitude, respectively, with all other settings unchanged. Because
the sticking coefficient enters Eq.~(\ref{eq:massaction}) as well as the absorption
probability of the ray tracer, this change affects the chemistry and the
transport. The step coverage falls to $66$ and then to $55\,\%$, and the void along the axis widens. The blanket rate rises from $7.178 \times 10^{-2}$ to $2.155$ and $4.949$\,nm/s, respectively. It grows by a smaller factor than $k_\mathrm{ads}$ because the free-site factor of Eq.~(\ref{eq:massaction}) decreases as the hydrogen coverage increases from $0.092$ to $0.502$ and $0.761$, respectively. The sticking coefficient also depends on temperature, through the $T^{-1/2}$ factor of Eq.~(\ref{eq:sticking}) and the barrier of Eq.~(\ref{eq:arrhenius}). At $1100$\,K, the sticking coefficient rises to $1.61 \times 10^{-3}$ and the step coverage falls from $79$ to
$75\,\%$.

\section{Selective epitaxy, \texttt{sige\_stack}}
\label{note:sige}

The three reactions of the silane case are applied to silicon, silicon germanium, the oxide, the hard mask and the deposited polysilicon, with both forms of a per-material block (Appendix~\ref{note:format}). Silicon and silicon germanium adsorb silane with
different sticking coefficients and barriers. The sticking coefficient is a dimensionless probability per incident silane molecule, and its prefactor and barrier are $1.00$ and $0.60$\,eV, respectively, on silicon and $0.80$ and $0.45$\,eV on silicon germanium. The deposited polysilicon grows at the silicon rate through the scalar form, in which the scalar $1.0$ multiplies the prefactor of the step. The oxide and the hard mask take the \texttt{default} prefactor of zero and do not grow. The two thermal steps have the prefactor $k_{0j} = 6.8 \times 10^{12}$ of the silane case. Hydrogen desorption has a block of its own, because surface germanium lowers its barrier from $2.0$\,eV on silicon to $1.6$\,eV on silicon germanium. All values are illustrative.

\begin{unbroken}
\begin{lstlisting}[style=code]
name: sige_si_selective_epi
temperature: 900.0        # K
species:
  SiH4:  {phase: gas,     flux: 1.0e3}
  H2:    {phase: gas,     reemit_only: true}
  SiH3*: {phase: surface}
  H*:    {phase: surface}
  Si:    {phase: solid,   rho: 5.02}
surface:
  - {eq: "SiH4 -> SiH3* + H*", type: adsorption, s0: 1.0, Ea: 0.60,
     materials: {default: 0.0,
                 SiGe: {s0: 0.80, Ea: 0.45},
                 Si:   {s0: 1.00, Ea: 0.60},
                 PolySi: 1.0}}
  - {eq: "SiH3* -> Si + H* + H2", type: surface, k0: 6.8e12, Ea: 1.5}
  - {eq: "2H* -> H2", type: desorption, k0: 6.8e12, Ea: 2.0,
     materials: {SiGe: {Ea: 1.6}}}
\end{lstlisting}
\end{unbroken}

\begin{figure}[!htbp]
\centering
\includegraphics[width=0.94\textwidth]{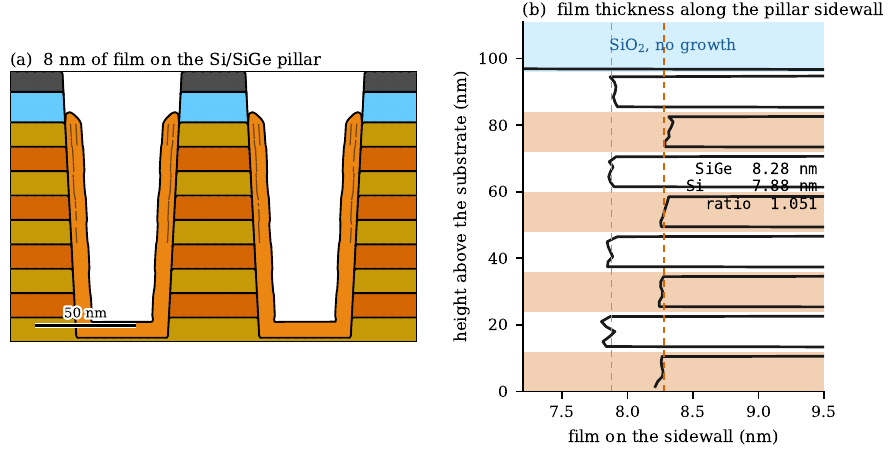}
\caption{Selective epitaxy on a Si/SiGe superlattice. (a) The pillar after
$8$\,nm of growth, with Si in gold, SiGe in dark orange, the oxide in blue and
the film in orange. (b) Film thickness along the pillar sidewall, with the SiGe
bands shaded and the band means averaged over three runs with different random seeds. Growth stops at the oxide, and the film on the SiGe bands is $5.1\,\%$ thicker than on the Si bands because it grew at the SiGe rate until the bands were covered.}
\label{fig:case2}
\end{figure}

The geometry is a superlattice of four SiGe and four Si layers, each
$12$\,nm thick, under a $15$\,nm oxide and a $10$\,nm hard mask. The stack is modelled on the SiGe geometry of the ViennaPS examples. Two trenches tapered at $3^\circ$ are cut through all layers, so that the pillar between them exposes every layer on its sidewall. Eight nanometres of film are grown, and Fig.~\ref{fig:case2} shows the result (a) and the film thickness along the
pillar sidewall (b).

The growth rate on the oxide is zero throughout the run, since the adsorption there has the \texttt{default} prefactor of zero. Because the ray tracer uses the same per-material sticking coefficient as the rate law, the oxide reflects all silane that reaches it, and this silane adds to the flux arriving at the exposed layers below. On bare surfaces, the growth rate is $7.178 \times 10^{-2}$\,nm/s on silicon, $4.644 \times 10^{-1}$\,nm/s on silicon germanium and the silicon rate on polysilicon. The rate ratio of $6.47$ between SiGe and Si comes mostly from the higher sticking of silane on SiGe. A smaller factor is due to the lower hydrogen desorption barrier on germanium-containing surfaces, on which more sites remain free.
The thickness ratio on the sidewall is much smaller, because the deposit is silicon and the film on a SiGe band grows at the SiGe rate only while the band is still exposed.
The film is $8.28$\,nm thick on the four SiGe bands and $7.88$\,nm on the four
Si bands, a ratio of $1.051$.

\section{Diamond growth, \texttt{diamond}}
\label{note:diamond}

Case~3 of Table~\ref{tab:mechanisms}, the standard growth model of May and Mankelevich~\cite{May2008}, is used to test the treatment of a reversible step. The reversible step is the hydrogen abstraction, the first step of the file, and setting its reverse constants to zero removes the last term of Eq.~(\ref{eq:diamondR}). Since every step of this model has a gas reactant, all its rate constants are dimensionless probabilities that an incident particle reacts on a surface fully covered by the surface reactant of the step (Eq.~(\ref{eq:kunits})). The barriers, in kcal/mol, are converted from the exponents of Eq.~(\ref{eq:diamondR}), $3430$\,K for the abstraction and $3430 + 4420$\,K for its reverse. The forward
constant of the first step, $1.0$ with a barrier of $6.816$\,kcal/mol, is the
probability that an incident hydrogen atom abstracts a surface hydrogen and
leaves a radical site. An incident hydrogen molecule terminates a radical site and releases a hydrogen atom with a probability given by the reverse constant $0.1\sqrt{2} = 0.141421$ and a barrier of $15.6$\,kcal/mol. The factor $\sqrt{2}$ is needed because Eq.~(\ref{eq:diamondR}) contains the ratio of the gas concentrations, while the rate laws of the file contain fluxes. At equal concentration, the flux of H$_2$ is smaller than that of H by the square root of their mass ratio. The constants of the second and third steps, $0.3$ and $0.33$, respectively, are the probabilities that an incident hydrogen atom terminates a
radical site and that an incident methyl radical adds to a radical site and is
incorporated into the film.

\begin{unbroken}
\begin{lstlisting}[style=code]
name: diamond_cvd_standard_growth_model
temperature: 1200.0       # K, substrate
energyUnit: kcal/mol
species:
  H:   {phase: gas,     flux: 1.0100e5}    # 1e15/cm2/s
  H2:  {phase: gas,     flux: 1.3998e7}
  CH3: {phase: gas,     flux: 2.6152e2}
  H*:  {phase: surface}                    # hydrogen-terminated site
  C:   {phase: solid,   rho: 17.6}         # 1e22 atoms/cm3, diamond
surface:
  - {eq: "H* + H <-> H2", type: adsorption, s0: 1.0, Ea: 6.816,
     k0_rev: 0.141421, Ea_rev: 15.600}
  - {eq: "H -> H*",       type: adsorption, s0: 0.3, Ea: 0.0}
  - {eq: "CH3 -> C + H* + H2", type: adsorption, s0: 0.33, Ea: 0.0}
\end{lstlisting}
\end{unbroken}

\begin{figure}[!htbp]
\centering
\includegraphics[width=0.82\textwidth]{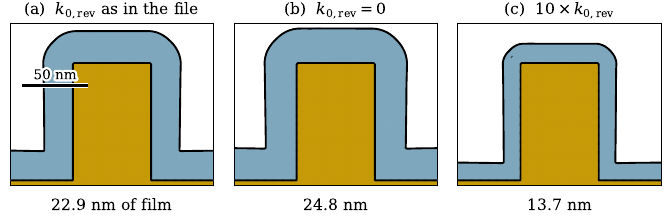}
\caption{Diamond over a silicon fin, $33.6$\,s in each panel, with the film in
blue. (a) The reverse constant of the hydrogen abstraction as given in the file. (b) The same file with the reverse step removed, which raises the radical
fraction and thickens the film. (c) The constant multiplied by ten, which lowers the radical fraction by $40\,\%$. The panels show the first of three runs, and the film thickness under each is the mean of the three.}
\label{fig:case3}
\end{figure}

Diamond is grown over a fin $60$\,nm wide and $90$\,nm high for $33.6$\,s. The
film in Fig.~\ref{fig:case3}(a) is nearly conformal, $22.0$\,nm on the sidewall at half height against $22.9$\,nm on the field, a ratio of $96\,\%$. This is because methyl is absorbed at few of its wall contacts. Its absorption probability, the sticking coefficient of $0.33$ times the radical fraction, is about $0.05$. Most of the methyl that reaches the floor beside the fin is therefore re-emitted, and part of it reaches the sidewall.

In Figs.~\ref{fig:case3}(b) and (c), only the reverse constant of the first step is
changed. This constant is compiled into a separate reaction. Without the reverse reaction, the film on the field reaches $24.8$\,nm, and with its prefactor raised tenfold it reaches $13.7$\,nm. The reverse constant changes the film thickness through the radical fraction of Eq.~(\ref{eq:diamondR}), because methyl adsorbs only on a free site. Without the reverse constants, Eq.~(\ref{eq:diamondR}) reduces to $R = [1 + 0.3 \exp(3430/T)]^{-1} = 0.1605$ at $1200$\,K, against $0.1601$ from the solver. Multiplying the constant by ten multiplies the last term by ten, and the closed form then gives $R = 0.0896$ and the solver $0.0895$.

\section{GaAs from trimethylgallium and arsine, \texttt{gaas\_cvd}}
\label{note:gaas}

After compilation, the model has thirty reactions and two solids, and its seven coverages lie on two site types. All constants are from Table~II of Mountziaris
and Jensen~\cite{Mountziaris1991}. As in the source, the steps with a gas reactant have the prefactor one, a dimensionless probability per incident particle. Of these steps, the abstraction of hydrogen from AsH$^\ast$ by methyl has a barrier of $20$\,kcal/mol, and the others have none. The source gives the constants of the thermal steps
for surface concentrations in mol\,cm$^{-2}$, the first-order desorption
constants in s$^{-1}$ and the second-order constants of the reactions between
adsorbates in cm$^2$\,mol$^{-1}$\,s$^{-1}$. They are converted into rates per
unit area at unit coverage in $10^{15}$\,cm$^{-2}$s$^{-1}$ with the density $S_0
= 2.2125 \times 10^{14}$\,cm$^{-2}$ of each site type, as $k^{(1)} S_0 /
10^{15}$ and $k^{(2)} S_0^2 / (N_A\, 10^{15})$, where $N_A$ is the Avogadro
constant. The desorption prefactors of $10^{12}$ and $10^{14}$\,s$^{-1}$ thus
become $2.2125 \times 10^{11}$ and $2.2125 \times 10^{13}$, respectively, and the second-order prefactors of $10^{17}$ to $5 \times
10^{18}$\,cm$^2$\,mol$^{-1}$\,s$^{-1}$ become about $8.13 \times 10^{6}$ to $4.06 \times 10^{8}$.

\begin{unbroken}
\begin{lstlisting}[style=code]
name: gaas_movpe_full
temperature: 1000.0       # K, near the growth-rate maximum
energyUnit: kcal/mol      # as published
species:
  H:       {phase: gas, flux: 0.0}          # [S1],[S2],[S16],[S17]
  CH3:     {phase: gas, flux: 0.0}          # [S3],[S4],[S14],[S15],[S24]
  GaCH3:   {phase: gas, flux: 1.653e4}      # [S5], from p(TMG) = 0.137 torr
  GaC2H6:  {phase: gas, flux: 0.0}          # [S6]  Ga(CH3)2
  GaC3H9:  {phase: gas, flux: 0.0}          # [S7]  Ga(CH3)3, TMG
  GaCH2:   {phase: gas, flux: 0.0}          # [S8]  gallium carbene
  GaC2H5:  {phase: gas, flux: 0.0}          # [S9]  Ga(CH3)CH2
  GaC3H8:  {phase: gas, flux: 0.0}          # [S10] Ga(CH3)2CH2
  AsH:     {phase: gas, flux: 3.187e5}      # [S11], from p(AsH3) = 2.50 torr
  AsH2:    {phase: gas, flux: 0.0}          # [S12]
  AsH3:    {phase: gas, flux: 0.0}          # [S13] arsine
  CH4:     {phase: gas, reemit_only: true}
  H2:      {phase: gas, reemit_only: true}
  C2H6:    {phase: gas, reemit_only: true}
  As2:     {phase: gas, reemit_only: true}
  H_G*:    {phase: surface, site: G, atoms: {H: 1}}
  H_A*:    {phase: surface, site: A, atoms: {H: 1}}
  CH3_G*:  {phase: surface, site: G, atoms: {C: 1, H: 3}}
  CH3_A*:  {phase: surface, site: A, atoms: {C: 1, H: 3}}
  GaCH3*:  {phase: surface, site: G}
  AsH*:    {phase: surface, site: A}
  As*:     {phase: surface, site: A}
  GaAs:    {phase: solid, rho: 2.2}         # 1e22 formula units/cm3
  GaC:     {phase: solid, rho: 2.2}         # carbon on an As lattice site
siteTypes:
  - {name: G, density: 2.2125e14}           # cm^-2, half of S0 = 4.425e14
  - {name: A, density: 2.2125e14}
surface:
  - {eq: "H -> H_G*", type: adsorption, s0: 1.0, Ea: 0.0}
  - {eq: "H -> H_A*", type: adsorption, s0: 1.0, Ea: 0.0}
  - {eq: "CH3 <-> CH3_G*", type: adsorption, s0: 1.0, Ea: 0.0,
     k0_rev: 2.2125e11, Ea_rev: 20.0}       # 1e12 /s
  - {eq: "CH3 <-> CH3_A*", type: adsorption, s0: 1.0, Ea: 0.0,
     k0_rev: 2.2125e11, Ea_rev: 20.0}
  - {eq: "GaCH3 <-> GaCH3*", type: adsorption, s0: 1.0, Ea: 0.0,
     k0_rev: 2.2125e13, Ea_rev: 40.0}       # 1e14 /s
  - {eq: "GaC2H6 -> GaCH3* + CH3",      type: adsorption, s0: 1.0, Ea: 0.0}
  - {eq: "GaC3H9 -> GaCH3* + 2CH3",     type: adsorption, s0: 1.0, Ea: 0.0}
  - {eq: "GaCH2 -> GaC + H2",           type: adsorption, s0: 1.0, Ea: 0.0,
     order: {"*G": 1, "*A": 1}}
  - {eq: "GaC2H5 -> GaC + CH3 + H2",    type: adsorption, s0: 1.0, Ea: 0.0,
     order: {"*G": 1, "*A": 1}}
  - {eq: "GaC3H8 -> GaC + 2CH3 + H2",   type: adsorption, s0: 1.0, Ea: 0.0,
     order: {"*G": 1, "*A": 1}}
  - {eq: "AsH <-> AsH*", type: adsorption, s0: 1.0, Ea: 0.0,
     k0_rev: 2.2125e13, Ea_rev: 40.0}       # 1e14 /s
  - {eq: "AsH2 -> AsH* + H",  type: adsorption, s0: 1.0, Ea: 0.0}
  - {eq: "AsH3 -> AsH* + H2", type: adsorption, s0: 1.0, Ea: 0.0}
  - {eq: "CH3 + H_G* -> CH4",   type: adsorption, s0: 1.0, Ea: 0.0}
  - {eq: "CH3 + H_A* -> CH4",   type: adsorption, s0: 1.0, Ea: 0.0}
  - {eq: "H + CH3_G* -> CH4",   type: adsorption, s0: 1.0, Ea: 0.0}
  - {eq: "H + CH3_A* -> CH4",   type: adsorption, s0: 1.0, Ea: 0.0}
  - {eq: "CH3 + AsH* -> As* + CH4", type: adsorption, s0: 1.0, Ea: 20.0}
  - {eq: "H_G* + CH3_A* -> CH4", type: surface, k0: 8.1285e6, Ea: 10.0}  # 1e17
  - {eq: "H_A* + CH3_G* -> CH4", type: surface, k0: 8.1285e6, Ea: 10.0}  # 1e17
  - {eq: "H_A* + H_G* -> H2",    type: surface, k0: 9.7542e6, Ea: 20.0}  # 1.2e17
  - {eq: "CH3_A* + CH3_G* -> C2H6", type: surface, k0: 8.1285e6, Ea: 20.0} # 1e17
  - {eq: "GaCH3* + AsH* -> GaAs + CH4", type: surface, k0: 4.0643e8, Ea: 29.3} # 5e18
  - {eq: "AsH* + AsH* -> As2 + H2", type: surface, k0: 8.1285e6, Ea: 35.0}  # 1e17
  - {eq: "As* + As* -> As2", type: surface, k0: 8.1285e7, Ea: 30.0}        # 1e18
  - {eq: "GaCH3* + As* -> GaAs + CH3", type: surface, k0: 4.0643e8, Ea: 20.0} # 5e18
\end{lstlisting}
\end{unbroken}

The simulation is run in a trench $80$\,nm wide and $120$\,nm deep on a
$1.5$\,nm grid, at $4000$ rays per surface point, until $20$\,nm have grown on
the open field (Fig.~\ref{fig:gaas}a). With $s_0 = 1$ for every adsorption in
the file, a TMG fragment is absorbed at a wall with a probability equal to the
local free fraction. The trench receives $5.6$\,nm on the floor and $3.1$\,nm on the sidewall at half depth, step coverages of $28$ and $15\,\%$, respectively. The net incorporation probability of this run is $2.5 \times 10^{-3}$, a factor of $400$ below the sticking coefficient in the file. The shape of the deposit is nevertheless set by the sticking coefficient, because GaCH$_3$ that desorbs again is not traced further.

The reduction of Figs.~\ref{fig:gaas}(b) and (c), \texttt{gaas\_reversible},
consists of the three steps that Mountziaris and Jensen identified as dominant
by sensitivity analysis. The full file contains twenty-six steps and the
reduction three, of which four and two, respectively, are reversible. Expanding each
reversible step into two reactions gives the thirty reactions of
Table~\ref{tab:mechanisms} for the full file and five for the reduction. The reduction lacks one growth channel of the full file. Methyl released by the pyrolysis of TMG removes the hydrogen from an adsorbed AsH through step [S24] and leaves a bare arsenic atom. The bare arsenic forms GaAs with adsorbed GaCH$_3$ through [S26], with a barrier of $20$\,kcal/mol compared with $29.3$\,kcal/mol for the main step [S22].

\section{Argon sputtering, \texttt{ar\_sputter}}
\label{note:sputter}

Since the model contains only physical sputtering and no adsorbates ($N_\theta = 0$), the velocity is proportional to the ion flux. The sputtering step has $k_j = 1$, and its rate is the yield-weighted ion flux, in which $Y$ of Eq.~(\ref{eq:yield}) is the number of atoms removed per incident ion. The yield prefactor $A$ is $0.02$\,eV$^{-1/2}$ on silicon and $0.002$\,eV$^{-1/2}$ on the mask, and the threshold energies are $20$ and $40$\,eV, respectively. These values are illustrative.

\begin{unbroken}
\begin{lstlisting}[style=code]
name: ar_physical_sputtering
temperature: 300.0        # K, irrelevant here: the step has no thermal barrier
species:
  Ar_ion: {phase: ion,   flux: 5.0e1,      # 1e15 /cm^2 /s
           meanEnergy: 100.0, sigmaEnergy: 10.0, exponent: 200.0}
  Si_s:   {phase: gas,   reemit_only: true,   # the sputtered atom leaves
           atoms: {Si: 1}}                    # 'Si' is the solid, so it needs a name
  Si:     {phase: solid, rho: 5.02}           # 1e22 atoms/cm^3
surface:
  - {eq: "Ar_ion + Si -> Si_s", type: sputtering,
     yield: {A: 0.02, Eth: 20.0,
             materials: {Mask: {A: 0.002, Eth: 40.0}}}}
\end{lstlisting}
\end{unbroken}

\begin{figure}[!htbp]
\centering
\includegraphics[width=0.62\textwidth]{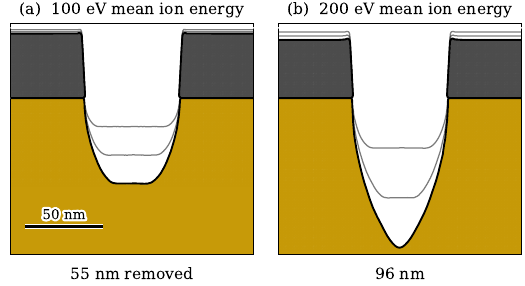}
\caption{Argon sputtering of silicon through a mask, $50$\,s in each panel. With no adsorbed species, the velocity is given by the ion flux and the yield of Eq.~(\ref{eq:yield}), and the profile is shaped by the angular dependence of the yield. Doubling the mean ion energy deepens the cavity by the factor that the threshold form predicts and reduces the selectivity, because the mask has the higher threshold. The panels show the first of three runs, and the depth under each is the mean of the three.}
\label{fig:case5}
\end{figure}

A $60$\,nm opening in a $45$\,nm mask is etched for $50$\,s. Figure~\ref{fig:case5}(a) shows the cavity produced by a threshold yield with
its angular dependence alone, $54.7$\,nm deep with a floor that rounds towards
the corners. The flat top of the mask loses $3.6$\,nm in the same time, a
selectivity of $15$ to $1$. The yield prefactors of silicon and mask alone would give $10$ to $1$, and with the higher threshold of the mask the yields of Eq.~(\ref{eq:yield}) at the mean ion energy give the same $15$ to $1$. In Fig.~\ref{fig:case5}(b), the mean ion energy is doubled, and the energy spread with it, so that both runs have a spread of one tenth of the mean. With all other inputs unchanged, the depth increases from $54.7$ to $95.8$\,nm, a factor of $1.75$. At the substrate threshold of $20$\,eV, Eq.~(\ref{eq:yield}) gives the same factor, $(\sqrt{200}-\sqrt{20})/(\sqrt{100}-\sqrt{20}) = 1.75$. The mask loses $7.7$\,nm against $3.6$\,nm at the lower energy, and the selectivity falls to $12$ to $1$, in agreement with the ratio of the yields at the mean ion energy. Both effects are a consequence of the threshold form, in which the square root of the
threshold is subtracted from that of the ion energy. With the larger threshold on the mask, the ratio of the two yields falls towards the ratio of the prefactors as the energy rises.

\section{Fluorocarbon etching of SiO$_2$, \texttt{polymer\_etch}}
\label{note:polymer}

In this model, a fluorocarbon polymer is deposited while SiO$_2$ is etched. For equipment and feature-scale models, Zhang and Kushner wrote mechanisms of this kind, in which the deposition of a passivating fluorocarbon film competes with etching~\cite{Zhang2000,Zhang2001}. The sign of each contribution to the velocity of Eq.~(\ref{eq:velocity}) is determined by the side of the reaction on which its solid appears. The sticking coefficients, the chemical etch and the ion yields are those of the fluorocarbon model of ViennaPS, which is based on the work of La Magna and Garozzo~\cite{LaMagna2003}. In that model, the adsorbed precursor is deposited directly as film. The polymer formation step of the file and its prefactor are therefore illustrative. The sticking coefficients are dimensionless probabilities per incident particle. For CF$_2$, the sticking coefficient is $0.01$ on the mask and $0.26$ elsewhere, and for F it is $0.9$ on the oxide, $0.1$ on the mask and zero elsewhere. The prefactors are $k_{0j} = 200$ for the polymer formation and $k_{0j} = 0.418$ for the chemical etch, which has a barrier of $0.168$\,eV. Both are rates per unit area at unit coverage in $10^{15}$\,cm$^{-2}$s$^{-1}$, and with the default density of both site types they are also rates per site in s$^{-1}$. Because the hand-written class evaluates the chemical etch with the etchant flux at the source, the prefactor of this step is the rate coefficient $K$ of the class times the etchant flux of $150 \times 10^{15}$\,cm$^{-2}$s$^{-1}$. The three ion-driven steps have $k_j = 1$
and the yield prefactors $0.0722$, $0.0361$ and $0.0139$\,eV$^{-1/2}$ with
thresholds of $4$, $4$ and $18$\,eV, for the removal of the adsorbed fragment, the ion-enhanced etch and the sputtering, respectively.

\begin{unbroken}
\begin{lstlisting}[style=code]
name: fluorocarbon_sio2_etch
temperature: 300.0        # K
species:
  CF2:    {phase: gas, flux: 5.0e0}        # polyFlux,    1e15 /cm2 /s
  F:      {phase: gas, flux: 1.5e2}        # etchantFlux, 1e15 /cm2 /s
  Ar_ion: {phase: ion, flux: 5.6e1,        # ionFlux,     1e15 /cm2 /s
           meanEnergy: 100.0, sigmaEnergy: 10.0, exponent: 100.0,
           inflectAngle: 89.0, n_l: 10.0, minAngle: 80.0}
  SiF2:   {phase: gas, reemit_only: true}
  O2:     {phase: gas, reemit_only: true}
  CF4:    {phase: gas, reemit_only: true}
  SiO2_s: {phase: gas, reemit_only: true, atoms: {Si: 1, O: 2}}
  F*:     {phase: surface, site: sub}      # etchant on the substrate
  CF2*:   {phase: surface, site: poly}     # adsorbed polymer fragment
  C:      {phase: solid, rho: 2.0}         # the polymer film
  SiO2:   {phase: solid, rho: {default: 2.2, Mask: 500.0}}
siteTypes:
  - {name: sub}
  - {name: poly}
\end{lstlisting}
\end{unbroken}

\begin{unbroken}
\begin{lstlisting}[style=code]
surface:
  - {eq: "CF2 -> CF2*", type: adsorption, s0: 0.26, Ea: 0.0,
     materials: {Mask: {s0: 0.01, Ea: 0.0}}}
  - {eq: "2CF2* -> C + CF4", type: surface, k0: 2.0e2, Ea: 0.0}
  - {eq: "Ar_ion + CF2* -> CF2", type: ion_enhanced,
     order: {CF2*: 1},
     yield: {A: 0.0722, Eth: 4.0}}
  - {eq: "F -> F*", type: adsorption, s0: 0.9, Ea: 0.0,
     materials: {default: 0.0, SiO2: 1.0, Mask: {s0: 0.1, Ea: 0.0}}}
  - {eq: "2F* + SiO2 -> SiF2 + O2", type: surface,
     k0: 0.41842375568174655, Ea: 0.168,
     order: {F*: 1},
     materials: {default: 0.0, SiO2: 1.0}}
  - {eq: "Ar_ion + 2F* + SiO2 -> SiF2 + O2", type: ion_enhanced,
     order: {F*: 1},
     yield: {A: 0.0361, Eth: 4.0,
             materials: {Mask: {A: 0.0, Eth: 4.0}}}}
  - {eq: "Ar_ion + SiO2 -> SiO2_s", type: sputtering,
     order: {"*sub": 1},
     yield: {A: 0.0139, Eth: 18.0, B: 9.3,
             materials: {Mask: {A: 0.0139, Eth: 20.0},
                         Polymer: {A: 0.0, Eth: 18.0}}}}
\end{lstlisting}
\end{unbroken}

\begin{figure}[!htbp]
\centering
\includegraphics[width=0.82\textwidth]{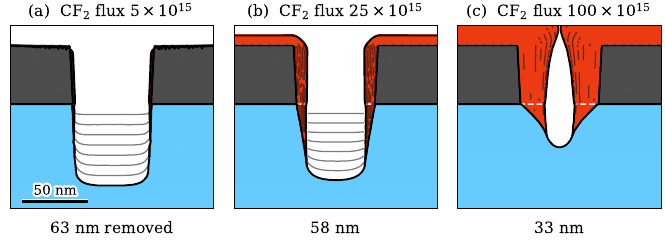}
\caption{SiO$_2$ etched through a mask while a fluorocarbon polymer is deposited, $10$\,s in each panel. The panel titles give the precursor flux $\Gamma_{\mathrm{CF}_2}$ in cm$^{-2}$s$^{-1}$. The mask is
grey, the oxide blue, the polymer red, and the thin lines are the surface after each eighth of the run. The velocity at a point is the sum of
the contributions of the two solids. The sidewall gains material while the floor loses it. At five times the precursor flux, the polymer covers the mask and the wall is tapered. At twenty times the precursor flux, the polymer covers the mask,
narrows the opening and tapers the wall, and the trench is about half as deep. A further increase of the flux reverses the sign of the open-field
velocity. The panels show the first of three runs, and the depth under each is the mean of the three.}
\label{fig:case7}
\end{figure}

A $60$\,nm opening in a $45$\,nm mask over SiO$_2$ is etched for $10$\,s with
the fluxes of the ViennaPS example, $5$, $150$ and $56 \times
10^{15}$\,cm$^{-2}$s$^{-1}$ of precursor, etchant and ions, respectively. In Fig.~\ref{fig:case7}(a), the sidewall gains polymer while the floor is etched, and both solids contribute to one velocity. The trench reaches $63$\,nm with a nearly vertical wall, while the mask hardly moves. Because the chemical and ion-enhanced etch steps are confined to the oxide, fluorine adsorbed on the mask is not consumed there. This fluorine fills the substrate sites that the sputtering step requires, and the mask is not sputtered. In Figs.~\ref{fig:case7}(b) and (c), the fluorocarbon flux is raised fivefold and twentyfold, to $25$ and $100 \times 10^{15}$\,cm$^{-2}$s$^{-1}$, respectively, with all other inputs unchanged. At five times the precursor flux, the polymer covers the mask and the upper part of the wall, the wall is tapered, and the trench reaches $58$\,nm. At the twentyfold flux, a thick film covers the mask and the upper sidewalls and narrows the mouth to a slit.
Ions still remove material from the floor. The trench below the slit becomes V-shaped, with a taper set by the balance between polymer deposition and oxide removal, and in the same time reaches $33$\,nm, compared with $63$\,nm at the lowest flux. At higher fluxes, the open-field velocity becomes positive near $170 \times
10^{15}$\,cm$^{-2}$s$^{-1}$, and at $200 \times 10^{15}$\,cm$^{-2}$s$^{-1}$ the net deposition rate is $0.98$\,nm/s.

\section{Atomic layer deposition of Al$_2$O$_3$, \texttt{al2o3}}
\label{note:al2o3}

The model is written as two files, \texttt{al2o3\_tma} and \texttt{al2o3\_h2o}, one for each half-cycle. The two files share the surface species and the two site types for the anion and cation positions of the lattice. An anion
position holds a bare bridging oxygen, a hydroxyl or an aluminium bonded to the
oxygen, and a cation position holds a bare, a methylated or a hydroxylated
lattice aluminium. TMA is written as the monomer Al(CH$_3$)$_3$, the species for which Travis and Adomaitis~\cite{Travis2014} formulated their kinetic model and Arts et al.~\cite{Arts2019} measured sticking probabilities. With the monomer, each adsorption event places one aluminium on the surface, as molecularly adsorbed TMA (adduct) on a hydroxyl or on a bare oxygen bridge. The ligand exchange, the dissociative reaction on a bare oxygen bridge and the water reaction are Eqs.~(15), (19) and (16) of Puurunen's review of this process~\cite{Puurunen2005}, respectively, with the activation energies of the first two from Travis and Adomaitis~\cite{Travis2014} and that of the water reaction from Sperling et al.~\cite{Sperling2020}. The condensation of two hydroxylated aluminium species into one formula unit of Al$_2$O$_3$ closes the cycle. The solid is Al$_2$O$_3$ at the formula-unit density of the amorphous oxide.

The steps with a gas reactant have dimensionless reaction probabilities per
incident molecule on a surface fully covered by the surface reactant. For TMA on hydroxyls, the probability is $10^{-3}$, the logarithmic midpoint of the range $(0.5$ to $2) \times 10^{-3}$ that Arts et al. extracted from thickness profiles in
lateral high-aspect-ratio structures at about $220\,^\circ$C~\cite{Arts2019}.
Since TMA adsorbs molecularly without a barrier, this value is also used at $450$\,K. The dissociative channel on bare oxygen bridges is given the same probability. For water on methyl groups, the probability is $s_0 \exp(-E_a/k_B T)$ with $E_a =
0.187$\,eV, the effective barrier that Sperling et al. find on a fully
methylated surface~\cite{Sperling2020}. The prefactor $s_0 = 3.4 \times 10^{-3}$ reproduces the probability of $2 \times 10^{-5}$ at $150\,^\circ$C that they take from earlier measurements~\cite{Sperling2020}. At $450$\,K, the probability is $2.7 \times 10^{-5}$. The three thermal steps proceed at the transition-state rate per site $(k_B T/h_{\mathrm{P}}) \exp(-E_{a,j}/k_B T)$, where $h_{\mathrm{P}}$ is the Planck constant. Since $k_j$ is a rate per unit area at unit coverage in $10^{15}$\,cm$^{-2}$s$^{-1}$, the prefactor in the files is $k_{0j} = (k_B/h_{\mathrm{P}})\, S_0/10^{15} \approx 2.084 \times 10^{10} \times 0.4 = 8.336 \times 10^{9}$ with $\beta_j = 1$, for the site density $S_0 = 4 \times 10^{14}$\,cm$^{-2}$ of both types. The barriers are $0.52$\,eV for the methane elimination from adsorbed TMA on a hydroxyl, $0.5$\,eV for the dissociation of adsorbed TMA on an oxygen bridge and $0.20$\,eV for the condensation. No published barrier is available for the condensation. At $0.20$\,eV, the step is fast compared with the water exposure, and any barrier up to $0.8$\,eV changes the growth per cycle by less than $0.01\,\%$.

The dissociative channel consumes a bare cation, and no step of either half-cycle consumes the hydroxylated cation. On a surface that starts with bare cations, as in the cavity runs, this channel therefore closes after the first cycle wherever the first dose
saturates the surface, and in later cycles the film grows by ligand exchange.

\begin{unbroken}
\begin{lstlisting}[style=code]
name: al2o3_ald_tma_dose
temperature: 450.0        # K, 177 C
energyUnit: eV

siteTypes:
  - {name: A, density: 4.0e14}      # /cm2, anion positions in the top layer
  - {name: C, density: 4.0e14}      # /cm2, cation positions
species:
  AlC3H9:   {phase: gas, pressure: 1.0e2}    # TMA monomer, mTorr
  CH4:      {phase: gas, reemit_only: true}
  H2O:      {phase: gas, reemit_only: true}  # released by the condensation
  OH:       {phase: surface, site: A}        # ||O-H
  O:        {phase: surface, site: A}        # ||O||, a bare bridging oxygen
  AlC3H10O: {phase: surface, site: A}        # ||O-H . Al(CH3)3, the R1 adduct
  AlC3H9O:  {phase: surface, site: A}        # ||O . Al(CH3)3, the R4 adduct
  AlC2H6O:  {phase: surface, site: A}        # ||O-Al(CH3)2
  AlCH4O2:  {phase: surface, site: A}        # ||O-Al(CH3)(OH)
  AlH2O3:   {phase: surface, site: A}        # ||O-Al(OH)2
  Al:       {phase: surface, site: C}        # a bare lattice aluminium
  AlCH3:    {phase: surface, site: C}        # ||Al-CH3
  AlHO:     {phase: surface, site: C}        # ||Al-OH
  Al2O3:    {phase: solid, rho: 1.77}        # 1e22 formula units /cm3

surface:
  - {eq: "AlC3H9 + OH -> AlC3H10O", type: adsorption, s0: 1.0e-3, Ea: 0.0}
  - {eq: "AlC3H10O -> AlC2H6O + CH4", type: surface,
     k0: 8.336e9, Ea: 0.52, beta: 1.0}
  - {eq: "AlC3H9 + O -> AlC3H9O", type: adsorption, s0: 1.0e-3, Ea: 0.0}
  - {eq: "AlC3H9O + Al -> AlC2H6O + AlCH3", type: surface,
     k0: 8.336e9, Ea: 0.5, beta: 1.0}
\end{lstlisting}
\end{unbroken}

\begin{unbroken}
\begin{lstlisting}[style=code]
name: al2o3_ald_h2o_dose
temperature: 450.0        # K, 177 C
energyUnit: eV

siteTypes:
  - {name: A, density: 4.0e14}      # /cm2
  - {name: C, density: 4.0e14}      # /cm2
species:
  H2O:      {phase: gas, pressure: 1.0e2}    # mTorr
  CH4:      {phase: gas, reemit_only: true}
  AlC3H9:   {phase: gas, reemit_only: true}  # shared with the dose
  OH:       {phase: surface, site: A}
  O:        {phase: surface, site: A}
  AlC3H10O: {phase: surface, site: A}
  AlC3H9O:  {phase: surface, site: A}
  AlC2H6O:  {phase: surface, site: A}
  AlCH4O2:  {phase: surface, site: A}
  AlH2O3:   {phase: surface, site: A}
  Al:       {phase: surface, site: C}
  AlCH3:    {phase: surface, site: C}
  AlHO:     {phase: surface, site: C}
  Al2O3:    {phase: solid, rho: 1.77}        # 1e22 formula units /cm3

surface:
  - {eq: "H2O + AlC2H6O -> AlCH4O2 + CH4", type: adsorption, s0: 3.4e-3, Ea: 0.187}
  - {eq: "H2O + AlCH4O2 -> AlH2O3 + CH4", type: adsorption, s0: 3.4e-3, Ea: 0.187}
  - {eq: "H2O + AlCH3 -> AlHO + CH4", type: adsorption, s0: 3.4e-3, Ea: 0.187}
  - {eq: "2 AlH2O3 -> Al2O3 + 2 OH + H2O", type: surface,
     k0: 8.336e9, Ea: 0.20, beta: 1.0}
\end{lstlisting}
\end{unbroken}

The two files are run as the four-stage cycle of Section~\ref{sec:setup} in the lateral cavity of Fig.~\ref{fig:pillarhall}(a), $0.5\,\upmu$m high and $100\,\upmu$m long. Both dimensions are within the range of the lateral high-aspect-ratio test structures of Gao et al.~\cite{Gao2015}, whose cavity height and length can be varied, for example from $200$ to $1000$\,nm and from $20$ to $5000\,\upmu$m. In these structures, a cavity of known height lies between
a silicon floor and a membrane supported by pillars. The membrane is peeled off after deposition, so that the film inside can be measured along its length. The thickness profile after a dose shows the conformality, and fitting a diffusion-reaction model to it, for instance the model of Ylilammi et al.~\cite{Ylilammi2018}, yields sticking coefficients~\cite{Arts2019,Cremers2019}.

\section{Symbols}
\label{app:symbols}

\setlength{\LTcapwidth}{\linewidth}
\begin{longtable}{@{}l>{\raggedright\arraybackslash}p{0.80\linewidth}@{}}
\caption{Symbols used in the text, in alphabetical order with Greek letters sorted by their names, followed by the indices and sets. A
symbol with more than one meaning is listed once per
meaning.}\label{tab:symbols}\\
\toprule
symbol & meaning \\
\midrule
\endfirsthead
\toprule
symbol & meaning \\
\midrule
\endhead
$A$        & prefactor of the ion yield, in eV$^{-1/2}$, Eq.~\eqref{eq:yield} \\
$A_j$      & the rate constant and gas factors of reaction $j$, Eq.~\eqref{eq:ratesplit} \\
$a$, $b$   & lengths of the dose and of the co-reactant pulse divided by the time constant $S_0/(s_0\Gamma)$, Appendix~\ref{note:solvers} \\
$B$        & angular parameter of the sputtering yield, which sets where it peaks and how sharply it rises, Eq.~\eqref{eq:angular} \\
$\beta_j$  & temperature exponent of a rate constant, Eq.~\eqref{eq:arrhenius} \\
$\bm{\delta}$ & Newton step in the coverages, Eq.~\eqref{eq:newton} \\
$E$        & energy of an incident ion, Eq.~\eqref{eq:yield} \\
$E_{\mathrm{th}}$ & threshold energy of an ion yield, Eq.~\eqref{eq:yield} \\
$E_{a,j}$  & activation energy of reaction $j$, Eq.~\eqref{eq:arrhenius} \\
$e$        & base of the natural logarithm, Eq.~\eqref{eq:diamondR} \\
$\epsilon$ & largest coverage change permitted in one sub-step, Eq.~\eqref{eq:substep} \\
$F_i$      & residual of the coverage balance of species $i$, Eq.~\eqref{eq:steady} \\
$f(\vartheta)$ & angular factor of an ion yield, Eqs.~\eqref{eq:yield} and \eqref{eq:angular} \\
$\Gamma_g$ & incident flux of gas species $g$, Eq.~\eqref{eq:massaction} \\
$\Gamma_{\mathrm{ie}}$, $\Gamma_{\mathrm{p}}$, $\Gamma_{\mathrm{sp}}$ & yield-weighted ion fluxes of the ion-enhanced etch, the removal of oxygen and the sputtering in the SF$_6$/O$_2$ model, Eq.~\eqref{eq:sf6o2balance} \\
$g_{ij}$   & gain weight of reaction $j$ on state $i$, Eq.~\eqref{eq:transfer} \\
$h$        & length of an integrator sub-step, Eqs.~\eqref{eq:patankar} and \eqref{eq:substep} \\
$h_{\mathrm{P}}$ & Planck constant, Appendix~\ref{note:al2o3} \\
$\mathbf{I}$ & identity matrix, Eq.~\eqref{eq:patankar} \\
$\mathbf{J}$, $J_{ik}$ & Jacobian of the residuals, Eq.~\eqref{eq:jacobian} \\
$k_j$      & rate constant of reaction $j$, Eqs.~\eqref{eq:massaction} and \eqref{eq:arrhenius}, with the unit of Eq.~\eqref{eq:kunits} \\
$k_{0j}$   & prefactor of that rate constant, Eq.~\eqref{eq:arrhenius} \\
$k_{\mathrm{ads}}$ & adsorption rate constant, Eq.~\eqref{eq:sticking} \\
$k_B$      & Boltzmann constant, Eq.~\eqref{eq:arrhenius} \\
$k^{(1)}$, $k^{(2)}$ & first- and second-order rate constant of a step among adsorbates, Eq.~\eqref{eq:kunits} \\
$L_k$      & total rate coefficient at which state $k$ is consumed, Eq.~\eqref{eq:patankar} \\
$l_{kj}$   & loss weight of reaction $j$ on state $k$, Eq.~\eqref{eq:transfer} \\
$\lambda$  & damping factor of the Newton step, Eqs.~\eqref{eq:newton} to \eqref{eq:dampsite} \\
$M$        & molar mass of a gas species, Eq.~\eqref{eq:sticking} \\
$N$        & number of rays per surface point, Fig.~\ref{fig:conv} \\
$N_A$      & Avogadro constant, Appendix~\ref{note:gaas} \\
$N_\theta$ & number of adsorbates, Eq.~\eqref{eq:steady} \\
$n$        & number density of a gas above the surface, Eq.~\eqref{eq:sticking} \\
$n_{ij}$, $n_{gj}$, $n_{\ast \alpha j}$ & order of reaction $j$ in a coverage, a gas and a free site, Eq.~\eqref{eq:massaction} \\
$n_{S,j}$  & atoms of solid $S$ produced per event of reaction $j$, Eq.~\eqref{eq:velocity} \\
$\Delta n_{\mathrm{cycle}}$ & distance a surface point advances along its normal in one cycle, Eq.~\eqref{eq:growthintegral} \\
$\nu_{ij}$ & stoichiometric coefficient of species $i$ in reaction $j$, Eq.~\eqref{eq:steady} \\
$\Phi_j$   & product of the free-site factors of reaction $j$, Eq.~\eqref{eq:ratesplit} \\
$\Pi$      & projection that clamps each coverage to $[0,1]$, Eq.~\eqref{eq:newton} \\
$\pi$      & the circle constant, Eq.~\eqref{eq:angular} \\
$q_{ik}$   & occupancy of state $k$ moving into state $i$ per unit of $k$ per unit time, Eq.~\eqref{eq:transfer} \\
$R$        & gas constant, Eq.~\eqref{eq:sticking} \\
$R$        & radical fraction of a diamond surface, Eq.~\eqref{eq:diamondR} \\
$r_j$      & rate of reaction $j$ per unit area and time, Eq.~\eqref{eq:massaction} \\
$\rho_{S}$ & number density of solid $S$, Eq.~\eqref{eq:velocity} \\
$S_0$      & areal density of the sites of one type, Eq.~\eqref{eq:closure} \\
$s_0$      & reactive sticking coefficient, Eq.~\eqref{eq:sticking} \\
$s_{\mathrm{eff},g}$ & probability that a neutral of species $g$ is absorbed at a surface hit, Eq.~\eqref{eq:reemission} \\
$\Sigma_\alpha$ & occupancy of site type $\alpha$, Eq.~\eqref{eq:dampsite} \\
$\sigma_\alpha$ & sum of the step components on site type $\alpha$, Eq.~\eqref{eq:dampsite} \\
$T$        & surface temperature, Eq.~\eqref{eq:arrhenius} \\
$t$        & time, Eq.~\eqref{eq:steady} \\
$\Delta t$ & interval over which the coverages are advanced, Eq.~\eqref{eq:substep} \\
$\theta_i$ & fractional coverage of surface species $i$, Eq.~\eqref{eq:massaction} \\
$\theta_{\ast \alpha}$ & free fraction of site type $\alpha$, Eq.~\eqref{eq:closure} \\
$\Theta_j$ & product of the coverage factors of reaction $j$, Eq.~\eqref{eq:ratesplit} \\
$\vartheta$ & angle of incidence of an ion, Eqs.~\eqref{eq:yield} and \eqref{eq:angular} \\
$\bar{v}$  & mean thermal speed of a gas species, Eq.~\eqref{eq:sticking} \\
$v_n$      & velocity of the surface along its normal, Eq.~\eqref{eq:velocity} \\
$x^\ast$   & fraction of the reactive termination at the start of a dose on the limit cycle, Appendix~\ref{note:solvers} \\
$Y$        & yield of an ion-driven reaction, Eq.~\eqref{eq:yield} \\
\midrule
\multicolumn{2}{@{}l}{\textit{Indices and sets}} \\
$i$, $k$   & surface species, and the states of one site type \\
$j$        & reaction \\
$g$        & gas species, Eqs.~\eqref{eq:massaction} and \eqref{eq:reemission} \\
$m$        & Newton iteration, Eq.~\eqref{eq:newton}, and a summation index, Eq.~\eqref{eq:transfer} \\
$\alpha$, $\tau$ & site type, Eqs.~\eqref{eq:closure} and \eqref{eq:drate} \\
$S$        & solid phase, Eq.~\eqref{eq:velocity} \\
$O$, $F$   & the oxygen and fluorine coverages of the SF$_6$/O$_2$ model, Eq.~\eqref{eq:sf6o2balance} \\
$\mathcal{A}$ & coverages that a full Newton step would make negative, Eq.~\eqref{eq:damping} \\
$\mathcal{R}_g$ & reactions in which gas species $g$ is the only gas reactant, Eq.~\eqref{eq:reemission} \\
\bottomrule
\end{longtable}

\section*{Code availability}

ViennaChem (\url{https://github.com/ViennaTools/ViennaChem}) contains the
reaction-file compiler, the symbolic formulation used to verify it and the
Python evaluator. The surface model is part of ViennaPS (\url{https://github.com/ViennaTools/ViennaPS}), branch \texttt{cpc-deposit}, based on the master branch of 18 September 2026. The two example programs and all reaction files are in its directory \texttt{examples/surfaceChemistry}, together with the compiled data and a reference test that solves every non-cyclic case at a single point and compares the result with stored output. ViennaPS uses ViennaRay 4.5.1. ViennaPS and ViennaRay are free software under the GNU General Public License, version 3, and ViennaChem under the same licence, version 3 or later. The versions used for this article, ViennaPS at commit \texttt{1a2dabde} and ViennaChem at commit \texttt{6d64b40}, are archived at TU Wien Research Data (\url{https://doi.org/10.48436/azegw-xq879})~\cite{FilipovicData2026}.

\section*{Data availability}

The reaction files, the data compiled from them and the stored output
of the reference test are in the example directory of ViennaPS (Code
availability). The scripts that run the simulations and produce all figures and computed numbers reported here, together with the output of every simulation and the figure data, are archived with the code at TU Wien Research Data (\url{https://doi.org/10.48436/azegw-xq879})~\cite{FilipovicData2026}.

\section*{CRediT authorship contribution statement}
\textbf{Lado Filipovic:} Conceptualization, Methodology, Software,
Validation, Formal analysis, Investigation, Visualization, Writing (original
draft), Writing (review and editing), Supervision, Project administration,
Funding acquisition.
\textbf{Tobias Reiter:} Software, Resources, Writing (review and editing).
\textbf{Philipp Haslhofer:} Methodology (atomic layer deposition), Writing (review and editing).

\section*{Declaration of competing interest}

The authors declare that they have no known competing financial interests or
personal relationships that could have appeared to influence the work reported
in this paper.

\section*{Funding}

This work was supported by the Christian Doppler Research Association, the
Austrian Federal Ministry of Labour and Economy and the National Foundation
for Research, Technology and Development. The funding sources had no
involvement in the study design, in the collection, analysis and
interpretation of the data, in the writing of the article or in the decision
to submit it for publication.

\section*{Declaration of generative AI and AI-assisted technologies in the manuscript preparation process}

During the preparation of this work, the authors used Claude (Anthropic) to
assist with drafting and language editing of the manuscript and with the
development of code. After using this tool, the authors reviewed and edited the
content as needed and take full responsibility for the content of the published
article.

\section*{Acknowledgements}

Financial support by the Federal Ministry of Labour and Economy, the National
Foundation for Research, Technology and Development, and the Christian Doppler
Research Association is gratefully acknowledged. The authors acknowledge TU
Wien Bibliothek for financial support through its Open Access Funding
Programme.


\begin{thebibliography}{99}

\bibitem{Osher1988} S.~Osher, J.A.~Sethian, J. Comput. Phys. 79 (1988) 12--49. \url{https://doi.org/10.1016/0021-9991(88)90002-2}.

\bibitem{Osher2003} S.~Osher, R.~Fedkiw, Level Set Methods and Dynamic Implicit Surfaces, Applied Mathematical Sciences, Springer, New York, 2003. \url{https://doi.org/10.1007/b98879}.

\bibitem{Cale1991} T.S.~Cale, T.H.~Gandy, G.B.~Raupp, J. Vac. Sci. Technol. A 9 (1991) 524--529. \url{https://doi.org/10.1116/1.577402}.

\bibitem{Reiter2025} T.~Reiter, L.~Filipovic, SoftwareX 32 (2025) 102453, \url{https://doi.org/10.1016/j.softx.2025.102453}.

\bibitem{BelenSF6O2} R.J.~Belen, S.~Gomez, D.~Cooperberg, M.~Kiehlbauch, E.S.~Aydil, J. Vac. Sci. Technol. A 23 (2005) 1430--1439. \url{https://doi.org/10.1116/1.2013317}.

\bibitem{Bobinac2023} J.~Bobinac, T.~Reiter, J.~Piso, X.~Klemenschits, O.~Baumgartner, Z.~Stanojevic, G.~Strof, M.~Karner, L.~Filipovic, Micromachines 14 (2023) 665, \url{https://doi.org/10.3390/mi14030665}.

\bibitem{Holzer2009} G.~Holzer, R.~Knechtel, S.~Breit, G.~Schr\"opfer, in: 2009 IEEE/SEMI Advanced Semiconductor Manufacturing Conference (ASMC), IEEE, 2009, pp.~200--205. \url{https://doi.org/10.1109/ASMC.2009.5155983}.

\bibitem{Hargrove2023} M.~Hargrove, S.~Wen, D.~Yim, K.~Egelhofer~Ruegger, P.~Nanja, S.~Sarkar, B.~Lowe, B.~Vincent, J.~Ervin, D.~Fried, J. Micro/Nanopatterning Mater. Metrol. 22 (2023) 031209, \url{https://doi.org/10.1117/1.JMM.22.3.031209}.

\bibitem{Sobieski2019} D.~Sobieski, R.~Wise, Y.~Pan, D.~Fried, J.~Yu, N.~Shamma, in: Advanced Etch Technology for Nanopatterning VIII, Proc. SPIE 10963, SPIE, 2019, 109630H. \url{https://doi.org/10.1117/12.2518553}.

\bibitem{Klemenschits2021} X.~Klemenschits, S.~Selberherr, L.~Filipovic, Comput. Methods Appl. Mech. Eng. 386 (2021) 114196, \url{https://doi.org/10.1016/j.cma.2021.114196}.

\bibitem{Kushner2009} M.J.~Kushner, J. Phys. D: Appl. Phys. 42 (2009) 194013, \url{https://doi.org/10.1088/0022-3727/42/19/194013}.

\bibitem{Hoekstra1997} R.J.~Hoekstra, M.J.~Grapperhaus, M.J.~Kushner, J. Vac. Sci. Technol. A 15 (1997) 1913--1921. \url{https://doi.org/10.1116/1.580659}.

\bibitem{Zhang2000} D.~Zhang, M.J.~Kushner, J. Appl. Phys. 87 (2000) 1060--1069. \url{https://doi.org/10.1063/1.371980}.

\bibitem{Zhang2001} D.~Zhang, M.J.~Kushner, J. Vac. Sci. Technol. A 19 (2001) 524--538. \url{https://doi.org/10.1116/1.1349728}.

\bibitem{Huang2019} S.~Huang, C.~Huard, S.~Shim, S.K.~Nam, I.-C.~Song, S.~Lu, M.J.~Kushner, J. Vac. Sci. Technol. A 37 (2019) 031304, \url{https://doi.org/10.1116/1.5090606}.

\bibitem{Huard2018} C.M.~Huard, S.~Sriraman, A.~Paterson, M.J.~Kushner, J. Vac. Sci. Technol. A 36 (2018) 06B101, \url{https://doi.org/10.1116/1.5049225}.

\bibitem{Belen2005} R.J.~Belen, S.~Gomez, M.~Kiehlbauch, D.~Cooperberg, E.S.~Aydil, J. Vac. Sci. Technol. A 23 (2005) 99--113. \url{https://doi.org/10.1116/1.1830495}.

\bibitem{Klemenschits2018} X.~Klemenschits, S.~Selberherr, L.~Filipovic, in: 2018 Joint International EUROSOI Workshop and International Conference on Ultimate Integration on Silicon (EUROSOI-ULIS), IEEE, 2018, pp.~1--4. \url{https://doi.org/10.1109/ULIS.2018.8354763}.

\bibitem{Aguinsky2022} L.F.~Aguinsky, F.~Rodrigues, G.~Wachter, M.~Trupke, U.~Schmid, A.~H\"ossinger, J.~Weinbub, Solid-State Electron. 191 (2022) 108262, \url{https://doi.org/10.1016/j.sse.2022.108262}.

\bibitem{Aguinsky2023} L.F.~Aguinsky, F.~Rodrigues, T.~Reiter, X.~Klemenschits, L.~Filipovic, A.~H\"ossinger, J.~Weinbub, Solid-State Electron. 201 (2023) 108584, \url{https://doi.org/10.1016/j.sse.2022.108584}.

\bibitem{Reiter2024} T.~Reiter, L.F.~Aguinsky, F.~Rodrigues, J.~Weinbub, A.~H\"ossinger, L.~Filipovic, Solid-State Electron. 211 (2024) 108816, \url{https://doi.org/10.1016/j.sse.2023.108816}.

\bibitem{Coltrin1991} M.E.~Coltrin, R.J.~Kee, F.M.~Rupley, Int. J. Chem. Kinet. 23 (1991) 1111--1128. \url{https://doi.org/10.1002/kin.550231205}.

\bibitem{Mountziaris1991} T.J.~Mountziaris, K.F.~Jensen, J. Electrochem. Soc. 138 (1991) 2426--2439. \url{https://doi.org/10.1149/1.2085990}.

\bibitem{Cantera2023} D.G.~Goodwin, H.K.~Moffat, I.~Schoegl, R.L.~Speth, B.W.~Weber, Cantera: An object-oriented software toolkit for chemical kinetics, thermodynamics, and transport processes, version 3.2.0 [software], Zenodo, 2025. \url{https://doi.org/10.5281/zenodo.17620923}.

\bibitem{Law1998} M.E.~Law, S.M.~Cea, Comput. Mater. Sci. 12 (1998) 289--308. \url{https://doi.org/10.1016/S0927-0256(98)00020-2}.

\bibitem{Labun2000} A.H.~Labun, H.K.~Moffat, T.S.~Cale, J. Vac. Sci. Technol. B 18 (2000) 267--278. \url{https://doi.org/10.1116/1.591182}.

\bibitem{Krueger2024} F.~Kr\"uger, D.~Zhang, P.~Luan, M.~Park, A.~Metz, M.J.~Kushner, J. Vac. Sci. Technol. A 42 (2024) 043008, \url{https://doi.org/10.1116/6.0003554}.

\bibitem{Lorensen1987} W.E.~Lorensen, H.E.~Cline, ACM SIGGRAPH Comput. Graph. 21 (1987) 163--169. \url{https://doi.org/10.1145/37402.37422}.

\bibitem{Kolasinski2012} K.W.~Kolasinski, Surface Science: Foundations of Catalysis and Nanoscience, third ed., Wiley, Chichester, 2012. \url{https://doi.org/10.1002/9781119941798}.

\bibitem{Steinbruchel1989} C.~Steinbr\"uchel, Appl. Phys. Lett. 55 (1989) 1960--1962. \url{https://doi.org/10.1063/1.102336}.

\bibitem{Belen2006} R.J.~Belen, S.~Gomez, M.~Kiehlbauch, E.S.~Aydil, J. Vac. Sci. Technol. A 24 (2006) 2176--2186. \url{https://doi.org/10.1116/1.2362725}.

\bibitem{Reiter2024IBE} T.~Reiter, A.~Toifl, S.W.~Kong, A.~H\"ossinger, L.~Filipovic, in: 2024 International Conference on Simulation of Semiconductor Processes and Devices (SISPAD), IEEE, 2024, pp.~1--4. \url{https://doi.org/10.1109/SISPAD62626.2024.10733316}.

\bibitem{Deuflhard2011} P.~Deuflhard, Newton Methods for Nonlinear Problems: Affine Invariance and Adaptive Algorithms, Springer, Berlin, 2011. \url{https://doi.org/10.1007/978-3-642-23899-4}.

\bibitem{Burchard2003} H.~Burchard, E.~Deleersnijder, A.~Meister, Appl. Numer. Math. 47 (2003) 1--30. \url{https://doi.org/10.1016/S0168-9274(03)00101-6}.

\bibitem{Meurer2017} A.~Meurer et al., PeerJ Comput. Sci. 3 (2017) e103, \url{https://doi.org/10.7717/peerj-cs.103}.

\bibitem{May2008} P.W.~May, Yu.A.~Mankelevich, J. Phys. Chem. C 112 (2008) 12432--12441. \url{https://doi.org/10.1021/jp803735a}.

\bibitem{LaMagna2003} A.~La~Magna, G.~Garozzo, J. Electrochem. Soc. 150 (2003) F178--F185. \url{https://doi.org/10.1149/1.1602084}.

\bibitem{Puurunen2005} R.L.~Puurunen, J. Appl. Phys. 97 (2005) 121301, \url{https://doi.org/10.1063/1.1940727}.

\bibitem{Travis2014} C.D.~Travis, R.A.~Adomaitis, Theor. Chem. Acc. 133 (2014) 1414, \url{https://doi.org/10.1007/s00214-013-1414-0}.

\bibitem{Sperling2020} B.A.~Sperling, B.~Kalanyan, J.E.~Maslar, J. Phys. Chem. C 124 (2020) 3410--3420. \url{https://doi.org/10.1021/acs.jpcc.9b11291}.

\bibitem{Coltrin1984} M.E.~Coltrin, R.J.~Kee, J.A.~Miller, J. Electrochem. Soc. 131 (1984) 425--434. \url{https://doi.org/10.1149/1.2115598}.

\bibitem{Coburn1979} J.W.~Coburn, H.F.~Winters, J. Appl. Phys. 50 (1979) 3189--3196. \url{https://doi.org/10.1063/1.326355}.

\bibitem{Arts2019} K.~Arts, V.~Vandalon, R.L.~Puurunen, M.~Utriainen, F.~Gao, W.M.M.~Kessels, H.C.M.~Knoops, J. Vac. Sci. Technol. A 37 (2019) 030908, \url{https://doi.org/10.1116/1.5093620}.

\bibitem{Gao2015} F.~Gao, S.~Arpiainen, R.L.~Puurunen, J. Vac. Sci. Technol. A 33 (2015) 010601, \url{https://doi.org/10.1116/1.4903941}.

\bibitem{Cremers2019} V.~Cremers, R.L.~Puurunen, J.~Dendooven, Appl. Phys. Rev. 6 (2019) 021302, \url{https://doi.org/10.1063/1.5060967}.

\bibitem{Ylilammi2018} M.~Ylilammi, O.M.E.~Ylivaara, R.L.~Puurunen, J. Appl. Phys. 123 (2018) 205301, \url{https://doi.org/10.1063/1.5028178}.

\bibitem{FilipovicData2026} L.~Filipovic, Code and data for ``Surface chemistry as data for level-set simulation of etching and deposition'' [dataset], TU Wien Research Data, 2026. \url{https://doi.org/10.48436/azegw-xq879}.

\end{thebibliography}
\end{document}